\documentclass{IEEEtran}

\usepackage[utf8]{inputenc}
\usepackage[T1]{fontenc}

\usepackage{amsmath}
\usepackage{amssymb}
\usepackage[linesnumbered]{algorithm2e}
\usepackage{booktabs}
\usepackage{dsfont}
\usepackage{fancyvrb}
\usepackage{manfnt}
\usepackage{pgfplots}
\usepackage[caption=no]{subfig}
\usepackage{tabularx}
\usepackage{tikz}

\usetikzlibrary{calc,fit,positioning}

\tikzset{flowbox/.style={draw,inner sep=2pt,align=center},
         flowtool/.style={inner sep=2pt,font=\footnotesize\itshape,align=center}}

\newcommand{\B}{\ensuremath{\mathds{B}}}
\newcommand{\N}{\ensuremath{\mathds{N}}}
\newcommand{\T}{\ensuremath{\mathrm{T}}}

\newcommand{\fanin}{\ensuremath{\operatorname{fanin}}}
\newcommand{\driver}{\ensuremath{\operatorname{driver}}}
\newcommand{\tfi}{\ensuremath{\operatorname{tfi}}}
\newcommand{\hexf}[1]{$^{\mathrm{\#}}${\fontfamily{cmtt}\selectfont#1}}

\newtheorem{lemma}{Lemma}
\newtheorem{example}{Example}

\SetKwInOut{Input}{Input}
\SetKwInOut{Output}{Output}
\SetKwProg{Fn}{function}{}{}

\def\hang{\hangindent19pt}
\def\d@anger{\medbreak\begingroup\clubpenalty=10000
 \def\par{\endgraf\endgroup\medbreak} \noindent\hang\hangafter=-2
 \hbox to0pt{\hskip-\hangindent\dbend\hfill}\small}
\outer\def\danger{\d@anger}

\pgfplotsset{
  /pgfplots/xlabel near ticks/.style={
     /pgfplots/every axis x label/.style={
        at={(ticklabel cs:0.5)},anchor=near ticklabel
     }
  },
  /pgfplots/ylabel near ticks/.style={
     /pgfplots/every axis y label/.style={
        at={(ticklabel cs:0.5)},rotate=90,anchor=near ticklabel}
     }
   }

\newsavebox{\qpicbbox}

\newcommand{\useqpicbox}{\input{\jobname.tikz}}

\newcommand{\subfloatqpicarg}[1]{}
{\VerbatimEnvironment
\renewcommand{\subfloatqpicarg}[1]{#1}
\begin{lrbox}{\qpicbbox}
\begin{VerbatimOut}{\jobname.qpic}}
{\end{VerbatimOut}
\immediate\write18{qpic \jobname.qpic --outfile \jobname.tikz}
\end{lrbox}
\subfloat[\subfloatqpicarg{}]{\useqpicbox}
}

\usetikzlibrary{positioning,shapes.geometric}

\newcommand{\tikzLUT}{%
\begin{tikzpicture}[scale=.3,font=\scriptsize]
  \begin{scope}[every node/.style={inner sep=.5pt}]
    \node (f0) at (260.0bp,306.0bp) [] {$y_1$};
    \node (f1) at (370.0bp,306.0bp) [] {$y_2$};
    \node (f2) at (609.0bp,306.0bp) [] {$y_3$};

    \node (x8) at (387.0bp,18.0bp) [] {$x_9$};
    \node (x9) at (531.0bp,18.0bp) [] {$x_{10}$};
    \node (x2) at (776.0bp,18.0bp) [] {$x_3$};
    \node (x3) at (171.0bp,18.0bp) [] {$x_4$};
    \node (x0) at (27.0bp,18.0bp) [] {$x_1$};
    \node (x1) at (700.0bp,18.0bp) [] {$x_2$};
    \node (x6) at (315.0bp,18.0bp) [] {$x_7$};
    \node (x7) at (459.0bp,18.0bp) [] {$x_8$};
    \node (x4) at (99.0bp,18.0bp) [] {$x_5$};
    \node (x5) at (243.0bp,18.0bp) [] {$x_6$};
    \node (x10) at (603.0bp,18.0bp) [] {$x_{11}$};
  \end{scope}

  \begin{scope}[every node/.style={draw,ellipse,inner sep=.5pt,minimum width=1.5em}]
    \node (24) at (679.0bp,162.0bp) [] {$10$};
    \node (25) at (609.0bp,234.0bp) [] {$13$};
    \node (26) at (480.0bp,162.0bp) [] {$9$};
    \node (27) at (459.0bp,90.0bp) [] {$5$};
    \node (20) at (569.0bp,90.0bp) [] {$6$};
    \node (21) at (222.0bp,162.0bp) [] {$7$};
    \node (22) at (169.0bp,90.0bp) [] {$1$};
    \node (23) at (370.0bp,234.0bp) [] {$12$};
    \node (15) at (260.0bp,234.0bp) [] {$11$};
    \node (17) at (243.0bp,90.0bp) [] {$2$};
    \node (16) at (370.0bp,162.0bp) [] {$8$};
    \node (19) at (315.0bp,90.0bp) [] {$3$};
    \node (18) at (387.0bp,90.0bp) [] {$4$};
  \end{scope}
  \draw [->] (x1) ..controls (693.89bp,60.33bp) and (687.31bp,104.79bp)  .. (24);
  \draw [->] (27) ..controls (466.48bp,115.94bp) and (469.4bp,125.67bp)  .. (26);
  \draw [->] (18) ..controls (403.55bp,125.31bp) and (413.66bp,155.08bp)  .. (406.0bp,180.0bp) .. controls (402.71bp,190.69bp) and (396.46bp,201.15bp)  .. (23);
  \draw [->] (x8) ..controls (411.75bp,43.062bp) and (425.52bp,56.454bp)  .. (27);
  \draw [->] (19) ..controls (318.89bp,126.35bp) and (323.83bp,156.17bp)  .. (334.0bp,180.0bp) .. controls (338.39bp,190.28bp) and (344.88bp,200.66bp)  .. (23);
  \draw [->] (x7) ..controls (413.4bp,41.168bp) and (373.28bp,60.669bp)  .. (19);
  \draw [->] (x6) ..controls (339.75bp,43.062bp) and (353.52bp,56.454bp)  .. (18);
  \draw [->] (20) ..controls (509.99bp,111.76bp) and (443.03bp,135.31bp)  .. (16);
  \draw [->] (x10) ..controls (591.1bp,43.509bp) and (586.11bp,53.768bp)  .. (20);
  \draw [->] (17) ..controls (268.89bp,103.97bp) and (274.06bp,106.2bp)  .. (279.0bp,108.0bp) .. controls (309.47bp,119.1bp) and (396.22bp,140.74bp)  .. (26);
  \draw [->] (x9) ..controls (506.25bp,43.062bp) and (492.48bp,56.454bp)  .. (27);
  \draw [->] (x8) ..controls (387.0bp,44.017bp) and (387.0bp,53.288bp)  .. (18);
  \draw [->] (x10) ..controls (624.89bp,59.899bp) and (649.97bp,106.76bp)  .. (24);
  \draw [->] (x6) ..controls (290.25bp,43.062bp) and (276.48bp,56.454bp)  .. (17);
  \draw [->] (22) ..controls (187.62bp,115.6bp) and (196.42bp,127.21bp)  .. (21);
  \draw [->] (17) ..controls (269.7bp,124.04bp) and (295.02bp,154.74bp)  .. (318.0bp,180.0bp) .. controls (327.58bp,190.53bp) and (338.58bp,201.85bp)  .. (23);
  \draw [->] (18) ..controls (418.52bp,114.73bp) and (438.42bp,129.7bp)  .. (26);
  \draw [->] (22) ..controls (165.4bp,128.46bp) and (166.49bp,161.71bp)  .. (186.0bp,180.0bp) .. controls (200.08bp,193.2bp) and (472.12bp,220.08bp)  .. (25);
  \draw [->] (16) ..controls (333.33bp,186.33bp) and (307.24bp,202.94bp)  .. (15);
  \draw [->] (x1) ..controls (660.56bp,67.546bp) and (592.12bp,144.37bp)  .. (516.0bp,180.0bp) .. controls (441.32bp,214.95bp) and (414.64bp,198.77bp)  .. (334.0bp,216.0bp) .. controls (321.21bp,218.73bp) and (307.22bp,221.9bp)  .. (15);
  \draw [->] (x0) ..controls (58.618bp,65.719bp) and (108.87bp,134.95bp)  .. (164.0bp,180.0bp) .. controls (183.64bp,196.05bp) and (208.83bp,209.74bp)  .. (15);
  \draw [->] (15) ..controls (260.0bp,260.02bp) and (260.0bp,269.29bp)  .. (f0);
  \draw [->] (18) ..controls (360.93bp,103.52bp) and (355.81bp,105.88bp)  .. (351.0bp,108.0bp) .. controls (318.39bp,122.39bp) and (280.63bp,137.75bp)  .. (21);
  \draw [->] (x4) ..controls (144.6bp,41.168bp) and (184.72bp,60.669bp)  .. (17);
  \draw [->] (25) ..controls (609.0bp,260.02bp) and (609.0bp,269.29bp)  .. (f2);
  \draw [->] (21) ..controls (235.42bp,187.72bp) and (241.16bp,198.29bp)  .. (15);
  \draw [->] (x3) ..controls (170.29bp,44.017bp) and (170.02bp,53.288bp)  .. (22);
  \draw [->] (x3) ..controls (195.75bp,43.062bp) and (209.52bp,56.454bp)  .. (17);
  \draw [->] (26) ..controls (521.76bp,185.66bp) and (555.4bp,203.91bp)  .. (25);
  \draw [->] (24) ..controls (651.05bp,175.53bp) and (644.34bp,178.08bp)  .. (638.0bp,180.0bp) .. controls (556.89bp,204.57bp) and (458.44bp,220.66bp)  .. (23);
  \draw [->] (19) ..controls (334.21bp,115.45bp) and (343.66bp,127.48bp)  .. (16);
  \draw [->] (x7) ..controls (459.0bp,44.017bp) and (459.0bp,53.288bp)  .. (27);
  \draw [->] (x2) ..controls (776.13bp,68.457bp) and (771.14bp,138.04bp)  .. (735.0bp,180.0bp) .. controls (712.03bp,206.66bp) and (673.4bp,220.22bp)  .. (25);
  \draw [->] (17) ..controls (235.52bp,115.94bp) and (232.6bp,125.67bp)  .. (21);
  \draw [->] (x5) ..controls (243.0bp,44.017bp) and (243.0bp,53.288bp)  .. (17);
  \draw [->] (x1) ..controls (720.73bp,66.513bp) and (743.34bp,134.15bp)  .. (715.0bp,180.0bp) .. controls (699.69bp,204.77bp) and (668.92bp,218.35bp)  .. (25);
  \draw [->] (17) ..controls (283.99bp,113.59bp) and (316.84bp,131.7bp)  .. (16);
  \draw [->] (x5) ..controls (288.6bp,41.168bp) and (328.72bp,60.669bp)  .. (18);
  \draw [->] (x7) ..controls (434.25bp,43.062bp) and (420.48bp,56.454bp)  .. (18);
  \draw [->] (23) ..controls (370.0bp,260.02bp) and (370.0bp,269.29bp)  .. (f1);
  \draw [->] (x9) ..controls (544.42bp,43.717bp) and (550.16bp,54.286bp)  .. (20);
  \draw [->] (x4) ..controls (124.87bp,32.017bp) and (130.05bp,34.233bp)  .. (135.0bp,36.0bp) .. controls (197.13bp,58.191bp) and (216.87bp,49.809bp)  .. (279.0bp,72.0bp) .. controls (280.86bp,72.663bp) and (282.74bp,73.388bp)  .. (19);
  \draw [->] (x10) ..controls (557.4bp,41.168bp) and (517.28bp,60.669bp)  .. (27);
  \draw [->] (x9) ..controls (521.73bp,55.169bp) and (517.86bp,86.906bp)  .. (533.0bp,108.0bp) .. controls (557.79bp,142.54bp) and (607.44bp,154.64bp)  .. (24);
  \draw [->] (x8) ..controls (362.25bp,43.062bp) and (348.48bp,56.454bp)  .. (19);
  \draw [->] (x3) ..controls (216.6bp,41.168bp) and (256.72bp,60.669bp)  .. (19);
  \draw [->] (18) ..controls (380.99bp,115.74bp) and (378.7bp,125.18bp)  .. (16);
  \draw [->] (x4) ..controls (123.27bp,43.265bp) and (136.31bp,56.31bp)  .. (22);
\end{tikzpicture}
}

\newcommand{\tikzExample}{%
\begin{tikzpicture}[scale=.25,font=\scriptsize]
  \begin{scope}[every node/.style={inner sep=.5pt}]
    \node (y2) at (243.0bp,306.0bp) [] {$y_2$};
    \node (y1) at (135.0bp,306.0bp) [] {$y_1$};
    \node (x2) at (171.0bp,18.0bp) [] {$x_3$};
    \node (x3) at (99.0bp,18.0bp) [] {$x_2$};
    \node (x1) at (27.0bp,18.0bp) [] {$x_1$};
    \node (x4) at (243.0bp,18.0bp) [] {$x_4$};
    \node (x5) at (315.0bp,18.0bp) [] {$x_5$};
  \end{scope}
  \begin{scope}[every node/.style={draw,ellipse,inner sep=.5pt,minimum width=1.5em}]
    \node (1) at (135.0bp,90.0bp) [] {$1$};
    \node (3) at (135.0bp,162.0bp) [] {$3$};
    \node (2) at (243.0bp,90.0bp) [] {$2$};
    \node (5) at (243.0bp,234.0bp) [] {$5$};
    \node (4) at (243.0bp,162.0bp) [] {$4$};
  \end{scope}
  \draw [->] (x3) ..controls (111.71bp,43.717bp) and (117.15bp,54.286bp)  .. (1);
  \draw [->] (x5) ..controls (290.25bp,43.062bp) and (276.48bp,56.454bp)  .. (2);
  \draw [->] (x1) ..controls (57.672bp,59.328bp) and (94.948bp,108.34bp)  .. (3);
  \draw [->] (x2) ..controls (158.29bp,43.717bp) and (152.85bp,54.286bp)  .. (1);
  \draw [->] (4) ..controls (243.0bp,188.02bp) and (243.0bp,197.29bp)  .. (5);
  \draw [->] (x5) ..controls (309.87bp,66.24bp) and (300.49bp,130.32bp)  .. (279.0bp,180.0bp) .. controls (274.56bp,190.26bp) and (268.06bp,200.63bp)  .. (5);
  \draw [->] (1) ..controls (171.0bp,114.33bp) and (196.62bp,130.94bp)  .. (4);
  \draw [->] (3) ..controls (135.0bp,204.33bp) and (135.0bp,248.79bp)  .. (y1);
  \draw [->] (x4) ..controls (243.0bp,44.017bp) and (243.0bp,53.288bp)  .. (2);
  \draw [->] (1) ..controls (135.0bp,116.02bp) and (135.0bp,125.29bp)  .. (3);
  \draw [->] (5) ..controls (243.0bp,260.02bp) and (243.0bp,269.29bp)  .. (y2);
  \draw [->] (2) ..controls (243.0bp,116.02bp) and (243.0bp,125.29bp)  .. (4);
\end{tikzpicture}
}

\title{Logic Synthesis for Quantum Computing}
\author{%
  Mathias Soeken, Martin Roetteler, Nathan Wiebe, and Giovanni De Micheli%
  \thanks{A preliminary version of this manuscript has been presented at the DAC 2017 conference~\cite{SRWM17b}.}
  \thanks{M.\ Soeken and G.\ De Micheli are with the Integrated Systems Laboratory, EPFL, Lausanne, Switzerland. M.\ Roetteler and N.\ Wiebe are with Microsoft Research, Redmond, USA.}
  \thanks{This research was supported by H2020-ERC-2014-ADG 669354 CyberCare, the Swiss National Science Foundation (200021-169084
    MAJesty), and the ICT COST Action IC1405.}
}

\begin{document}

\maketitle

\begin{abstract}
  Today's rapid advances in the physical implementation of quantum
  computers call for scalable synthesis methods to map practical logic
  designs to quantum architectures.  We present a synthesis framework
  to map logic networks into quantum circuits for quantum computing.
  The synthesis framework is based on LUT networks (lookup-table
  networks), which play a key role in state-of-the-art conventional
  logic synthesis.  Establishing a connection between LUTs in a LUT
  network and reversible single-target gates in a reversible network
  allows us to bridge conventional logic synthesis with logic
  synthesis for quantum computing---despite several fundamental
  differences.  As a result, our proposed synthesis framework directly
  benefits from the scientific achievements that were made in logic
  synthesis during the past decades.

  We call our synthesis framework \emph{LUT-based Hierarchical
    Reversible Logic Synthesis} (LHRS).  Input to LHRS is a classical
  logic network, e.g., represented as Verilog description; output is a
  quantum network (realized in terms of Clifford+$T$ gates, the most
  frequently used gate library in quantum computing).  The framework
  offers to trade-off the number of qubits for the number of quantum
  gates.  In a first step, an initial network is derived that only
  consists of single-target gates and already completely determines
  the number of qubits in the final quantum network.  Different
  methods are then used to map each single-target gate into
  Clifford+$T$ gates, while aiming at optimally using available
  resources.

  We demonstrate the effectiveness of our method in automatically
  synthesizing IEEE compliant floating point networks up to double
  precision.  As many quantum algorithms target scientific simulation
  applications, they can make rich use of floating point arithmetic
  components.  But due to the lack of quantum circuit descriptions for
  those components, it can be difficult to find a realistic cost
  estimation for the algorithms.  Our synthesized benchmarks provide
  cost estimates that allow quantum algorithm designers to provide the
  first complete cost estimates for a host of quantum algorithms.
  Thus, the benchmarks and, more generally, the LHRS framework are an
  essential step towards the goal of understanding which quantum
  algorithms will be practical in the first generations of quantum
  computers.
\end{abstract}

\section{Introduction}
\IEEEPARstart{R}{ecent} progress in fabrication makes the practical
application of quantum computers a tangible
prospect~\cite{DLF+16,LMR+17,MMS+16,MBK+16}. However, as quantum
computers scale up to tackle problems in computational chemistry,
machine learning, and cryptoanalysis, design automation will be
necessary to fully leverage the power of this emerging computational
model.

Quantum circuits differ significantly in comparison to classical
circuits.  This needs to be addressed by design automation tools:
\begin{enumerate}
\item Quantum computers process \emph{qubits} instead of classical
  bits.  A qubit can be in superposition and several qubits can be
  entangled.  We target purely Boolean functions as input to our
  synthesis algorithms.  At design state, it is sufficient to assume
  that all input values are Boolean, even though entangled qubits in
  superposition are eventually acted upon by the quantum hardware.
\item All operations on qubits besides measurement, called quantum
  gates, must be \emph{reversible}.  Gates with multiple fanout known
  from classical circuits are therefore not possible.  Temporarily
  computed values must be stored on additional helper qubits, called
  ancillae.  An intensive use of intermediate results therefore
  increases the qubit requirements of the resulting quantum circuit.
  Since qubits are a limited resource, the aim is to find circuits
  with a possibly small number of ancillae.  Quantum circuits that compute a
  purely Boolean function are often referred to as reversible networks.
\item The quantum gates that can be implemented by current quantum
  computers can act on a single or at most two qubits~\cite{LMR+17}.
  Something as simple as an AND operation can therefore not be
  expressed by a single quantum gate.  A universal fault-tolerant quantum gate
  library is the Clifford+$T$ gate set~\cite{LMR+17}.  In this gate set, the
  $T$ gate is sufficiently expensive in most approaches to fault
  tolerant quantum computing such that it is customary to neglect all
  other gates when costing a quantum circuit~\cite{AMMR13}.  Mapping
  reversible functions into networks that minimize $T$ gates is
  therefore a central challenge in quantum computing~\cite{Maslov16}.
\item When executing a quantum circuit on a quantum computer, all
  qubits must eventually hold either a primary input value, a primary output
  value, or a constant. A circuit should not expose  intermediate results to output lines as this can potentially destroy wanted interference effects, in particular if the circuit is used as a subroutine in a larger quantum computation. Qubits that nevertheless expose intermediate results are sometimes referred to as garbage outputs.
\end{enumerate}

It has recently been shown~\cite{Rawski15,SC16,SRWM17} that
hierarchical reversible logic synthesis methods based on logic network
representations are able to synthesize large arithmetic designs.  The
underlying idea is to map subnetworks into reversible networks.
\emph{Hierarchical} refers to the fact that intermediate results
computed by the subnetworks must be stored on additional ancilla
qubits.  If the subnetworks are small enough, one can locally apply
less efficient reversible synthesis methods that do not require
ancilla qubits and are based on Boolean satisfiability~\cite{GWDD09},
truth tables~\cite{MMD03}, or decision diagrams~\cite{SWD10}.
However, state-of-the-art hierarchical synthesis methods mainly suffer
from two disadvantages.  First, they do not explicitly
\emph{uncompute} the temporary values from the subnetworks and leave
garbage outputs.  In order to use the network in a quantum computer,
one can apply a technique called ``Bennett trick''~\cite{Bennett73},
which requires to double the number of gates and add one further
ancilla for each primary output.  Second, current algorithms do not
offer satisfying solutions to trade the number of qubits for the
number of $T$ gates.  In contrast, many algorithms optimize towards
the direction of one extreme~\cite{SRWM17}, i.e., the number of qubits
is very small for the cost of a very high number of $T$ gates or vice
versa.

This paper presents a hierarchical synthesis framework based on
$k$-feasible Boolean logic networks, which find use in conventional
logic synthesis.  These are logic networks in which every gate has at
most $k$ inputs.  They are often referred to as $k$-LUT (lookup table)
networks.  We show that there is a one-to-one correspondence between a
$k$-input LUT in a logic network and a reversible single-target gate
with $k$ control lines in a reversible network.  A single-target gate
has a $k$-input control function and a single target line that is
inverted if and only if the control function evaluates to 1.  The
initial reversible network with single-target gates can be derived
quickly and provides a skeleton for subsequent synthesis that already
fixes the number of qubits in the final quantum network.  As a second
step, each single-target gate is mapped into a Clifford+$T$ network.
We propose different methods for the mapping.  A direct method makes
use of the exclusive-sum-of-product (ESOP) representation of the
control function that can be directly translated into
multiple-controlled Toffoli gates~\cite{FTR07}.  Multiple-controlled
Toffoli gates are a specialization of single-target gates for which
automated translations into Clifford+$T$ networks exist.  Another
method tries to remap a single-target gate into a LUT network with
fewer number of inputs in the LUTs, by making use of temporarily
unused qubits in the overall quantum network.  
We show that
near-optimal Clifford+$T$ networks can be precomputed and stored in a
database if such LUT networks require sufficiently few gates. 

The presented LHRS algorithm is evaluated both on academic and industrial benchmarks.
On the academic EPFL arithmetic benchmarks, we show how the various
parameters effect the number of qubits and the number of $T$ gates in
the final quantum network as well as the algorithm's runtime.  We also
used the algorithm to find quantum networks for several industrial
floating point arithmetic networks up to double precision.  From these
networks we can derive cost estimates for their use in quantum
algorithms.  This has been a missing information in many proposed
algorithms, and arithmetic computation has often not been explicitly
taken into account.  Our cost estimates show that this is misleading
as for some algorithms the arithmetic computation accounts for the
dominant cost.

Quantum programming frameworks such as LIQ$Ui|\rangle$ \cite{WS14} or ProjectQ \cite{HST:2016} can link in the Clifford$+T$ circuits that are automatically generated by LHRS. 

The paper is structured as follows.  The next section introduces
definitions and notations.  Section~\ref{sec:outline} provides the
problem definition and gives a coarse outline of the algorithm,
separating it into two steps: synthesizing the mapping, described in
Sect.~\ref{sec:synth-mapping} and mapping single-target gates,
described in Sect.~\ref{sec:synth-gates}.  Section~\ref{sec:results}
discusses the results of the experimental evaluation and
Sect.~\ref{sec:conclusion} concludes.

\section{Preliminaries}
\label{sec:prelims}

\subsection{Some Notation}
\label{sec:notation}
A digraph $G = (V, A)$ is called \emph{simple}, if
$A \subseteq V \times V$, i.e., there can be at most one arc between
two vertices for each direction.  An acyclic digraph is called a
\emph{dag}.  We refer to $d^-(v) = \#\{w \mid (w, v) \in A\}$ and
$d^+(v) = \#\{w \mid (v, w) \in A\}$ as \emph{in-degree} and
\emph{out-degree} of $v$, respectively.

\subsection{Boolean Logic Networks}
\label{sec:networks}
A Boolean \emph{logic network} is a simple dag whose vertices are
primary inputs, primary outputs, and gates and whose arcs connect
gates to inputs, outputs, and other gates.  Formally, a Boolean logic
network $N=(V,A,F)$ consists of a simple dag $(V,A)$ and a function
mapping $F$.  It has vertices $V = X \cup Y \cup G$ for primary inputs
$X$, primary outputs $Y$, and gates $G$.  We have $d^-(x) = 0$ for all
$x \in X$ and $d^+(y) = 0$ for all $y\in Y$.  Arcs
$A \subseteq (X \cup G \times G \cup Y)$ connect primary inputs and
gates to other gates and primary outputs.  Each gate $g \in G$
realizes a Boolean function $F(g) : \B^{d^-(g)} \to \B$, i.e., the
number of inputs in $F(g)$ coincides with the number of ingoing arcs
of $g$.
\begin{figure}[t]
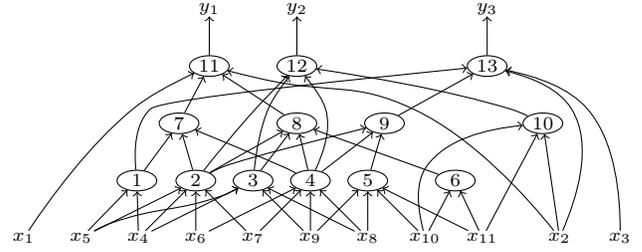

  \centering
  \tikzLUT
  \caption{A 4-feasible network with 11 inputs, 3 outputs, and 13 gates.}
  \label{fig:lut}
\end{figure}

\begin{example}
  Fig.~\ref{fig:lut} shows a logic network of the benchmark
  \emph{cm85a} obtained using ABC~\cite{BM10}.  It has 11 inputs, 3
  outputs, and 13 gates.  The gate functions are not shown but it can
  easily be checked that each gate has at most 4 inputs.
\end{example}

The \emph{fanin} of a gate or output $v \in G \cup Y$, denoted
$\fanin(v)$, is the set of source vertices of ingoing arcs:
\begin{equation}
  \label{eq:fanin}
  \fanin(v) = \{w \mid (w,v) \in A\}
\end{equation}
For a gate $g \in G$, this set is ordered according to the position of
variables in $F(g)$.  For a primary output $y \in Y$, we have
$d^-(y) = 1$, i.e., $\fanin(y) = \{v\}$ for some $v \in X \cup G$.
The vertex $v$ is called \emph{driver} of $y$ and we introduce the
notation $\driver(y) = v$.  The \emph{transitive fan-in} of a vertex
$v \in V$, denoted $\tfi(v)$, is the set containing $v$ itself, all
primary inputs that can be reached from $v$, and all gates which are
on any path from $v$ to the primary inputs.  The transitive fan-in can
be constructed using the following recursive definition:
\begin{equation}
  \label{eq:tfi}
  \tfi(v) =
  \begin{cases}
    \{v\} & \text{if $v \in X$,} \\
    \{v\} \cup \bigcup\limits_{w \in \fanin(v)}\tfi(w) & \text{otherwise.}
  \end{cases}
\end{equation}
\begin{example}
  The transitive fan-in of output $y_3$ in the logic network in
  Fig.~\ref{fig:lut} contains
  $\{y_3, 1, 2, 4, 5, 9, 13, x_2, x_3, x_4, x_5, x_6, x_7, x_8, x_9,
  x_{10}, x_{11}\}$.  The driver of $y_3$ is gate $13$.
\end{example}

We call a network \emph{$k$-feasible} if $d^-(g) \le k$ for all
$g \in G$.  Sometimes $k$-feasible networks are referred to as $k$-LUT
networks (LUT is a shorthand for lookup-table) and LUT mapping (see,
e.g.,~\cite{CD94,CC04,MCCB07,MCB07,RME+12}) refers to a family of
algorithms that obtain $k$-feasible networks, e.g., from homogeneous
logic representations such as And-inverter graphs
(AIGs,~\cite{KPKG02}) or Majority-inverter graphs
(MIGs,~\cite{AGM16}).
\begin{example}
  The logic network in Fig.~\ref{fig:lut} is 4-feasible.
\end{example}

\subsection{Reversible Logic Networks}
\label{sec:reversible-networks}
A reversible logic network realizes a reversible function, which makes
it very different from conventional logic networks.  The number of
lines, which correspond to logical qubits, remains the same for the
whole network, such that reversible networks are cascades of
reversible gates and each gate is applied to the current qubit
assignment.  The most general reversible gate we consider in this
paper is a \emph{single-target gate}.  A single-target gate
$\T_c(\{x_1, \dots, x_k\}, x_{k+1})$ has an ordered set of
\emph{control lines} $\{x_1, \dots, x_k\}$, a \emph{target line}
$x_{k+1}$, and a \emph{control function} $c : \B^k \to \B$.  It
realizes the reversible function $f : \B^{k+1} \to \B^{k+1}$ with
$f : x_i \mapsto x_i$ for $i \le k$ and
$f : x_{k+1} \mapsto x_{k+1} \oplus c(x_1, \dots, x_k)$. It is known that all
reversible functions can be realized by cascades of single-target
gates~\cite{VR08}.  We use the `$\circ$' operator for concatenation of
gates.

\begin{figure}[t]
  \footnotesize
  \centering
  \subfloat[]{%
\begin{tikzpicture}[scale=0.900000,x=1pt,y=1pt]
\filldraw[color=white] (0.000000, -6.500000) rectangle (32.000000, 58.500000);
\draw[color=black] (0.000000,52.000000) -- (32.000000,52.000000);
\draw[color=black] (0.000000,52.000000) node[left] {$a$};
\draw[color=black] (0.000000,39.000000) -- (32.000000,39.000000);
\draw[color=black] (0.000000,39.000000) node[left] {$b$};
\draw[color=black] (0.000000,26.000000) -- (32.000000,26.000000);
\draw[color=black] (0.000000,26.000000) node[left] {$c$};
\draw[color=black] (0.000000,13.000000) -- (32.000000,13.000000);
\draw[color=black] (0.000000,13.000000) node[left] {$0$};
\draw[color=black] (0.000000,0.000000) -- (32.000000,0.000000);
\draw[color=black] (0.000000,0.000000) node[left] {$0$};
\draw (8.000000,52.000000) -- (8.000000,13.000000);
\begin{scope}[rounded corners=2pt]
\begin{scope}
\draw[fill=white] (8.000000, 39.000000) +(-45.000000:8.485281pt and 26.870058pt) -- +(45.000000:8.485281pt and 26.870058pt) -- +(135.000000:8.485281pt and 26.870058pt) -- +(225.000000:8.485281pt and 26.870058pt) -- cycle;
\clip (8.000000, 39.000000) +(-45.000000:8.485281pt and 26.870058pt) -- +(45.000000:8.485281pt and 26.870058pt) -- +(135.000000:8.485281pt and 26.870058pt) -- +(225.000000:8.485281pt and 26.870058pt) -- cycle;
\draw (8.000000, 39.000000) node {{\rotatebox{-90}{$a{\oplus}b{\oplus}c$}}};
\end{scope}
\end{scope}
\begin{scope}
\draw[fill=white] (8.000000, 13.000000) circle(3.000000pt);
\clip (8.000000, 13.000000) circle(3.000000pt);
\draw (5.000000, 13.000000) -- (11.000000, 13.000000);
\draw (8.000000, 10.000000) -- (8.000000, 16.000000);
\end{scope}
\draw (24.000000,52.000000) -- (24.000000,0.000000);
\begin{scope}[rounded corners=2pt]
\begin{scope}
\draw[fill=white] (24.000000, 39.000000) +(-45.000000:8.485281pt and 26.870058pt) -- +(45.000000:8.485281pt and 26.870058pt) -- +(135.000000:8.485281pt and 26.870058pt) -- +(225.000000:8.485281pt and 26.870058pt) -- cycle;
\clip (24.000000, 39.000000) +(-45.000000:8.485281pt and 26.870058pt) -- +(45.000000:8.485281pt and 26.870058pt) -- +(135.000000:8.485281pt and 26.870058pt) -- +(225.000000:8.485281pt and 26.870058pt) -- cycle;
\draw (24.000000, 39.000000) node {{\rotatebox{-90}{$\langle abc\rangle$}}};
\end{scope}
\end{scope}
\begin{scope}
\draw[fill=white] (24.000000, 0.000000) circle(3.000000pt);
\clip (24.000000, 0.000000) circle(3.000000pt);
\draw (21.000000, 0.000000) -- (27.000000, 0.000000);
\draw (24.000000, -3.000000) -- (24.000000, 3.000000);
\end{scope}
\draw[color=black] (32.000000,52.000000) node[right] {$a$};
\draw[color=black] (32.000000,39.000000) node[right] {$b$};
\draw[color=black] (32.000000,26.000000) node[right] {$c$};
\draw[color=black] (32.000000,13.000000) node[right] {$s$};
\draw[color=black] (32.000000,0.000000) node[right] {${c'}$};
\end{tikzpicture}}
%
%
  \hfil
  \subfloat[]{%
\begin{tikzpicture}[scale=0.900000,x=1pt,y=1pt]
\filldraw[color=white] (0.000000, -6.500000) rectangle (56.000000, 58.500000);
\draw[color=black] (0.000000,52.000000) -- (56.000000,52.000000);
\draw[color=black] (0.000000,52.000000) node[left] {$a$};
\draw[color=black] (0.000000,39.000000) -- (56.000000,39.000000);
\draw[color=black] (0.000000,39.000000) node[left] {$b$};
\draw[color=black] (0.000000,26.000000) -- (56.000000,26.000000);
\draw[color=black] (0.000000,26.000000) node[left] {$c$};
\draw[color=black] (0.000000,13.000000) -- (56.000000,13.000000);
\draw[color=black] (0.000000,13.000000) node[left] {$0$};
\draw[color=black] (0.000000,0.000000) -- (56.000000,0.000000);
\draw[color=black] (0.000000,0.000000) node[left] {$0$};
\draw (5.000000,52.000000) -- (5.000000,13.000000);
\filldraw (5.000000, 52.000000) circle(1.500000pt);
\begin{scope}
\draw[fill=white] (5.000000, 13.000000) circle(3.000000pt);
\clip (5.000000, 13.000000) circle(3.000000pt);
\draw (2.000000, 13.000000) -- (8.000000, 13.000000);
\draw (5.000000, 10.000000) -- (5.000000, 16.000000);
\end{scope}
\draw (15.000000,39.000000) -- (15.000000,13.000000);
\filldraw (15.000000, 39.000000) circle(1.500000pt);
\begin{scope}
\draw[fill=white] (15.000000, 13.000000) circle(3.000000pt);
\clip (15.000000, 13.000000) circle(3.000000pt);
\draw (12.000000, 13.000000) -- (18.000000, 13.000000);
\draw (15.000000, 10.000000) -- (15.000000, 16.000000);
\end{scope}
\draw (25.000000,26.000000) -- (25.000000,13.000000);
\filldraw (25.000000, 26.000000) circle(1.500000pt);
\begin{scope}
\draw[fill=white] (25.000000, 13.000000) circle(3.000000pt);
\clip (25.000000, 13.000000) circle(3.000000pt);
\draw (22.000000, 13.000000) -- (28.000000, 13.000000);
\draw (25.000000, 10.000000) -- (25.000000, 16.000000);
\end{scope}
\draw (31.000000,52.000000) -- (31.000000,0.000000);
\draw[fill=white] (31.000000, 52.000000) circle(2.250000pt);
\filldraw (31.000000, 39.000000) circle(1.500000pt);
\begin{scope}
\draw[fill=white] (31.000000, 0.000000) circle(3.000000pt);
\clip (31.000000, 0.000000) circle(3.000000pt);
\draw (28.000000, 0.000000) -- (34.000000, 0.000000);
\draw (31.000000, -3.000000) -- (31.000000, 3.000000);
\end{scope}
\draw (41.000000,52.000000) -- (41.000000,0.000000);
\filldraw (41.000000, 52.000000) circle(1.500000pt);
\filldraw (41.000000, 26.000000) circle(1.500000pt);
\begin{scope}
\draw[fill=white] (41.000000, 0.000000) circle(3.000000pt);
\clip (41.000000, 0.000000) circle(3.000000pt);
\draw (38.000000, 0.000000) -- (44.000000, 0.000000);
\draw (41.000000, -3.000000) -- (41.000000, 3.000000);
\end{scope}
\draw (51.000000,39.000000) -- (51.000000,0.000000);
\filldraw (51.000000, 39.000000) circle(1.500000pt);
\draw[fill=white] (51.000000, 26.000000) circle(2.250000pt);
\begin{scope}
\draw[fill=white] (51.000000, 0.000000) circle(3.000000pt);
\clip (51.000000, 0.000000) circle(3.000000pt);
\draw (48.000000, 0.000000) -- (54.000000, 0.000000);
\draw (51.000000, -3.000000) -- (51.000000, 3.000000);
\end{scope}
\draw[color=black] (56.000000,52.000000) node[right] {$a$};
\draw[color=black] (56.000000,39.000000) node[right] {$b$};
\draw[color=black] (56.000000,26.000000) node[right] {$c$};
\draw[color=black] (56.000000,13.000000) node[right] {$s$};
\draw[color=black] (56.000000,0.000000) node[right] {${c'}$};
\end{tikzpicture}}
%
%
  \hfil
  \subfloat[]{%
\begin{tikzpicture}[scale=0.900000,x=1pt,y=1pt]
\filldraw[color=white] (0.000000, -6.500000) rectangle (70.000000, 58.500000);
\draw[color=black] (0.000000,52.000000) -- (70.000000,52.000000);
\draw[color=black] (0.000000,52.000000) node[left] {$a$};
\draw[color=black] (0.000000,39.000000) -- (70.000000,39.000000);
\draw[color=black] (0.000000,39.000000) node[left] {$b$};
\draw[color=black] (0.000000,26.000000) -- (70.000000,26.000000);
\draw[color=black] (0.000000,26.000000) node[left] {$c$};
\draw[color=black] (0.000000,13.000000) -- (70.000000,13.000000);
\draw[color=black] (0.000000,13.000000) node[left] {$0$};
\draw[color=black] (0.000000,0.000000) -- (70.000000,0.000000);
\draw[color=black] (0.000000,0.000000) node[left] {$0$};
\draw (5.000000,52.000000) -- (5.000000,39.000000);
\filldraw (5.000000, 52.000000) circle(1.500000pt);
\begin{scope}
\draw[fill=white] (5.000000, 39.000000) circle(3.000000pt);
\clip (5.000000, 39.000000) circle(3.000000pt);
\draw (2.000000, 39.000000) -- (8.000000, 39.000000);
\draw (5.000000, 36.000000) -- (5.000000, 42.000000);
\end{scope}
\draw (15.000000,52.000000) -- (15.000000,13.000000);
\filldraw (15.000000, 52.000000) circle(1.500000pt);
\begin{scope}
\draw[fill=white] (15.000000, 13.000000) circle(3.000000pt);
\clip (15.000000, 13.000000) circle(3.000000pt);
\draw (12.000000, 13.000000) -- (18.000000, 13.000000);
\draw (15.000000, 10.000000) -- (15.000000, 16.000000);
\end{scope}
\draw (25.000000,26.000000) -- (25.000000,13.000000);
\filldraw (25.000000, 26.000000) circle(1.500000pt);
\begin{scope}
\draw[fill=white] (25.000000, 13.000000) circle(3.000000pt);
\clip (25.000000, 13.000000) circle(3.000000pt);
\draw (22.000000, 13.000000) -- (28.000000, 13.000000);
\draw (25.000000, 10.000000) -- (25.000000, 16.000000);
\end{scope}
\draw (35.000000,39.000000) -- (35.000000,0.000000);
\filldraw (35.000000, 39.000000) circle(1.500000pt);
\filldraw (35.000000, 13.000000) circle(1.500000pt);
\begin{scope}
\draw[fill=white] (35.000000, 0.000000) circle(3.000000pt);
\clip (35.000000, 0.000000) circle(3.000000pt);
\draw (32.000000, 0.000000) -- (38.000000, 0.000000);
\draw (35.000000, -3.000000) -- (35.000000, 3.000000);
\end{scope}
\draw (45.000000,52.000000) -- (45.000000,0.000000);
\filldraw (45.000000, 52.000000) circle(1.500000pt);
\begin{scope}
\draw[fill=white] (45.000000, 0.000000) circle(3.000000pt);
\clip (45.000000, 0.000000) circle(3.000000pt);
\draw (42.000000, 0.000000) -- (48.000000, 0.000000);
\draw (45.000000, -3.000000) -- (45.000000, 3.000000);
\end{scope}
\draw (55.000000,52.000000) -- (55.000000,39.000000);
\filldraw (55.000000, 52.000000) circle(1.500000pt);
\begin{scope}
\draw[fill=white] (55.000000, 39.000000) circle(3.000000pt);
\clip (55.000000, 39.000000) circle(3.000000pt);
\draw (52.000000, 39.000000) -- (58.000000, 39.000000);
\draw (55.000000, 36.000000) -- (55.000000, 42.000000);
\end{scope}
\draw (65.000000,39.000000) -- (65.000000,13.000000);
\filldraw (65.000000, 39.000000) circle(1.500000pt);
\begin{scope}
\draw[fill=white] (65.000000, 13.000000) circle(3.000000pt);
\clip (65.000000, 13.000000) circle(3.000000pt);
\draw (62.000000, 13.000000) -- (68.000000, 13.000000);
\draw (65.000000, 10.000000) -- (65.000000, 16.000000);
\end{scope}
\draw[color=black] (70.000000,52.000000) node[right] {$a$};
\draw[color=black] (70.000000,39.000000) node[right] {$b$};
\draw[color=black] (70.000000,26.000000) node[right] {$c$};
\draw[color=black] (70.000000,13.000000) node[right] {$s$};
\draw[color=black] (70.000000,0.000000) node[right] {${c'}$};
\end{tikzpicture}}
%
  \caption{Reversible circuit for a full adder using (a) 2 single-target gates, (b) 3 Toffoli gates and 3 CNOT gates, and (c) 1 Toffoli gate and 6 CNOT gates.}
  \label{fig:fadd}
\end{figure}
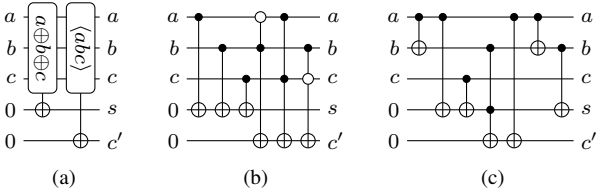

\begin{example}
  Fig.~\ref{fig:fadd}(a) shows a reversible circuit that realizes a full adder
  using two single-target gates, one for each output.  Two additional lines,
  called \emph{ancilla} and initialized with 0, are added to the network to
  store the result of the outputs.  All inputs are kept as output.
\end{example}

A \emph{multiple-controlled Toffoli gate} is a single-target gate in
which the control function is $1$ (tautology) or can be expressed in
terms of a single product term.  One can always decompose a
single-target gate $\T_c(\{x_1, \dots, x_k\}, x_{k+1})$ into a cascade
of Toffoli gates
\begin{equation}
  \label{eq:esop}
  \T_{c_1}(X_1, x_{k+1}) \circ \T_{c_2}(X_2, x_{k+1}) \circ \cdots \circ \T_{c_l}(X_l, x_{k+1}),
\end{equation}
where $c = c_1 \oplus c_2 \oplus \cdots \oplus c_l$, each $c_i$ is a
product term or $1$, and $X_i \subseteq \{x_1, \dots, x_k\}$ is the
support of $c_i$.  This decomposition of $c$ is also referred to as
ESOP decomposition~\cite{BDD73,SDP04,MP01}.  ESOP minimization
algorithms try to reduce $l$, i.e., the number of product terms in the
ESOP expression.  Smaller ESOP expressions lead to fewer
multiple-controlled Toffoli gates in the decomposition of a
single-target gate.  If $c = 1$, we refer to
$\T_c(\emptyset, x_{k+1})$ as NOT gate, and if $c = x_i$, we refer to
$\T_c(\{x_i\}, x_{k+1})$ as CNOT gate.

\begin{example}
  Fig.~\ref{fig:fadd}(b) shows the full adder circuit from the
  previous example in terms of Toffoli gates.  Each single-target gate
  is expressed in terms of $3$ Toffoli gates.  Positive and negative
  control lines of the Toffoli gates are drawn as solid and white
  dots, respectively.  Fig.~\ref{fig:fadd}(c) shows an alternative
  realization of the same output function, albeit with $1$ Toffoli
  gate.
\end{example}

\subsection{Mapping to Quantum Networks}
\label{sec:mapping}
Quantum networks are described in terms of a small library of gates
that interact with one or two qubits.  One of the most frequently
considered libraries is the so-called Clifford+$T$ gate library that
consists of the reversible CNOT gate, the Hadamard gate, abbreviated
$H$, as well as the $T$ gate, and its inverse $T^\dagger$.  Quantum
gates on $n$ qubits are represented as $2^n \times 2^n$ unitary
matrices.  We write $T^\dagger$ to mean the complex conjugate of $T$,
and use the symbol `$\dagger$' also for other quantum gates.  The $T$
gate is sufficiently expensive in most approaches to fault tolerant
quantum computing~\cite{AMMR13} that it is customary to neglect all
other gates when costing a quantum algorithm.  For more details on
quantum gates we refer the reader to~\cite{NC00}.

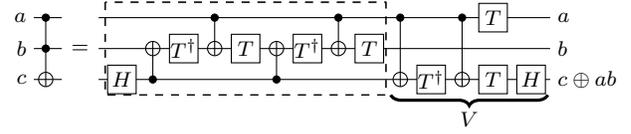
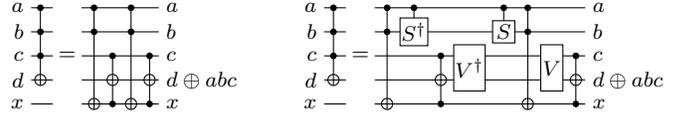
\begin{figure}
  \footnotesize
  \centering
  \subfloat[2-control Toffoli gate.]{%
\begin{tikzpicture}[scale=0.900000,x=1pt,y=1pt]
\filldraw[color=white] (0.000000, -6.500000) rectangle (217.000000, 32.500000);
\draw[color=black] (0.000000,26.000000) -- (217.000000,26.000000);
\draw[color=black] (0.000000,26.000000) node[left] {$a$};
\draw[color=black] (0.000000,13.000000) -- (217.000000,13.000000);
\draw[color=black] (0.000000,13.000000) node[left] {$b$};
\draw[color=black] (0.000000,0.000000) -- (217.000000,0.000000);
\draw[color=black] (0.000000,0.000000) node[left] {$c$};
\draw (5.000000,26.000000) -- (5.000000,0.000000);
\filldraw (5.000000, 26.000000) circle(1.500000pt);
\filldraw (5.000000, 13.000000) circle(1.500000pt);
\begin{scope}
\draw[fill=white] (5.000000, 0.000000) circle(3.000000pt);
\clip (5.000000, 0.000000) circle(3.000000pt);
\draw (2.000000, 0.000000) -- (8.000000, 0.000000);
\draw (5.000000, -3.000000) -- (5.000000, 3.000000);
\end{scope}
\draw[fill=white,color=white] (12.000000, -6.000000) rectangle (27.000000, 32.000000);
\draw (19.500000, 13.000000) node {$=$};
\begin{scope}
\draw[fill=white] (37.000000, -0.000000) +(-45.000000:8.485281pt and 8.485281pt) -- +(45.000000:8.485281pt and 8.485281pt) -- +(135.000000:8.485281pt and 8.485281pt) -- +(225.000000:8.485281pt and 8.485281pt) -- cycle;
\clip (37.000000, -0.000000) +(-45.000000:8.485281pt and 8.485281pt) -- +(45.000000:8.485281pt and 8.485281pt) -- +(135.000000:8.485281pt and 8.485281pt) -- +(225.000000:8.485281pt and 8.485281pt) -- cycle;
\draw (37.000000, -0.000000) node {$H$};
\end{scope}
\draw (50.000000,13.000000) -- (50.000000,0.000000);
\filldraw (50.000000, 0.000000) circle(1.500000pt);
\begin{scope}
\draw[fill=white] (50.000000, 13.000000) circle(3.000000pt);
\clip (50.000000, 13.000000) circle(3.000000pt);
\draw (47.000000, 13.000000) -- (53.000000, 13.000000);
\draw (50.000000, 10.000000) -- (50.000000, 16.000000);
\end{scope}
\begin{scope}
\draw[fill=white] (63.000000, 13.000000) +(-45.000000:8.485281pt and 8.485281pt) -- +(45.000000:8.485281pt and 8.485281pt) -- +(135.000000:8.485281pt and 8.485281pt) -- +(225.000000:8.485281pt and 8.485281pt) -- cycle;
\clip (63.000000, 13.000000) +(-45.000000:8.485281pt and 8.485281pt) -- +(45.000000:8.485281pt and 8.485281pt) -- +(135.000000:8.485281pt and 8.485281pt) -- +(225.000000:8.485281pt and 8.485281pt) -- cycle;
\draw (63.000000, 13.000000) node {{$T^\dagger$}};
\end{scope}
\draw (76.000000,26.000000) -- (76.000000,13.000000);
\filldraw (76.000000, 26.000000) circle(1.500000pt);
\begin{scope}
\draw[fill=white] (76.000000, 13.000000) circle(3.000000pt);
\clip (76.000000, 13.000000) circle(3.000000pt);
\draw (73.000000, 13.000000) -- (79.000000, 13.000000);
\draw (76.000000, 10.000000) -- (76.000000, 16.000000);
\end{scope}
\begin{scope}
\draw[fill=white] (89.000000, 13.000000) +(-45.000000:8.485281pt and 8.485281pt) -- +(45.000000:8.485281pt and 8.485281pt) -- +(135.000000:8.485281pt and 8.485281pt) -- +(225.000000:8.485281pt and 8.485281pt) -- cycle;
\clip (89.000000, 13.000000) +(-45.000000:8.485281pt and 8.485281pt) -- +(45.000000:8.485281pt and 8.485281pt) -- +(135.000000:8.485281pt and 8.485281pt) -- +(225.000000:8.485281pt and 8.485281pt) -- cycle;
\draw (89.000000, 13.000000) node {{$T$}};
\end{scope}
\draw (102.000000,13.000000) -- (102.000000,0.000000);
\filldraw (102.000000, 0.000000) circle(1.500000pt);
\begin{scope}
\draw[fill=white] (102.000000, 13.000000) circle(3.000000pt);
\clip (102.000000, 13.000000) circle(3.000000pt);
\draw (99.000000, 13.000000) -- (105.000000, 13.000000);
\draw (102.000000, 10.000000) -- (102.000000, 16.000000);
\end{scope}
\begin{scope}
\draw[fill=white] (115.000000, 13.000000) +(-45.000000:8.485281pt and 8.485281pt) -- +(45.000000:8.485281pt and 8.485281pt) -- +(135.000000:8.485281pt and 8.485281pt) -- +(225.000000:8.485281pt and 8.485281pt) -- cycle;
\clip (115.000000, 13.000000) +(-45.000000:8.485281pt and 8.485281pt) -- +(45.000000:8.485281pt and 8.485281pt) -- +(135.000000:8.485281pt and 8.485281pt) -- +(225.000000:8.485281pt and 8.485281pt) -- cycle;
\draw (115.000000, 13.000000) node {{$T^\dagger$}};
\end{scope}
\draw (128.000000,26.000000) -- (128.000000,13.000000);
\filldraw (128.000000, 26.000000) circle(1.500000pt);
\begin{scope}
\draw[fill=white] (128.000000, 13.000000) circle(3.000000pt);
\clip (128.000000, 13.000000) circle(3.000000pt);
\draw (125.000000, 13.000000) -- (131.000000, 13.000000);
\draw (128.000000, 10.000000) -- (128.000000, 16.000000);
\end{scope}
\begin{scope}
\draw[fill=white] (141.000000, 13.000000) +(-45.000000:8.485281pt and 8.485281pt) -- +(45.000000:8.485281pt and 8.485281pt) -- +(135.000000:8.485281pt and 8.485281pt) -- +(225.000000:8.485281pt and 8.485281pt) -- cycle;
\clip (141.000000, 13.000000) +(-45.000000:8.485281pt and 8.485281pt) -- +(45.000000:8.485281pt and 8.485281pt) -- +(135.000000:8.485281pt and 8.485281pt) -- +(225.000000:8.485281pt and 8.485281pt) -- cycle;
\draw (141.000000, 13.000000) node {{$T$}};
\end{scope}
\draw (154.000000,26.000000) -- (154.000000,0.000000);
\filldraw (154.000000, 26.000000) circle(1.500000pt);
\begin{scope}
\draw[fill=white] (154.000000, 0.000000) circle(3.000000pt);
\clip (154.000000, 0.000000) circle(3.000000pt);
\draw (151.000000, 0.000000) -- (157.000000, 0.000000);
\draw (154.000000, -3.000000) -- (154.000000, 3.000000);
\end{scope}
\begin{scope}
\draw[fill=white] (167.000000, -0.000000) +(-45.000000:8.485281pt and 8.485281pt) -- +(45.000000:8.485281pt and 8.485281pt) -- +(135.000000:8.485281pt and 8.485281pt) -- +(225.000000:8.485281pt and 8.485281pt) -- cycle;
\clip (167.000000, -0.000000) +(-45.000000:8.485281pt and 8.485281pt) -- +(45.000000:8.485281pt and 8.485281pt) -- +(135.000000:8.485281pt and 8.485281pt) -- +(225.000000:8.485281pt and 8.485281pt) -- cycle;
\draw (167.000000, -0.000000) node {{$T^\dagger$}};
\end{scope}
\draw (180.000000,26.000000) -- (180.000000,0.000000);
\filldraw (180.000000, 26.000000) circle(1.500000pt);
\begin{scope}
\draw[fill=white] (180.000000, 0.000000) circle(3.000000pt);
\clip (180.000000, 0.000000) circle(3.000000pt);
\draw (177.000000, 0.000000) -- (183.000000, 0.000000);
\draw (180.000000, -3.000000) -- (180.000000, 3.000000);
\end{scope}
\begin{scope}
\draw[fill=white] (193.000000, 26.000000) +(-45.000000:8.485281pt and 8.485281pt) -- +(45.000000:8.485281pt and 8.485281pt) -- +(135.000000:8.485281pt and 8.485281pt) -- +(225.000000:8.485281pt and 8.485281pt) -- cycle;
\clip (193.000000, 26.000000) +(-45.000000:8.485281pt and 8.485281pt) -- +(45.000000:8.485281pt and 8.485281pt) -- +(135.000000:8.485281pt and 8.485281pt) -- +(225.000000:8.485281pt and 8.485281pt) -- cycle;
\draw (193.000000, 26.000000) node {{$T$}};
\end{scope}
\begin{scope}
\draw[fill=white] (193.000000, -0.000000) +(-45.000000:8.485281pt and 8.485281pt) -- +(45.000000:8.485281pt and 8.485281pt) -- +(135.000000:8.485281pt and 8.485281pt) -- +(225.000000:8.485281pt and 8.485281pt) -- cycle;
\clip (193.000000, -0.000000) +(-45.000000:8.485281pt and 8.485281pt) -- +(45.000000:8.485281pt and 8.485281pt) -- +(135.000000:8.485281pt and 8.485281pt) -- +(225.000000:8.485281pt and 8.485281pt) -- cycle;
\draw (193.000000, -0.000000) node {{$T$}};
\end{scope}
\begin{scope}
\draw[fill=white] (209.000000, -0.000000) +(-45.000000:8.485281pt and 8.485281pt) -- +(45.000000:8.485281pt and 8.485281pt) -- +(135.000000:8.485281pt and 8.485281pt) -- +(225.000000:8.485281pt and 8.485281pt) -- cycle;
\clip (209.000000, -0.000000) +(-45.000000:8.485281pt and 8.485281pt) -- +(45.000000:8.485281pt and 8.485281pt) -- +(135.000000:8.485281pt and 8.485281pt) -- +(225.000000:8.485281pt and 8.485281pt) -- cycle;
\draw (209.000000, -0.000000) node {$H$};
\end{scope}
\draw[color=black] (217.000000,26.000000) node[right] {$a$};
\draw[color=black] (217.000000,13.000000) node[right] {$b$};
\draw[color=black] (217.000000,0.000000) node[right] {${c\oplus ab}$};
\draw[draw opacity=1.000000,fill opacity=0.200000,color=black,dashed] (30.000000,32.500000) rectangle (148.000000,-6.500000);
\draw[draw opacity=1.000000,fill opacity=0.200000,color=black,dashed] (30.000000,32.500000) rectangle (148.000000,-6.500000);
\draw[decorate,decoration={brace,mirror,amplitude = 4.000000pt},very thick] (150.000000,-6.500000) -- (216.000000,-6.500000);
\draw (183.000000, -10.500000) node[text width=144pt,below,text centered] {$V$};
\end{tikzpicture}}
%

\subfloat[3-control Toffoli gate (28 $T$ gates).]{%
\begin{tikzpicture}[scale=0.700000,x=1pt,y=1pt]
\filldraw[color=white] (0.000000, -6.500000) rectangle (69.000000, 58.500000);
\draw[color=black] (0.000000,52.000000) -- (69.000000,52.000000);
\draw[color=black] (0.000000,52.000000) node[left] {$a$};
\draw[color=black] (0.000000,39.000000) -- (69.000000,39.000000);
\draw[color=black] (0.000000,39.000000) node[left] {$b$};
\draw[color=black] (0.000000,26.000000) -- (69.000000,26.000000);
\draw[color=black] (0.000000,26.000000) node[left] {$c$};
\draw[color=black] (0.000000,13.000000) -- (69.000000,13.000000);
\draw[color=black] (0.000000,13.000000) node[left] {$d$};
\draw[color=black] (0.000000,0.000000) -- (69.000000,0.000000);
\draw[color=black] (0.000000,0.000000) node[left] {$x$};
\draw (5.000000,52.000000) -- (5.000000,13.000000);
\filldraw (5.000000, 52.000000) circle(1.500000pt);
\filldraw (5.000000, 39.000000) circle(1.500000pt);
\filldraw (5.000000, 26.000000) circle(1.500000pt);
\begin{scope}
\draw[fill=white] (5.000000, 13.000000) circle(3.000000pt);
\clip (5.000000, 13.000000) circle(3.000000pt);
\draw (2.000000, 13.000000) -- (8.000000, 13.000000);
\draw (5.000000, 10.000000) -- (5.000000, 16.000000);
\end{scope}
\draw[fill=white,color=white] (12.000000, -6.000000) rectangle (27.000000, 58.000000);
\draw (19.500000, 26.000000) node {$=$};
\draw (34.000000,52.000000) -- (34.000000,0.000000);
\filldraw (34.000000, 52.000000) circle(1.500000pt);
\filldraw (34.000000, 39.000000) circle(1.500000pt);
\begin{scope}
\draw[fill=white] (34.000000, 0.000000) circle(3.000000pt);
\clip (34.000000, 0.000000) circle(3.000000pt);
\draw (31.000000, 0.000000) -- (37.000000, 0.000000);
\draw (34.000000, -3.000000) -- (34.000000, 3.000000);
\end{scope}
\draw (44.000000,26.000000) -- (44.000000,0.000000);
\filldraw (44.000000, 26.000000) circle(1.500000pt);
\filldraw (44.000000, 0.000000) circle(1.500000pt);
\begin{scope}
\draw[fill=white] (44.000000, 13.000000) circle(3.000000pt);
\clip (44.000000, 13.000000) circle(3.000000pt);
\draw (41.000000, 13.000000) -- (47.000000, 13.000000);
\draw (44.000000, 10.000000) -- (44.000000, 16.000000);
\end{scope}
\draw (54.000000,52.000000) -- (54.000000,0.000000);
\filldraw (54.000000, 52.000000) circle(1.500000pt);
\filldraw (54.000000, 39.000000) circle(1.500000pt);
\begin{scope}
\draw[fill=white] (54.000000, 0.000000) circle(3.000000pt);
\clip (54.000000, 0.000000) circle(3.000000pt);
\draw (51.000000, 0.000000) -- (57.000000, 0.000000);
\draw (54.000000, -3.000000) -- (54.000000, 3.000000);
\end{scope}
\draw (64.000000,26.000000) -- (64.000000,0.000000);
\filldraw (64.000000, 26.000000) circle(1.500000pt);
\filldraw (64.000000, 0.000000) circle(1.500000pt);
\begin{scope}
\draw[fill=white] (64.000000, 13.000000) circle(3.000000pt);
\clip (64.000000, 13.000000) circle(3.000000pt);
\draw (61.000000, 13.000000) -- (67.000000, 13.000000);
\draw (64.000000, 10.000000) -- (64.000000, 16.000000);
\end{scope}
\draw[color=black] (69.000000,52.000000) node[right] {$a$};
\draw[color=black] (69.000000,39.000000) node[right] {$b$};
\draw[color=black] (69.000000,26.000000) node[right] {$c$};
\draw[color=black] (69.000000,13.000000) node[right] {${d\oplus abc}$};
\draw[color=black] (69.000000,0.000000) node[right] {$x$};
\end{tikzpicture}}
%
\hfill  \subfloat[3-control Toffoli gate (16 $T$ gates).]{%
\begin{tikzpicture}[scale=0.700000,x=1pt,y=1pt]
\filldraw[color=white] (0.000000, -6.500000) rectangle (141.000000, 58.500000);
\draw[color=black] (0.000000,52.000000) -- (141.000000,52.000000);
\draw[color=black] (0.000000,52.000000) node[left] {$a$};
\draw[color=black] (0.000000,39.000000) -- (141.000000,39.000000);
\draw[color=black] (0.000000,39.000000) node[left] {$b$};
\draw[color=black] (0.000000,26.000000) -- (141.000000,26.000000);
\draw[color=black] (0.000000,26.000000) node[left] {$c$};
\draw[color=black] (0.000000,13.000000) -- (141.000000,13.000000);
\draw[color=black] (0.000000,13.000000) node[left] {$d$};
\draw[color=black] (0.000000,0.000000) -- (141.000000,0.000000);
\draw[color=black] (0.000000,0.000000) node[left] {$x$};
\draw (5.000000,52.000000) -- (5.000000,13.000000);
\filldraw (5.000000, 52.000000) circle(1.500000pt);
\filldraw (5.000000, 39.000000) circle(1.500000pt);
\filldraw (5.000000, 26.000000) circle(1.500000pt);
\begin{scope}
\draw[fill=white] (5.000000, 13.000000) circle(3.000000pt);
\clip (5.000000, 13.000000) circle(3.000000pt);
\draw (2.000000, 13.000000) -- (8.000000, 13.000000);
\draw (5.000000, 10.000000) -- (5.000000, 16.000000);
\end{scope}
\draw[fill=white,color=white] (12.000000, -6.000000) rectangle (27.000000, 58.000000);
\draw (19.500000, 26.000000) node {$=$};
\draw (34.000000,52.000000) -- (34.000000,0.000000);
\filldraw (34.000000, 52.000000) circle(1.500000pt);
\filldraw (34.000000, 39.000000) circle(1.500000pt);
\begin{scope}
\draw[fill=white] (34.000000, 0.000000) circle(3.000000pt);
\clip (34.000000, 0.000000) circle(3.000000pt);
\draw (31.000000, 0.000000) -- (37.000000, 0.000000);
\draw (34.000000, -3.000000) -- (34.000000, 3.000000);
\end{scope}
\draw (48.500000,52.000000) -- (48.500000,39.000000);
\begin{scope}
\draw[fill=white] (48.500000, 39.000000) +(-45.000000:10.606602pt and 10.606602pt) -- +(45.000000:10.606602pt and 10.606602pt) -- +(135.000000:10.606602pt and 10.606602pt) -- +(225.000000:10.606602pt and 10.606602pt) -- cycle;
\clip (48.500000, 39.000000) +(-45.000000:10.606602pt and 10.606602pt) -- +(45.000000:10.606602pt and 10.606602pt) -- +(135.000000:10.606602pt and 10.606602pt) -- +(225.000000:10.606602pt and 10.606602pt) -- cycle;
\draw (48.500000, 39.000000) node {{$S^\dagger$}};
\end{scope}
\filldraw (48.500000, 52.000000) circle(1.500000pt);
\draw (63.000000,26.000000) -- (63.000000,0.000000);
\filldraw (63.000000, 26.000000) circle(1.500000pt);
\filldraw (63.000000, 0.000000) circle(1.500000pt);
\begin{scope}
\draw[fill=white] (63.000000, 13.000000) circle(3.000000pt);
\clip (63.000000, 13.000000) circle(3.000000pt);
\draw (60.000000, 13.000000) -- (66.000000, 13.000000);
\draw (63.000000, 10.000000) -- (63.000000, 16.000000);
\end{scope}
\draw (78.500000,26.000000) -- (78.500000,13.000000);
\begin{scope}
\draw[fill=white] (78.500000, 19.500000) +(-45.000000:12.020815pt and 17.677670pt) -- +(45.000000:12.020815pt and 17.677670pt) -- +(135.000000:12.020815pt and 17.677670pt) -- +(225.000000:12.020815pt and 17.677670pt) -- cycle;
\clip (78.500000, 19.500000) +(-45.000000:12.020815pt and 17.677670pt) -- +(45.000000:12.020815pt and 17.677670pt) -- +(135.000000:12.020815pt and 17.677670pt) -- +(225.000000:12.020815pt and 17.677670pt) -- cycle;
\draw (78.500000, 19.500000) node {{$V^\dagger$}};
\end{scope}
\draw (97.000000,52.000000) -- (97.000000,39.000000);
\begin{scope}
\draw[fill=white] (97.000000, 39.000000) +(-45.000000:8.485281pt and 8.485281pt) -- +(45.000000:8.485281pt and 8.485281pt) -- +(135.000000:8.485281pt and 8.485281pt) -- +(225.000000:8.485281pt and 8.485281pt) -- cycle;
\clip (97.000000, 39.000000) +(-45.000000:8.485281pt and 8.485281pt) -- +(45.000000:8.485281pt and 8.485281pt) -- +(135.000000:8.485281pt and 8.485281pt) -- +(225.000000:8.485281pt and 8.485281pt) -- cycle;
\draw (97.000000, 39.000000) node {{$S$}};
\end{scope}
\filldraw (97.000000, 52.000000) circle(1.500000pt);
\draw (110.000000,52.000000) -- (110.000000,0.000000);
\filldraw (110.000000, 52.000000) circle(1.500000pt);
\filldraw (110.000000, 39.000000) circle(1.500000pt);
\begin{scope}
\draw[fill=white] (110.000000, 0.000000) circle(3.000000pt);
\clip (110.000000, 0.000000) circle(3.000000pt);
\draw (107.000000, 0.000000) -- (113.000000, 0.000000);
\draw (110.000000, -3.000000) -- (110.000000, 3.000000);
\end{scope}
\draw (123.000000,26.000000) -- (123.000000,13.000000);
\begin{scope}
\draw[fill=white] (123.000000, 19.500000) +(-45.000000:8.485281pt and 17.677670pt) -- +(45.000000:8.485281pt and 17.677670pt) -- +(135.000000:8.485281pt and 17.677670pt) -- +(225.000000:8.485281pt and 17.677670pt) -- cycle;
\clip (123.000000, 19.500000) +(-45.000000:8.485281pt and 17.677670pt) -- +(45.000000:8.485281pt and 17.677670pt) -- +(135.000000:8.485281pt and 17.677670pt) -- +(225.000000:8.485281pt and 17.677670pt) -- cycle;
\draw (123.000000, 19.500000) node {{$V$}};
\end{scope}
\draw (136.000000,26.000000) -- (136.000000,0.000000);
\filldraw (136.000000, 26.000000) circle(1.500000pt);
\filldraw (136.000000, 0.000000) circle(1.500000pt);
\begin{scope}
\draw[fill=white] (136.000000, 13.000000) circle(3.000000pt);
\clip (136.000000, 13.000000) circle(3.000000pt);
\draw (133.000000, 13.000000) -- (139.000000, 13.000000);
\draw (136.000000, 10.000000) -- (136.000000, 16.000000);
\end{scope}
\draw[color=black] (141.000000,52.000000) node[right] {$a$};
\draw[color=black] (141.000000,39.000000) node[right] {$b$};
\draw[color=black] (141.000000,26.000000) node[right] {$c$};
\draw[color=black] (141.000000,13.000000) node[right] {${d\oplus abc}$};
\draw[color=black] (141.000000,0.000000) node[right] {$x$};
\end{tikzpicture}}
%

  \caption{Mapping Toffoli gates into Clifford+$T$ networks.}
  \label{fig:cliffordt}
\end{figure}

Fig.~\ref{fig:cliffordt}(a) shows one of the many realizations of the
2-control Toffoli gate, which can be found in~\cite{Maslov16}.  It
requires 7 $T$ gates which is optimum~\cite{AMMR13,GKMR14}.  Several works
from the literature describe how to map larger multiple-controlled
Toffoli gates into Clifford+$T$ gates (see,
e.g.,~\cite{AMMR13,Selinger13,AASD16,Maslov16}).
Fig.~\ref{fig:cliffordt}(b) shows one way to map the 3-control Toffoli
gate using a direct method as proposed by Barenco \emph{et
  al.}~\cite{BBC+95} Given a free ancilla line (that does not need to
be initialized to 0), it allows to map any multiple-controlled Toffoli
gate into a sequence of 2-control Toffoli gates which can then each be
mapped into the optimum network with $T$-count $7$.  However, the
number of $T$ gates can be reduced by modifying the Toffoli gates
slightly.  It can easily be seen that the network in
Fig.~\ref{fig:cliffordt}(c) is the same as in
Fig.~\ref{fig:cliffordt}(b), since the controlled $S^\dagger$ gate
cancels the controlled $S$ gate and the $V^\dagger$ gate cancels the
$V$ gate. However, the Toffoli gate combined with a controlled $S$
gate can be realized using only 4 $T$ gates~\cite{Selinger13}, and
applying the $V$ to the Clifford+$T$ realization cancels another 3 $T$
gates (see Fig.~\ref{fig:cliffordt}(a) and~\cite{Jones13,Maslov16}).
In general, a $k$-controlled Toffoli gate can be realized with at most
$16(k - 1)$ $T$ gates.  If the number of ancilla lines is larger or
equal to $\lfloor \frac{k-1}{2}\rfloor$, then $8(k-1)$ $T$ gates
suffice~\cite{BBC+95,Maslov16}.  Future improvements to the
decomposition of multiple-controlled Toffoli gates into Clifford+$T$
networks will have an immediate positive effect on our proposed
synthesis method.

\begin{algorithm}[t]
  \small
\Input{Logic network $N$, parameters $p_{\mathrm{Q}}$, parameters $p_{\mathrm{T}}$}
\Output{Clifford+$T$ network $R$}

set $N \gets \operatorname{lut\_mapping}(N, p_{\mathrm{Q}})$\;
set $R \gets \operatorname{synthesize\_mapping}(N, p_{\mathrm{Q}})$\;
set $R \gets \operatorname{synthesize\_gates}(R, p_{\mathrm{T}})$\;
\KwRet{$R$}\;

\medskip
\caption{Overview of the LHRS algorithm.}
\label{alg:lhrs}
\end{algorithm}

\section{Motivation and Problem Definition}
\label{sec:outline}
A major problem facing quantum computing is the inability of existing
hand crafted approaches to generate networks for scientific operations
that require a reasonable number of quantum bits and gates.  As an
example, the quantum linear systems algorithm requires only $100$
(logical) quantum bits to encode a $2^{100}\times 2^{100}$ matrix
inversion problem~\cite{HHL09,CJS13}, clearly demonstrating the
advantage that can be gained by using a quantum computer.  However, in
prior approaches, the reciprocal step ($1/x$) that is part of the
calculation can require in excess of $500$ quantum bits.  This means
that arithmetic may dominate the number of qubits of that
algorithm~\cite{WR16}, diminishing the potential improvements of a
quantum algorithmic implementation.  Similarly, recent quantum
chemistry simulation algorithms can provide improved scaling over the
best known methods but at the price of requiring the molecular
integrals that define the problem to be computed using floating point
arithmetic~\cite{BBK+16b}. While floating point addition was studied
before~\cite{NM14,TAV06}, currently networks do not exist for more
complex floating point operations such as exponential, reciprocal
square root, multiplication, and squaring.  Without the ability to
automatically generate circuits for these operations it will be a
difficult task to implement such algorithms on a quantum computer, to
estimate their full costs, or to verify that the underlying circuitry
is correct.

In this paper we tackle this challenge and address the following
problem: \textsl{Given a conventional combinational logic network that
  represents a desired target functionality, find a quantum circuit
  with a reasonable number of qubits and number of $T$ gates.}  The
algorithm should be highly configurable such that instead of a single
quantum circuit a whole design space of circuits with several
Pareto-optimal solutions can be explored.

\subsubsection*{Algorithm outline}
Alg.~\ref{alg:lhrs} illustrates the general outline of the algorithm.
The following subsections provide further details.  Input to the
algorithm is a logic network $N$ and it outputs a Clifford+$T$ quantum
network $R$.  In addition to $N$, two sets of parameters
$p_{\mathrm{Q}}$ and $p_{\mathrm{T}}$ are provided that control
detailed behavior of the algorithm.  The parameters will be introduced
in the following sections and are summarized in
Sect.~\ref{sec:params}; but for now it is sufficient to emphasize the
role of the parameters.  Parameters in $p_{\mathrm{Q}}$ can influence
both the number of qubits and $T$ gates in $R$, however, their main
purpose is to control the number of qubits.  Parameters in
$p_{\mathrm{T}}$ only affect the number of $T$ gates.

The first step in Alg.~\ref{alg:lhrs} is to derive a LUT mapping from
the input logic network.  As we will see in
Sect.~\ref{sec:synth-mapping}, one parameter in $p_{\mathrm{Q}}$ is
the LUT size for the mapping which has the strongest influence on the
number of qubits in $R$.  Given the LUT mapping, one can derive a
reversible logic network in which each LUT is translated into one or
two single-target gates.  In the last step, each of the gates is
mapped into Clifford+$T$ gates (Sect.~\ref{sec:synth-gates}).

It is important to know that most of the runtime is consumed by the
last step in Alg.~\ref{alg:lhrs}, and that after the first two steps
the number of qubits for the final Clifford+$T$ network is already
known.  This allows us to use the algorithm in an incremental way as
follows.  First, one explores assignments to parameters in
$p_{\mathrm{Q}}$ that lead to a desired number of qubits, particularly
by evaluating different LUT sizes.  This can be done by calling the
first two steps of the algorihm with different values for the
parameters in $p_{\mathrm{Q}}$.  Afterwards, one can optimize for the
number of $T$ gates by calling the last step by sampling the
parameters for $p_{\mathrm{T}}$.

\section{Synthesizing the Mapping}
\label{sec:synth-mapping}
This section describes how a LUT mapping can be translated into a
reversible network.  This is the second step of Alg.~\ref{alg:lhrs}.
The first step in Alg.~\ref{alg:lhrs} applies conventional LUT mapping
algorithms and is not further explained in this paper.  The interested
reader is referred to the
literature~\cite{CD94,CC04,MCCB07,MCB07,RME+12}.

\begin{figure}[t]
  \centering
  \hfil\subfloat[LUT network]{\tikzExample}\hfil\scriptsize
  \subfloat[Reversible network.]{%
\begin{tikzpicture}[scale=0.500000,x=1pt,y=1pt]
\filldraw[color=white] (0.000000, -7.500000) rectangle (108.000000, 142.500000);
\draw[color=black] (0.000000,135.000000) -- (108.000000,135.000000);
\draw[color=black] (0.000000,135.000000) node[left] {$x_1$};
\draw[color=black] (0.000000,120.000000) -- (108.000000,120.000000);
\draw[color=black] (0.000000,120.000000) node[left] {$x_2$};
\draw[color=black] (0.000000,105.000000) -- (108.000000,105.000000);
\draw[color=black] (0.000000,105.000000) node[left] {$x_3$};
\draw[color=black] (0.000000,90.000000) -- (108.000000,90.000000);
\draw[color=black] (0.000000,90.000000) node[left] {$x_4$};
\draw[color=black] (0.000000,75.000000) -- (108.000000,75.000000);
\draw[color=black] (0.000000,75.000000) node[left] {$x_5$};
\draw[color=black] (0.000000,60.000000) -- (108.000000,60.000000);
\draw[color=black] (0.000000,60.000000) node[left] {$0$};
\draw[color=black] (0.000000,45.000000) -- (108.000000,45.000000);
\draw[color=black] (0.000000,45.000000) node[left] {$0$};
\draw[color=black] (0.000000,30.000000) -- (108.000000,30.000000);
\draw[color=black] (0.000000,30.000000) node[left] {$0$};
\draw[color=black] (0.000000,15.000000) -- (108.000000,15.000000);
\draw[color=black] (0.000000,15.000000) node[left] {$0$};
\draw[color=black] (0.000000,0.000000) -- (108.000000,0.000000);
\draw[color=black] (0.000000,0.000000) node[left] {$0$};
\draw (12.000000,120.000000) -- (12.000000,60.000000);
\begin{scope}[rounded corners=2pt]
\begin{scope}
\draw[fill=white] (12.000000, 112.500000) +(-45.000000:8.485281pt and 19.091883pt) -- +(45.000000:8.485281pt and 19.091883pt) -- +(135.000000:8.485281pt and 19.091883pt) -- +(225.000000:8.485281pt and 19.091883pt) -- cycle;
\clip (12.000000, 112.500000) +(-45.000000:8.485281pt and 19.091883pt) -- +(45.000000:8.485281pt and 19.091883pt) -- +(135.000000:8.485281pt and 19.091883pt) -- +(225.000000:8.485281pt and 19.091883pt) -- cycle;
\draw (12.000000, 112.500000) node {{1}};
\end{scope}
\end{scope}
\begin{scope}
\draw[fill=white] (12.000000, 60.000000) circle(3.000000pt);
\clip (12.000000, 60.000000) circle(3.000000pt);
\draw (9.000000, 60.000000) -- (15.000000, 60.000000);
\draw (12.000000, 57.000000) -- (12.000000, 63.000000);
\end{scope}
\draw (24.000000,90.000000) -- (24.000000,45.000000);
\begin{scope}[rounded corners=2pt]
\begin{scope}
\draw[fill=white] (24.000000, 82.500000) +(-45.000000:8.485281pt and 19.091883pt) -- +(45.000000:8.485281pt and 19.091883pt) -- +(135.000000:8.485281pt and 19.091883pt) -- +(225.000000:8.485281pt and 19.091883pt) -- cycle;
\clip (24.000000, 82.500000) +(-45.000000:8.485281pt and 19.091883pt) -- +(45.000000:8.485281pt and 19.091883pt) -- +(135.000000:8.485281pt and 19.091883pt) -- +(225.000000:8.485281pt and 19.091883pt) -- cycle;
\draw (24.000000, 82.500000) node {{2}};
\end{scope}
\end{scope}
\begin{scope}
\draw[fill=white] (24.000000, 45.000000) circle(3.000000pt);
\clip (24.000000, 45.000000) circle(3.000000pt);
\draw (21.000000, 45.000000) -- (27.000000, 45.000000);
\draw (24.000000, 42.000000) -- (24.000000, 48.000000);
\end{scope}
\draw (48.000000,135.000000) -- (48.000000,30.000000);
\begin{scope}[rounded corners=2pt]
\begin{scope}
\draw[fill=white] (48.000000, 97.500000) +(-45.000000:8.485281pt and 61.518290pt) -- +(45.000000:8.485281pt and 61.518290pt) -- +(135.000000:8.485281pt and 61.518290pt) -- +(225.000000:8.485281pt and 61.518290pt) -- cycle;
\clip (48.000000, 97.500000) +(-45.000000:8.485281pt and 61.518290pt) -- +(45.000000:8.485281pt and 61.518290pt) -- +(135.000000:8.485281pt and 61.518290pt) -- +(225.000000:8.485281pt and 61.518290pt) -- cycle;
\draw (48.000000, 97.500000) node {{3}};
\end{scope}
\end{scope}
\draw[color=black,dash pattern=on 2pt off 1pt] (42.000000, 120.000000) -- (54.000000, 120.000000);
\draw[color=black,dash pattern=on 2pt off 1pt] (42.000000, 105.000000) -- (54.000000, 105.000000);
\draw[color=black,dash pattern=on 2pt off 1pt] (42.000000, 90.000000) -- (54.000000, 90.000000);
\draw[color=black,dash pattern=on 2pt off 1pt] (42.000000, 75.000000) -- (54.000000, 75.000000);
\begin{scope}
\draw[fill=white] (48.000000, 30.000000) circle(3.000000pt);
\clip (48.000000, 30.000000) circle(3.000000pt);
\draw (45.000000, 30.000000) -- (51.000000, 30.000000);
\draw (48.000000, 27.000000) -- (48.000000, 33.000000);
\end{scope}
\draw (72.000000,60.000000) -- (72.000000,15.000000);
\begin{scope}[rounded corners=2pt]
\begin{scope}
\draw[fill=white] (72.000000, 52.500000) +(-45.000000:8.485281pt and 19.091883pt) -- +(45.000000:8.485281pt and 19.091883pt) -- +(135.000000:8.485281pt and 19.091883pt) -- +(225.000000:8.485281pt and 19.091883pt) -- cycle;
\clip (72.000000, 52.500000) +(-45.000000:8.485281pt and 19.091883pt) -- +(45.000000:8.485281pt and 19.091883pt) -- +(135.000000:8.485281pt and 19.091883pt) -- +(225.000000:8.485281pt and 19.091883pt) -- cycle;
\draw (72.000000, 52.500000) node {{4}};
\end{scope}
\end{scope}
\begin{scope}
\draw[fill=white] (72.000000, 15.000000) circle(3.000000pt);
\clip (72.000000, 15.000000) circle(3.000000pt);
\draw (69.000000, 15.000000) -- (75.000000, 15.000000);
\draw (72.000000, 12.000000) -- (72.000000, 18.000000);
\end{scope}
\draw (96.000000,75.000000) -- (96.000000,0.000000);
\begin{scope}[rounded corners=2pt]
\begin{scope}
\draw[fill=white] (96.000000, 45.000000) +(-45.000000:8.485281pt and 50.911688pt) -- +(45.000000:8.485281pt and 50.911688pt) -- +(135.000000:8.485281pt and 50.911688pt) -- +(225.000000:8.485281pt and 50.911688pt) -- cycle;
\clip (96.000000, 45.000000) +(-45.000000:8.485281pt and 50.911688pt) -- +(45.000000:8.485281pt and 50.911688pt) -- +(135.000000:8.485281pt and 50.911688pt) -- +(225.000000:8.485281pt and 50.911688pt) -- cycle;
\draw (96.000000, 45.000000) node {{5}};
\end{scope}
\end{scope}
\draw[color=black,dash pattern=on 2pt off 1pt] (90.000000, 60.000000) -- (102.000000, 60.000000);
\draw[color=black,dash pattern=on 2pt off 1pt] (90.000000, 45.000000) -- (102.000000, 45.000000);
\draw[color=black,dash pattern=on 2pt off 1pt] (90.000000, 30.000000) -- (102.000000, 30.000000);
\begin{scope}
\draw[fill=white] (96.000000, 0.000000) circle(3.000000pt);
\clip (96.000000, 0.000000) circle(3.000000pt);
\draw (93.000000, 0.000000) -- (99.000000, 0.000000);
\draw (96.000000, -3.000000) -- (96.000000, 3.000000);
\end{scope}
\draw[color=black] (108.000000,135.000000) node[right] {$x_1$};
\draw[color=black] (108.000000,120.000000) node[right] {$x_2$};
\draw[color=black] (108.000000,105.000000) node[right] {$x_3$};
\draw[color=black] (108.000000,90.000000) node[right] {$x_4$};
\draw[color=black] (108.000000,75.000000) node[right] {$x_5$};
\draw[color=black] (108.000000,60.000000) node[right] {$\text{---}$};
\draw[color=black] (108.000000,45.000000) node[right] {$\text{---}$};
\draw[color=black] (108.000000,30.000000) node[right] {$y_1$};
\draw[color=black] (108.000000,15.000000) node[right] {$\text{---}$};
\draw[color=black] (108.000000,0.000000) node[right] {$y_2$};
\end{tikzpicture}}



\subfloat[Order: 1, 2, 3, 4, 5.]{%
\begin{tikzpicture}[scale=0.500000,x=1pt,y=1pt]
\filldraw[color=white] (0.000000, -7.500000) rectangle (168.000000, 142.500000);
\draw[color=black] (0.000000,135.000000) -- (168.000000,135.000000);
\draw[color=black] (0.000000,135.000000) node[left] {$x_1$};
\draw[color=black] (0.000000,120.000000) -- (168.000000,120.000000);
\draw[color=black] (0.000000,120.000000) node[left] {$x_2$};
\draw[color=black] (0.000000,105.000000) -- (168.000000,105.000000);
\draw[color=black] (0.000000,105.000000) node[left] {$x_3$};
\draw[color=black] (0.000000,90.000000) -- (168.000000,90.000000);
\draw[color=black] (0.000000,90.000000) node[left] {$x_4$};
\draw[color=black] (0.000000,75.000000) -- (168.000000,75.000000);
\draw[color=black] (0.000000,75.000000) node[left] {$x_5$};
\draw[color=black] (0.000000,60.000000) -- (168.000000,60.000000);
\draw[color=black] (0.000000,60.000000) node[left] {$0$};
\draw[color=black] (0.000000,45.000000) -- (168.000000,45.000000);
\draw[color=black] (0.000000,45.000000) node[left] {$0$};
\draw[color=black] (0.000000,30.000000) -- (168.000000,30.000000);
\draw[color=black] (0.000000,30.000000) node[left] {$0$};
\draw[color=black] (0.000000,15.000000) -- (168.000000,15.000000);
\draw[color=black] (0.000000,15.000000) node[left] {$0$};
\draw[color=black] (0.000000,0.000000) -- (168.000000,0.000000);
\draw[color=black] (0.000000,0.000000) node[left] {$0$};
\draw (12.000000,120.000000) -- (12.000000,60.000000);
\begin{scope}[rounded corners=2pt]
\begin{scope}
\draw[fill=white] (12.000000, 112.500000) +(-45.000000:8.485281pt and 19.091883pt) -- +(45.000000:8.485281pt and 19.091883pt) -- +(135.000000:8.485281pt and 19.091883pt) -- +(225.000000:8.485281pt and 19.091883pt) -- cycle;
\clip (12.000000, 112.500000) +(-45.000000:8.485281pt and 19.091883pt) -- +(45.000000:8.485281pt and 19.091883pt) -- +(135.000000:8.485281pt and 19.091883pt) -- +(225.000000:8.485281pt and 19.091883pt) -- cycle;
\draw (12.000000, 112.500000) node {{1}};
\end{scope}
\end{scope}
\begin{scope}
\draw[fill=white] (12.000000, 60.000000) circle(3.000000pt);
\clip (12.000000, 60.000000) circle(3.000000pt);
\draw (9.000000, 60.000000) -- (15.000000, 60.000000);
\draw (12.000000, 57.000000) -- (12.000000, 63.000000);
\end{scope}
\draw (24.000000,90.000000) -- (24.000000,45.000000);
\begin{scope}[rounded corners=2pt]
\begin{scope}
\draw[fill=white] (24.000000, 82.500000) +(-45.000000:8.485281pt and 19.091883pt) -- +(45.000000:8.485281pt and 19.091883pt) -- +(135.000000:8.485281pt and 19.091883pt) -- +(225.000000:8.485281pt and 19.091883pt) -- cycle;
\clip (24.000000, 82.500000) +(-45.000000:8.485281pt and 19.091883pt) -- +(45.000000:8.485281pt and 19.091883pt) -- +(135.000000:8.485281pt and 19.091883pt) -- +(225.000000:8.485281pt and 19.091883pt) -- cycle;
\draw (24.000000, 82.500000) node {{2}};
\end{scope}
\end{scope}
\begin{scope}
\draw[fill=white] (24.000000, 45.000000) circle(3.000000pt);
\clip (24.000000, 45.000000) circle(3.000000pt);
\draw (21.000000, 45.000000) -- (27.000000, 45.000000);
\draw (24.000000, 42.000000) -- (24.000000, 48.000000);
\end{scope}
\draw (48.000000,135.000000) -- (48.000000,30.000000);
\begin{scope}[rounded corners=2pt]
\begin{scope}
\draw[fill=white] (48.000000, 97.500000) +(-45.000000:8.485281pt and 61.518290pt) -- +(45.000000:8.485281pt and 61.518290pt) -- +(135.000000:8.485281pt and 61.518290pt) -- +(225.000000:8.485281pt and 61.518290pt) -- cycle;
\clip (48.000000, 97.500000) +(-45.000000:8.485281pt and 61.518290pt) -- +(45.000000:8.485281pt and 61.518290pt) -- +(135.000000:8.485281pt and 61.518290pt) -- +(225.000000:8.485281pt and 61.518290pt) -- cycle;
\draw (48.000000, 97.500000) node {{3}};
\end{scope}
\end{scope}
\draw[color=black,dash pattern=on 2pt off 1pt] (42.000000, 120.000000) -- (54.000000, 120.000000);
\draw[color=black,dash pattern=on 2pt off 1pt] (42.000000, 105.000000) -- (54.000000, 105.000000);
\draw[color=black,dash pattern=on 2pt off 1pt] (42.000000, 90.000000) -- (54.000000, 90.000000);
\draw[color=black,dash pattern=on 2pt off 1pt] (42.000000, 75.000000) -- (54.000000, 75.000000);
\begin{scope}
\draw[fill=white] (48.000000, 30.000000) circle(3.000000pt);
\clip (48.000000, 30.000000) circle(3.000000pt);
\draw (45.000000, 30.000000) -- (51.000000, 30.000000);
\draw (48.000000, 27.000000) -- (48.000000, 33.000000);
\end{scope}
\draw (72.000000,60.000000) -- (72.000000,15.000000);
\begin{scope}[rounded corners=2pt]
\begin{scope}
\draw[fill=white] (72.000000, 52.500000) +(-45.000000:8.485281pt and 19.091883pt) -- +(45.000000:8.485281pt and 19.091883pt) -- +(135.000000:8.485281pt and 19.091883pt) -- +(225.000000:8.485281pt and 19.091883pt) -- cycle;
\clip (72.000000, 52.500000) +(-45.000000:8.485281pt and 19.091883pt) -- +(45.000000:8.485281pt and 19.091883pt) -- +(135.000000:8.485281pt and 19.091883pt) -- +(225.000000:8.485281pt and 19.091883pt) -- cycle;
\draw (72.000000, 52.500000) node {{4}};
\end{scope}
\end{scope}
\begin{scope}
\draw[fill=white] (72.000000, 15.000000) circle(3.000000pt);
\clip (72.000000, 15.000000) circle(3.000000pt);
\draw (69.000000, 15.000000) -- (75.000000, 15.000000);
\draw (72.000000, 12.000000) -- (72.000000, 18.000000);
\end{scope}
\draw (96.000000,75.000000) -- (96.000000,0.000000);
\begin{scope}[rounded corners=2pt]
\begin{scope}
\draw[fill=white] (96.000000, 45.000000) +(-45.000000:8.485281pt and 50.911688pt) -- +(45.000000:8.485281pt and 50.911688pt) -- +(135.000000:8.485281pt and 50.911688pt) -- +(225.000000:8.485281pt and 50.911688pt) -- cycle;
\clip (96.000000, 45.000000) +(-45.000000:8.485281pt and 50.911688pt) -- +(45.000000:8.485281pt and 50.911688pt) -- +(135.000000:8.485281pt and 50.911688pt) -- +(225.000000:8.485281pt and 50.911688pt) -- cycle;
\draw (96.000000, 45.000000) node {{5}};
\end{scope}
\end{scope}
\draw[color=black,dash pattern=on 2pt off 1pt] (90.000000, 60.000000) -- (102.000000, 60.000000);
\draw[color=black,dash pattern=on 2pt off 1pt] (90.000000, 45.000000) -- (102.000000, 45.000000);
\draw[color=black,dash pattern=on 2pt off 1pt] (90.000000, 30.000000) -- (102.000000, 30.000000);
\begin{scope}
\draw[fill=white] (96.000000, 0.000000) circle(3.000000pt);
\clip (96.000000, 0.000000) circle(3.000000pt);
\draw (93.000000, 0.000000) -- (99.000000, 0.000000);
\draw (96.000000, -3.000000) -- (96.000000, 3.000000);
\end{scope}
\draw (120.000000,60.000000) -- (120.000000,15.000000);
\begin{scope}[rounded corners=2pt]
\begin{scope}
\draw[fill=white] (120.000000, 52.500000) +(-45.000000:8.485281pt and 19.091883pt) -- +(45.000000:8.485281pt and 19.091883pt) -- +(135.000000:8.485281pt and 19.091883pt) -- +(225.000000:8.485281pt and 19.091883pt) -- cycle;
\clip (120.000000, 52.500000) +(-45.000000:8.485281pt and 19.091883pt) -- +(45.000000:8.485281pt and 19.091883pt) -- +(135.000000:8.485281pt and 19.091883pt) -- +(225.000000:8.485281pt and 19.091883pt) -- cycle;
\draw (120.000000, 52.500000) node {{4}};
\end{scope}
\end{scope}
\begin{scope}
\draw[fill=white] (120.000000, 15.000000) circle(3.000000pt);
\clip (120.000000, 15.000000) circle(3.000000pt);
\draw (117.000000, 15.000000) -- (123.000000, 15.000000);
\draw (120.000000, 12.000000) -- (120.000000, 18.000000);
\end{scope}
\draw (144.000000,90.000000) -- (144.000000,45.000000);
\begin{scope}[rounded corners=2pt]
\begin{scope}
\draw[fill=white] (144.000000, 82.500000) +(-45.000000:8.485281pt and 19.091883pt) -- +(45.000000:8.485281pt and 19.091883pt) -- +(135.000000:8.485281pt and 19.091883pt) -- +(225.000000:8.485281pt and 19.091883pt) -- cycle;
\clip (144.000000, 82.500000) +(-45.000000:8.485281pt and 19.091883pt) -- +(45.000000:8.485281pt and 19.091883pt) -- +(135.000000:8.485281pt and 19.091883pt) -- +(225.000000:8.485281pt and 19.091883pt) -- cycle;
\draw (144.000000, 82.500000) node {{2}};
\end{scope}
\end{scope}
\begin{scope}
\draw[fill=white] (144.000000, 45.000000) circle(3.000000pt);
\clip (144.000000, 45.000000) circle(3.000000pt);
\draw (141.000000, 45.000000) -- (147.000000, 45.000000);
\draw (144.000000, 42.000000) -- (144.000000, 48.000000);
\end{scope}
\draw (156.000000,120.000000) -- (156.000000,60.000000);
\begin{scope}[rounded corners=2pt]
\begin{scope}
\draw[fill=white] (156.000000, 112.500000) +(-45.000000:8.485281pt and 19.091883pt) -- +(45.000000:8.485281pt and 19.091883pt) -- +(135.000000:8.485281pt and 19.091883pt) -- +(225.000000:8.485281pt and 19.091883pt) -- cycle;
\clip (156.000000, 112.500000) +(-45.000000:8.485281pt and 19.091883pt) -- +(45.000000:8.485281pt and 19.091883pt) -- +(135.000000:8.485281pt and 19.091883pt) -- +(225.000000:8.485281pt and 19.091883pt) -- cycle;
\draw (156.000000, 112.500000) node {{1}};
\end{scope}
\end{scope}
\begin{scope}
\draw[fill=white] (156.000000, 60.000000) circle(3.000000pt);
\clip (156.000000, 60.000000) circle(3.000000pt);
\draw (153.000000, 60.000000) -- (159.000000, 60.000000);
\draw (156.000000, 57.000000) -- (156.000000, 63.000000);
\end{scope}
\draw[color=black] (168.000000,135.000000) node[right] {$x_1$};
\draw[color=black] (168.000000,120.000000) node[right] {$x_2$};
\draw[color=black] (168.000000,105.000000) node[right] {$x_3$};
\draw[color=black] (168.000000,90.000000) node[right] {$x_4$};
\draw[color=black] (168.000000,75.000000) node[right] {$x_5$};
\draw[color=black] (168.000000,60.000000) node[right] {$0$};
\draw[color=black] (168.000000,45.000000) node[right] {$0$};
\draw[color=black] (168.000000,30.000000) node[right] {$y_1$};
\draw[color=black] (168.000000,15.000000) node[right] {$0$};
\draw[color=black] (168.000000,0.000000) node[right] {$y_2$};
\end{tikzpicture}}
%
%
\subfloat[Order: 1, 2, 4, 5, 3.]{
\begin{tikzpicture}[scale=0.500000,x=1pt,y=1pt]
\filldraw[color=white] (0.000000, -7.500000) rectangle (180.000000, 127.500000);
\draw[color=black] (0.000000,120.000000) -- (180.000000,120.000000);
\draw[color=black] (0.000000,120.000000) node[left] {$x_1$};
\draw[color=black] (0.000000,105.000000) -- (180.000000,105.000000);
\draw[color=black] (0.000000,105.000000) node[left] {$x_2$};
\draw[color=black] (0.000000,90.000000) -- (180.000000,90.000000);
\draw[color=black] (0.000000,90.000000) node[left] {$x_3$};
\draw[color=black] (0.000000,75.000000) -- (180.000000,75.000000);
\draw[color=black] (0.000000,75.000000) node[left] {$x_4$};
\draw[color=black] (0.000000,60.000000) -- (180.000000,60.000000);
\draw[color=black] (0.000000,60.000000) node[left] {$x_5$};
\draw[color=black] (0.000000,45.000000) -- (180.000000,45.000000);
\draw[color=black] (0.000000,45.000000) node[left] {$0$};
\draw[color=black] (0.000000,30.000000) -- (180.000000,30.000000);
\draw[color=black] (0.000000,30.000000) node[left] {$0$};
\draw[color=black] (0.000000,15.000000) -- (180.000000,15.000000);
\draw[color=black] (0.000000,15.000000) node[left] {$0$};
\draw[color=black] (0.000000,0.000000) -- (180.000000,0.000000);
\draw[color=black] (0.000000,0.000000) node[left] {$0$};
\draw (12.000000,105.000000) -- (12.000000,45.000000);
\begin{scope}[rounded corners=2pt]
\begin{scope}
\draw[fill=white] (12.000000, 97.500000) +(-45.000000:8.485281pt and 19.091883pt) -- +(45.000000:8.485281pt and 19.091883pt) -- +(135.000000:8.485281pt and 19.091883pt) -- +(225.000000:8.485281pt and 19.091883pt) -- cycle;
\clip (12.000000, 97.500000) +(-45.000000:8.485281pt and 19.091883pt) -- +(45.000000:8.485281pt and 19.091883pt) -- +(135.000000:8.485281pt and 19.091883pt) -- +(225.000000:8.485281pt and 19.091883pt) -- cycle;
\draw (12.000000, 97.500000) node {{1}};
\end{scope}
\end{scope}
\begin{scope}
\draw[fill=white] (12.000000, 45.000000) circle(3.000000pt);
\clip (12.000000, 45.000000) circle(3.000000pt);
\draw (9.000000, 45.000000) -- (15.000000, 45.000000);
\draw (12.000000, 42.000000) -- (12.000000, 48.000000);
\end{scope}
\draw (24.000000,75.000000) -- (24.000000,30.000000);
\begin{scope}[rounded corners=2pt]
\begin{scope}
\draw[fill=white] (24.000000, 67.500000) +(-45.000000:8.485281pt and 19.091883pt) -- +(45.000000:8.485281pt and 19.091883pt) -- +(135.000000:8.485281pt and 19.091883pt) -- +(225.000000:8.485281pt and 19.091883pt) -- cycle;
\clip (24.000000, 67.500000) +(-45.000000:8.485281pt and 19.091883pt) -- +(45.000000:8.485281pt and 19.091883pt) -- +(135.000000:8.485281pt and 19.091883pt) -- +(225.000000:8.485281pt and 19.091883pt) -- cycle;
\draw (24.000000, 67.500000) node {{2}};
\end{scope}
\end{scope}
\begin{scope}
\draw[fill=white] (24.000000, 30.000000) circle(3.000000pt);
\clip (24.000000, 30.000000) circle(3.000000pt);
\draw (21.000000, 30.000000) -- (27.000000, 30.000000);
\draw (24.000000, 27.000000) -- (24.000000, 33.000000);
\end{scope}
\draw (48.000000,45.000000) -- (48.000000,15.000000);
\begin{scope}[rounded corners=2pt]
\begin{scope}
\draw[fill=white] (48.000000, 37.500000) +(-45.000000:8.485281pt and 19.091883pt) -- +(45.000000:8.485281pt and 19.091883pt) -- +(135.000000:8.485281pt and 19.091883pt) -- +(225.000000:8.485281pt and 19.091883pt) -- cycle;
\clip (48.000000, 37.500000) +(-45.000000:8.485281pt and 19.091883pt) -- +(45.000000:8.485281pt and 19.091883pt) -- +(135.000000:8.485281pt and 19.091883pt) -- +(225.000000:8.485281pt and 19.091883pt) -- cycle;
\draw (48.000000, 37.500000) node {{4}};
\end{scope}
\end{scope}
\begin{scope}
\draw[fill=white] (48.000000, 15.000000) circle(3.000000pt);
\clip (48.000000, 15.000000) circle(3.000000pt);
\draw (45.000000, 15.000000) -- (51.000000, 15.000000);
\draw (48.000000, 12.000000) -- (48.000000, 18.000000);
\end{scope}
\draw (72.000000,30.000000) -- (72.000000,0.000000);
\begin{scope}[rounded corners=2pt]
\begin{scope}
\draw[fill=white] (72.000000, 22.500000) +(-45.000000:8.485281pt and 19.091883pt) -- +(45.000000:8.485281pt and 19.091883pt) -- +(135.000000:8.485281pt and 19.091883pt) -- +(225.000000:8.485281pt and 19.091883pt) -- cycle;
\clip (72.000000, 22.500000) +(-45.000000:8.485281pt and 19.091883pt) -- +(45.000000:8.485281pt and 19.091883pt) -- +(135.000000:8.485281pt and 19.091883pt) -- +(225.000000:8.485281pt and 19.091883pt) -- cycle;
\draw (72.000000, 22.500000) node {{5}};
\end{scope}
\end{scope}
\begin{scope}
\draw[fill=white] (72.000000, 0.000000) circle(3.000000pt);
\clip (72.000000, 0.000000) circle(3.000000pt);
\draw (69.000000, 0.000000) -- (75.000000, 0.000000);
\draw (72.000000, -3.000000) -- (72.000000, 3.000000);
\end{scope}
\draw (96.000000,45.000000) -- (96.000000,15.000000);
\begin{scope}[rounded corners=2pt]
\begin{scope}
\draw[fill=white] (96.000000, 37.500000) +(-45.000000:8.485281pt and 19.091883pt) -- +(45.000000:8.485281pt and 19.091883pt) -- +(135.000000:8.485281pt and 19.091883pt) -- +(225.000000:8.485281pt and 19.091883pt) -- cycle;
\clip (96.000000, 37.500000) +(-45.000000:8.485281pt and 19.091883pt) -- +(45.000000:8.485281pt and 19.091883pt) -- +(135.000000:8.485281pt and 19.091883pt) -- +(225.000000:8.485281pt and 19.091883pt) -- cycle;
\draw (96.000000, 37.500000) node {{4}};
\end{scope}
\end{scope}
\begin{scope}
\draw[fill=white] (96.000000, 15.000000) circle(3.000000pt);
\clip (96.000000, 15.000000) circle(3.000000pt);
\draw (93.000000, 15.000000) -- (99.000000, 15.000000);
\draw (96.000000, 12.000000) -- (96.000000, 18.000000);
\end{scope}
\draw (120.000000,75.000000) -- (120.000000,30.000000);
\begin{scope}[rounded corners=2pt]
\begin{scope}
\draw[fill=white] (120.000000, 67.500000) +(-45.000000:8.485281pt and 19.091883pt) -- +(45.000000:8.485281pt and 19.091883pt) -- +(135.000000:8.485281pt and 19.091883pt) -- +(225.000000:8.485281pt and 19.091883pt) -- cycle;
\clip (120.000000, 67.500000) +(-45.000000:8.485281pt and 19.091883pt) -- +(45.000000:8.485281pt and 19.091883pt) -- +(135.000000:8.485281pt and 19.091883pt) -- +(225.000000:8.485281pt and 19.091883pt) -- cycle;
\draw (120.000000, 67.500000) node {{2}};
\end{scope}
\end{scope}
\begin{scope}
\draw[fill=white] (120.000000, 30.000000) circle(3.000000pt);
\clip (120.000000, 30.000000) circle(3.000000pt);
\draw (117.000000, 30.000000) -- (123.000000, 30.000000);
\draw (120.000000, 27.000000) -- (120.000000, 33.000000);
\end{scope}
\draw (144.000000,120.000000) -- (144.000000,30.000000);
\begin{scope}[rounded corners=2pt]
\begin{scope}
\draw[fill=white] (144.000000, 82.500000) +(-45.000000:8.485281pt and 61.518290pt) -- +(45.000000:8.485281pt and 61.518290pt) -- +(135.000000:8.485281pt and 61.518290pt) -- +(225.000000:8.485281pt and 61.518290pt) -- cycle;
\clip (144.000000, 82.500000) +(-45.000000:8.485281pt and 61.518290pt) -- +(45.000000:8.485281pt and 61.518290pt) -- +(135.000000:8.485281pt and 61.518290pt) -- +(225.000000:8.485281pt and 61.518290pt) -- cycle;
\draw (144.000000, 82.500000) node {{3}};
\end{scope}
\end{scope}
\draw[color=black,dash pattern=on 2pt off 1pt] (138.000000, 105.000000) -- (150.000000, 105.000000);
\draw[color=black,dash pattern=on 2pt off 1pt] (138.000000, 90.000000) -- (150.000000, 90.000000);
\draw[color=black,dash pattern=on 2pt off 1pt] (138.000000, 75.000000) -- (150.000000, 75.000000);
\draw[color=black,dash pattern=on 2pt off 1pt] (138.000000, 60.000000) -- (150.000000, 60.000000);
\begin{scope}
\draw[fill=white] (144.000000, 30.000000) circle(3.000000pt);
\clip (144.000000, 30.000000) circle(3.000000pt);
\draw (141.000000, 30.000000) -- (147.000000, 30.000000);
\draw (144.000000, 27.000000) -- (144.000000, 33.000000);
\end{scope}
\draw (168.000000,105.000000) -- (168.000000,45.000000);
\begin{scope}[rounded corners=2pt]
\begin{scope}
\draw[fill=white] (168.000000, 97.500000) +(-45.000000:8.485281pt and 19.091883pt) -- +(45.000000:8.485281pt and 19.091883pt) -- +(135.000000:8.485281pt and 19.091883pt) -- +(225.000000:8.485281pt and 19.091883pt) -- cycle;
\clip (168.000000, 97.500000) +(-45.000000:8.485281pt and 19.091883pt) -- +(45.000000:8.485281pt and 19.091883pt) -- +(135.000000:8.485281pt and 19.091883pt) -- +(225.000000:8.485281pt and 19.091883pt) -- cycle;
\draw (168.000000, 97.500000) node {{1}};
\end{scope}
\end{scope}
\begin{scope}
\draw[fill=white] (168.000000, 45.000000) circle(3.000000pt);
\clip (168.000000, 45.000000) circle(3.000000pt);
\draw (165.000000, 45.000000) -- (171.000000, 45.000000);
\draw (168.000000, 42.000000) -- (168.000000, 48.000000);
\end{scope}
\draw[color=black] (180.000000,120.000000) node[right] {$x_1$};
\draw[color=black] (180.000000,105.000000) node[right] {$x_2$};
\draw[color=black] (180.000000,90.000000) node[right] {$x_3$};
\draw[color=black] (180.000000,75.000000) node[right] {$x_4$};
\draw[color=black] (180.000000,60.000000) node[right] {$x_5$};
\draw[color=black] (180.000000,45.000000) node[right] {$0$};
\draw[color=black] (180.000000,30.000000) node[right] {$y_1$};
\draw[color=black] (180.000000,15.000000) node[right] {$0$};
\draw[color=black] (180.000000,0.000000) node[right] {$y_2$};
\end{tikzpicture}}
%
%
  \caption{Simple LUT network to illustrate order heuristics (dashed lines in
    the single-target gates mean that the line is not input to the gate).}
  \label{fig:example}
\end{figure}
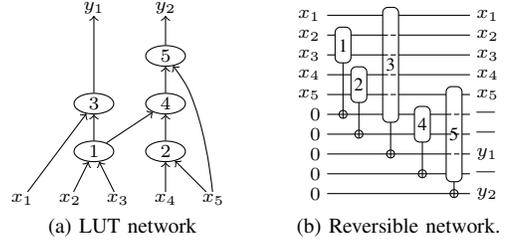
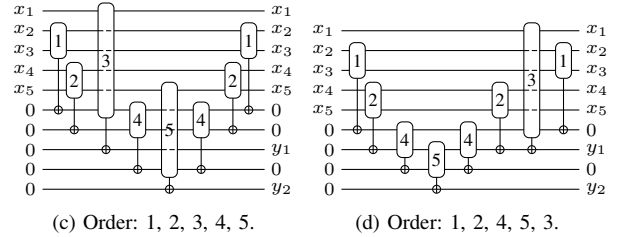

\subsection{Mapping $k$-LUTs into Single-target Gates}
Fig.~\ref{fig:example} illustrates the general idea how $k$-LUT
networks are mapped into reversible logic networks composed of
single-target gates with control functions with up to $k$ variables.
Fig.~\ref{fig:example}(a) shows a 2-LUT network with 5 inputs
$x_1, \dots, x_5$ and 5 LUTs with names $1$ to $5$.  It has two
outputs, $y_1$ and $y_2$, which functions are computed by LUT $3$ and
LUT $5$, respectively.

A straightforward way to translate the LUT network is by using one
single-target gate for each LUT in topological order.  The target of
each single-target gate is a 0-initialized new ancilla line.  The
reversible circuit in Fig.~\ref{fig:example}(b) results when applying
such a procedure.  With these five gates, the outputs $y_1$ and $y_2$
are realized at line 8 and 10 of the reversible circuit.  But after
these first five gates, the reversible circuit has garbage outputs on
lines 6, 7, and 9, indicated by \hbox{`---'}, which compute the
functions of the inner LUTs of the network.  The circuit must be free
of garbage outputs in order to be implemented on a quantum computer.
This is because the result of the calculation is entangled with the
intermediate results and so they cannot be discarded and recycled
without damaging the results they are entangled with~\cite{NC00}.  To
avoid the garbage outputs, we can \emph{uncompute} the intermediate
results by re-applying the single-target gates for the LUTs in reverse
topological order.  This disentangles the qubits, reverting them all
to constant 0s.  Fig.~\ref{fig:example}(c) shows the complete
reversible circuit; the last 3 gates uncompute intermediate results at
lines 6, 7, and 9.

But we can do better!  Once we have computed the LUT for a primary
output that does not fan in to another LUT, we can uncompute LUTs that
are not used any longer by other outputs.  The uncomputed lines
restore a 0 that can be used to store the intermediate results of
other LUTs instead of creating a new ancilla.  For the running
example, as shown in Fig.~\ref{fig:example}(d), we can first compute
output $y_2$ and then uncompute LUTs 4 and 2, as they are not in the
logic cone of output $y_1$.  The freed ancilla can be used for the
single-target gate realizing LUT 3.  Compared to the reversible
network in Fig.~\ref{fig:example}(c), this network requires one qubit
less by having the exact same gates.

\subsection{Bounds on the Number of Ancillae}
As we have seen in the previous section, the order in which LUTs are
traversed in the LUT network and translated into single-target gates
affects the number of qubits.  Two questions arise: (i) how many
ancillae do we need at least and at most, and (ii) what is a good
strategy?  We will answer the first question, and then discuss the
second one.

The example order that was used in the previous example leading to the
network in Fig.~\ref{fig:example}(c) illustrates an upper bound.  We
can always use one ancilla for each LUT in the LUT network, postulated
in the following lemma.

\begin{lemma}
  \label{lem:upper}
  When realizing a LUT network $N=(X \cup Y \cup G, A, F)$ by a
  reversible circuit that uses single-target gates for each LUT, one
  needs at most $|G|$ ancilla lines.
\end{lemma}

The optimized order in Fig.~\ref{fig:example}(d) used the fact, that
one can uncompute gates in the transitive fan-in cone of an output,
once the output has been computed.

This observation leads to a lemma
providing a lower bound.

\begin{lemma}
  \label{lem:lower}
  Given a LUT network $N = (X \cup Y \cup G, A, F)$, let
  \[
    l = \max \{\#\tfi(y) \mid y \in Y\}
  \]
  be the maximum cone size over all outputs.  When realizing the LUT
  network by a reversible circuit that uses single-target gates for
  each LUT, we need at least $l$ ancilla lines.
\end{lemma}

\begin{algorithm}[t]
\small
\Fn{$\operatorname{synthesize\_mapping}(N = (X \cup Y \cup G, A, F), p_{\mathrm{Q}})$\label{alg:l-main}}{%
  set $R \gets$ empty reversible network\;
  set $l \gets 1$\; \label{alg:l-init1}
  initialize empty stack $C$\;
  initialize empty map $m$\;
  set $S \gets \emptyset$\;\label{alg:l-init2}
  set $D \gets \{\driver(y) \mid y \in Y\}$\;\label{alg:l-drivers} \smallskip
  \lFor{$g \in G$}{set $r(g) \gets d^+(g) - [g \in D]$\label{alg:l-ref}}
  \For{$x \in X$\label{alg:l-inpb}}{%
    add input line with name $x$ to $R$\;
    set $m(x) \gets l$\;
    set $l \gets l + 1$\;
  }\label{alg:l-inpe}
  \For{$g \in \operatorname{topo\_order}(G, p_{\mathrm{Q}})$\label{alg:l-lutb}}{%
    set $t \gets \operatorname{request\_constant}()$\;\label{alg:l-reqc}
    append $\T_{F(g)}(m(\fanin(g)), t)$ to $R$\;\label{alg:l-comp}
    set $m(g) \gets t$\;\label{alg:l-mapt}
    \If{$r(g) = 0$\label{alg:l-checkunc}}{%
      set $S \gets \emptyset$\;
      $\operatorname{uncompute\_children(g)}$\;\label{alg:l-startunc}
    }
  }\label{alg:l-lute}
  \For{$y \in Y$\label{alg:l-outb}}{%
    rename output of line $m(\driver(y))$ in $R$ to $y$\;
  }\label{alg:l-oute}
  \KwRet{$R$}\;
}

\medskip
\Fn{$\operatorname{request\_constant}()$}{%
  \uIf{$C$ is not empty}{%
    \KwRet{$C$.$\operatorname{pop}()$}\;
  }\Else{%
    set $l \gets l + 1$\;
    \KwRet{$l$}\;
  }
}

\medskip
\Fn{$\operatorname{uncompute\_children}(g)$}{%
  \lFor{$g' \in \fanin(g) \cap G$}{set $r(g') \gets r(g') - 1$\label{alg:l-decref}}
  \For{$g' \in \fanin(g)$ such that $r(g') = 0$}{%
    $\operatorname{uncompute\_gate}(g')$\;\label{alg:l-uncgate}
  }
}

\medskip
\Fn{$\operatorname{uncompute\_gate}(g)$}{%
  \lIf{$g \in S$}{\KwRet{}\label{alg:l-checks}}
  \uIf{$g \notin D$}{%
    set $t \gets m(g)$\;
    append $\T_{F(g)}(m(\fanin(g)), m(g))$ to $R$\;\label{alg:l-unc}
    $C$.$\operatorname{push}(t)$\;\label{alg:l-pushc}
    set $m(t) \gets 0$\;\label{alg:l-freem}
  }
  set $S \gets S \cup \{g\}$\;\label{alg:l-adds}
  $\operatorname{uncompute\_children}(g)$\;\label{alg:l-recur}
}

\medskip
\caption{Synthesizing a LUT mapping into a reversible network with
  single-target gates.}
\label{alg:synth-seq}
\end{algorithm}

\begin{figure*}[t!]
\captionsetup{margin=10pt}
\subfloat[Max: the actual number of additional lines often matches the upper bound; after $k=19$ increasing the LUT size has no strong effect on the number of additional lines.]{\includegraphics{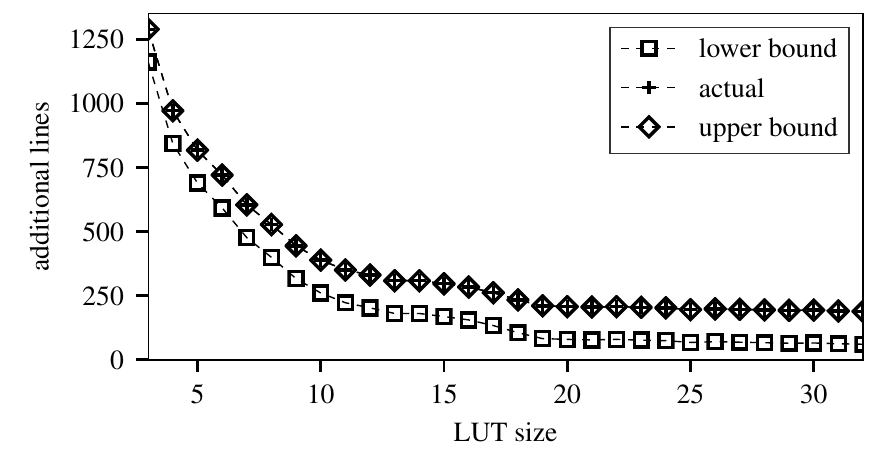}}\hfill%
\subfloat[Adder: the actual number of additional lines often matches the upper bound; the additional lines decrease almost linearly when increasing the LUT size.]{\includegraphics{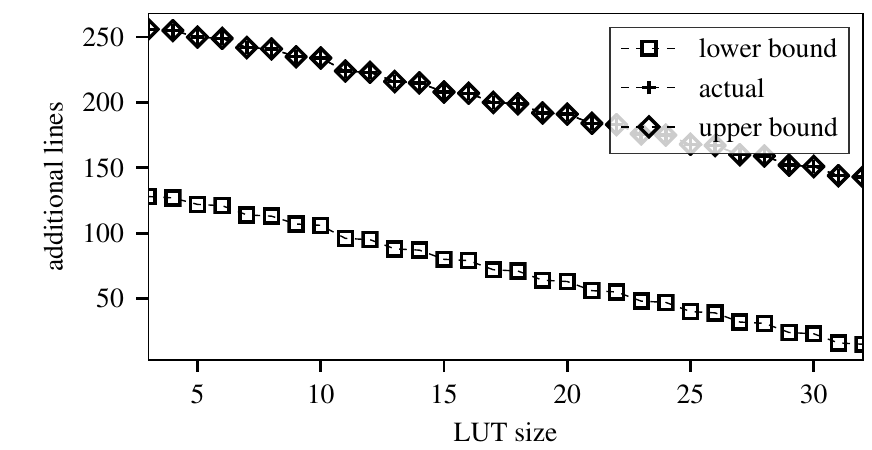}}

\subfloat[Divisor: the actual number of additional lines often matches the lower bound.]{\includegraphics{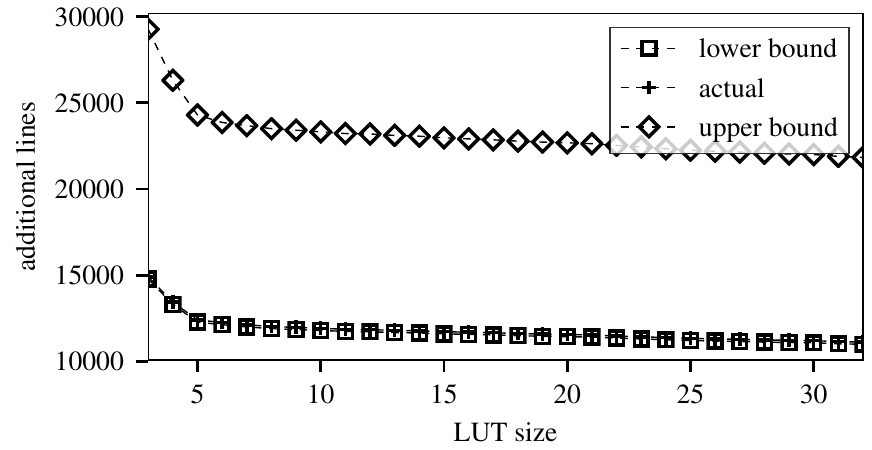}}\hfill%
\subfloat[Log2: the upper and lower bound are very close to each other.  Note that for $k=32$ the optimum number of additional lines is achieved, since the function has 32 inputs.]{\includegraphics{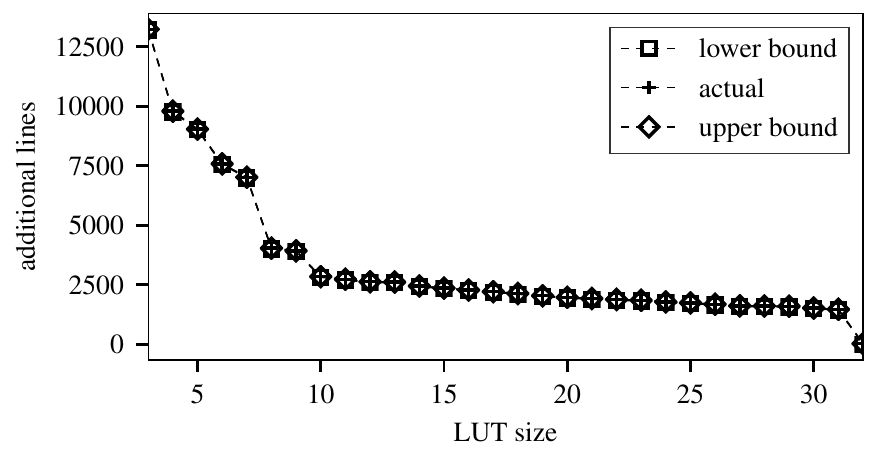}}
\caption{The plots show the upper and lower bound according to
  Lemmas~\ref{lem:upper} and~\ref{lem:lower} as well as the actual
  number of additional lines when synthesizing different arithmetic
  benchmarks with LUT sizes ranging from 3 to 32.  The $x$-axis shows
  LUT size and the $y$-axis shows the number of additional lines.}
\label{fig:lut-size-exp}
\end{figure*}

The lower bound inspires the following synthesis strategy that
minimizes the number of additional lines.  One starts by synthesizing
a circuit for the output with the maximum cone. Let's assume that this
cone contains $l$ LUTs.  These LUTs can be synthesized using $l$
single-target gates. Note that all of these are in fact needed,
because in order to uncompute a gate, the intermediate values of
children need to be available.  From these $l$ gates, $l-1$ gates can
be uncomputed (all except the LUT computing the output), and therefore
restores $l-1$ lines which hold a constant 0 value.  We can easily see
that the exact number of required lines may be a bit larger, since all
output values need to be kept.  Note that this strategy uncomputes all
LUTs in the transitive fan-in cone of an output---even if it is part
of a fan-in cone of another output.  Therefore, some LUTs will lead to
more than two single-target gates in the reversible network.

For a good tradeoff between the number of qubits and $T$-count one is
interested in the minimum number of qubits such that each LUT is
translated into at most two single-target gates in the reversible
network.  Finding the minimum number of ancillae under such
constraints relates to playing the \emph{reversible pebble
  game}~\cite{Bennett89} in minimum time using minimum number of
pebbles.  More details can be found
in~\cite{Kralovic01,Chan13,KSS15,PRS15}.

\subsection{Synthesizing a LUT Network}
\label{sec:implementation}
Alg.~\ref{alg:synth-seq} describes in detail how a $k$-LUT
network $N = (X \cup Y \cup G, A, F)$ is mapped into a reversible
network $R$ that consists of single-target gates with at most $k$
controls.  The main entry point is the function
`$\operatorname{synthesize\_mapping}$' (line~\ref{alg:l-main}).  This
function keeps track of the current number of lines $l$, available
ancillae in a stack $C$, a LUT-to-line mapping $m : G \to \N$ that
stores which LUT gates are computed on which lines in $R$, and a
visited list $S$ (lines~\ref{alg:l-init1}--\ref{alg:l-init2}).  The
reference counter $r(g)$ checks for each LUT $g$ how often it is
required as input to other LUTs.  For driving LUTs, stored in $D$
(line~\ref{alg:l-drivers}), the reference counter is decreased by 1.
It is initialized with the fan-out size and allows us to check if $g$
is not needed any longer such that it can be uncomputed
(line~\ref{alg:l-ref}).  This is the case whenever the reference
counter is $0$, and therefore the process of uncomputing is triggered
by driving LUTs.

\begin{figure*}[t]
  \centering
  \subfloat[Direct mapping.]{%
\begin{tikzpicture}[font=\footnotesize,node distance=.5cm]
  \node[flowtool] (aige) {AIG extract};
  \node[below=.2cm of aige,flowtool] (bdde) {BDD extract};

  \coordinate (c1) at (bdde.west |- aige.west);
  \coordinate (c2) at ($(c1)!.5!(bdde.west)$);
  \coordinate (c3) at (c2 -| bdde.east);

  \begin{scope}[every node/.style={flowbox}]
    \node[left=of c2] (aig) {$k$-LUT (AIG)};
    \node[right=of c3] (esop1) {ESOP};
  \end{scope}

  \coordinate (c4) at (esop1.east);
  \coordinate (c5) at ([xshift=.5cm] c4 |- aige.east);

  \node[right,flowtool] (min) at (c5) {ESOP minimization};

  \coordinate (c6) at (min.south |- esop1.east);

  \node[flowbox] (esop2) at ($(esop1)!2!(c6)$) {ESOP};
  \node[right=of esop2,flowtool] (esopbs) {ESOP-based \\ synthesis};
  \node[right=of esopbs,flowbox] (revc) {Reversible \\ network};
  \node[right=of revc,flowtool] (map) {Clifford+$T$ \\ mapping};
  \node[right=of map,flowbox] (ctn) {Clifford+$T$ \\ network};

  \begin{scope}[->]
    \draw (aig.east) to[out=0,in=180] (aige.west);
    \draw (aig.east) to[out=0,in=180] (bdde.west);
    \draw (aige.east) to[out=0,in=180] (esop1.west);
    \draw (bdde.east) to[out=0,in=180] (esop1.west);
    \draw (esop1.east) to[out=0,in=180] (min.west);
    \draw (min.east) to[out=0,in=180] (esop2.west);
    \draw (esop1.east) to (esop2.west);
    \draw (esop2.east) to (esopbs.west);
    \draw (esopbs.east) to (revc.west);
    \draw (revc.east) to (map.west);
    \draw (map.east) to (ctn.west);
  \end{scope}
\end{tikzpicture}}

\subfloat[LUT-based mapping.]{%
\begin{tikzpicture}[font=\footnotesize,node distance=.5cm]
  \node[flowbox] (aig) {$k$-LUT (AIG)};

  \node[flowtool,right=of aig] (4map) {$4$-LUT \\ mapping};

  \node[flowbox,right=of 4map] (4lut) {$4$-LUT \\ network};

  \node[flowbox,right=1cm of 4lut,yshift=.5cm] (lut) {$4$-LUT};

  \node[flowtool,right=of lut] (cfy) {AN-classification};

  \node[flowbox,right=of cfy] (class) {AN-class};

  \node[flowtool,right=of class] (dbl) {DB lookup};

  \node[flowbox,right=of dbl] (cts) {opt.\ Clifford+$T$ \\ network};

  \node[flowbox,right=.9cm of cts,yshift=-.5cm] (ctn) {Clifford+$T$ \\ network};

  \node[fit=(lut) (cts),draw] (loop) {};
  \node[anchor=west,xshift=3pt,draw,rounded corners=2pt,inner sep=1pt,font=\scriptsize\itshape,fill=white] at (loop.north west) {for each $4$-LUT};

  \node[flowtool,below=.3 of loop] (direct) {direct mapping of $k$-LUT network};

  \begin{scope}[->]
    \draw (aig.east) to[out=0,in=180] (4map.west);
    \draw (4map.east) to[out=0,in=180] (4lut.west);
    \draw (4lut.east) to[out=0,in=180] node[pos=.65,sloped,above,font=\scriptsize\itshape,align=center,inner sep=1pt] {enough \\ ancilla} (loop.west);
    \draw (loop.east) to[out=0,in=180] (ctn.west);

    \coordinate (a) at ([xshift=50pt] 4lut.east);
    \coordinate (b) at (a |- direct.west);
    \draw (4lut.east) to[out=0,in=180] node[pos=.4,sloped,below,font=\scriptsize\itshape,align=center,inner sep=1pt] {not enough \\ ancilla} (b) -- (direct.west);
    \coordinate (a) at ([xshift=-50pt] ctn.west);
    \coordinate (b) at (a |- direct.east);
    \draw (direct.east) -- (b) to[out=0,in=180] (ctn.west);
    \draw (lut.east) to[out=0,in=180] (cfy.west);
    \draw (cfy.east) to[out=0,in=180] (class.west);
    \draw (class.east) to[out=0,in=180] (dbl.west);
    \draw (dbl.east) to[out=0,in=180] (cts.west);
  \end{scope}








\end{tikzpicture}}

  \caption{Algorithms to map a single-target gate into a Clifford+$T$
    network.}
  \label{fig:decomp}
\end{figure*}
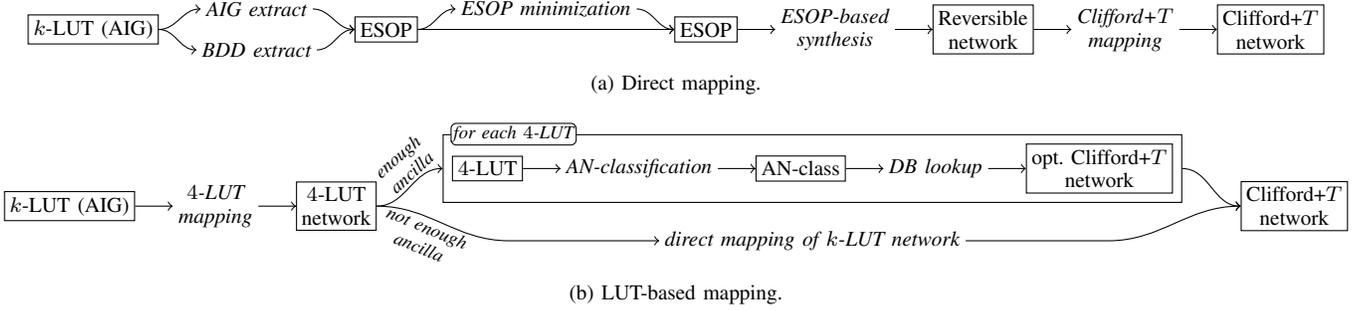

Input lines are added to $R$ in
lines~\ref{alg:l-inpb}--\ref{alg:l-inpe}.  Input vertices are mapped
to their line in $R$ using $m$.  In
lines~\ref{alg:l-lutb}--\ref{alg:l-lute} single-target gates to
compute and uncompute LUTs are added to $R$.  Each gate $g$ is visited
in topological order (details on `$\operatorname{topo\_order}$' follow
later).  First, a 0-initialized line $t$ is requested
(line~\ref{alg:l-reqc}).  Either there is one in $C$ or we get a new
line by incrementing $l$.  Given $t$, a single-target gate with
control function $F(g)$, controls
\begin{equation}
  m(\fanin(g)) = \{m(g') \mid g' \in \fanin(g)\},
\end{equation}
and target line $t$ is added $R$ (line~\ref{alg:l-comp}).  The
LUT-to-line map is updated according to the newly added gate
(line~\ref{alg:l-mapt}).  Then, if $g$ is driving an output, i.e.,
$r(g) = 0$ (line~\ref{alg:l-checkunc}), we try to uncompute the
children recursively by calling $\operatorname{uncompute\_children}$
(line~\ref{alg:l-startunc}).  In that function, first the reference
counter is decremented for each child $g'$ that is not a primary input
(line~\ref{alg:l-decref}).  Then, each child $g'$ that afterwards has
a reference count of $0$, is uncomputed using
$\operatorname{uncompute\_gate}$ (line~\ref{alg:l-uncgate}).  In
there, first it is checked whether the child has already been visited
(line~\ref{alg:l-checks}).  If the child is not driving an output, a
single-target gate to uncompute the line is added to $R$
(line~\ref{alg:l-unc}).  The freed line $t$ is added to $C$
(line~\ref{alg:l-pushc}) and the mapping is cleared accordingly
(line~\ref{alg:l-freem}).  The child is stored as visited
(line~\ref{alg:l-adds}), and the function recurs
(line~\ref{alg:l-recur}).  After all gates have been computed, outputs
are added to lines in $R$ (lines~\ref{alg:l-outb}--\ref{alg:l-oute}).
This procedure is simplified: two or more outputs may share the same
driving LUT.  In this case, one needs additional lines and copy the
output result using a CNOT gate.

With a given topological order of LUTs, the time complexity of
Alg.~\ref{alg:synth-seq} is linear in the number of LUTs.  As seen in
the beginning of this section, the order in which LUTs are visited has
an effect on the number of qubits.  Therefore, there can be several
strategies to compute a topological order on the gates.  This is
handled by the function `$\mathrm{topo\_order}$' that is configured by
a parameter in $p_{\mathrm{Q}}$.  Besides the default topological
order implied by the implementation of $N$ (referred to as
\emph{topo\_def} in the following), we also implemented the order
\emph{topo\_tfi\_sort}, which is inspired by Lemma~\ref{lem:lower}:
compute the transitive fan-in cone for each primary output and order
them by size in descending order.  The topological order is obtained
using a depth-first search for each cone by not including duplicates
when traversing a cone.

\subsection{The Role of the LUT Size}
\label{sec:role-lut-size}
As can be seen from previous discussions, the number of additional
lines roughly corresponds to the number of LUTs.  Hence, we are
interested in logic synthesis algorithms that minimize the number of
LUTs.  In classical logic synthesis the number of LUT-inputs $k$ needs
to be selected according to some target architecture.  For example in
FPGA mapping, its value is typically 6 or 7.  But for our algorithm,
we can use $k$ as a parameter that trades off the number of qubits to
the number of $T$ gates: If $k$ is small, one needs many LUTs to
realize the function, but the small number of inputs also limits the
number of control lines when mapping the single-target gates into
multiple-controlled Toffoli gates.  On the contrary, when $k$ is
large, one needs fewer LUTs but the resulting Toffoli gates are larger
and therefore require more $T$ gates.  Further, since for larger $k$
the LUT functions are getting more complex, the runtime to map a
single-target gate into multiple-controlled Toffoli gates increases.

To illustrate the influence of the LUT size we performed the following
experiment, illustrated in Fig.~\ref{fig:lut-size-exp}(a).  For four
benchmarks and for LUT sizes $k$ from 3 to 32, we computed a LUT
mapping using ABC's~\cite{BM10} command `\emph{if -K $k$ -a}'.  The
resulting network was used to compute both the upper and lower bound
on the number of additional lines according to Lemmas~\ref{lem:upper}
and~\ref{lem:lower}, and to compute the actual number of lines
according to Alg.~\ref{alg:synth-seq} with `\emph{topo\_def}' as
topological order.  It can be noted that the actual bound often either
matches the upper bound or the lower bound.  In some cases the bounds
are very close to each other, leaving not much flexibility to improve
on the number of additional lines.  Further, after larger LUT sizes
the gain in reducing the number of lines decreases when increasing the
LUT size.  It should be pointed out that for benchmark `\emph{Log2}'
an optimum number of additional lines can be achieved for $k = 32$,
because in this case $k$ matches the number of inputs of the function.
Consequently, the LUT mapping has as many gates as the number of
outputs.

\section{Mapping Single-target Gates}
\label{sec:synth-gates}
For the following discussion it is important to understand the
representation of the logic network that is given as input to
Alg.~\ref{alg:lhrs} and the $k$-LUT network that from results the
first step.  The input network is given as a gate-level logic network,
i.e., all gates are simple logic gates.  In our experimental
evaluation and current implementation the logic network is given as
AIG, i.e., a logic network composed of AND gates and inverters.  The
LUT network is represented by annotating in the gate-level netlist (i)
which nodes are LUT outputs and (ii) which nodes are LUT inputs for
each LUT.  As a result, the function of a LUT is \emph{implicitly}
represented as subnetwork in the gate-level logic network.

\subsection{Direct Mapping}
\label{sec:direct-decomp}
The idea of direct mapping is to represent the LUT function as ESOP
expression, optimize it, and then translate the optimized ESOP into a
Clifford+$T$ network.  Fig.~\ref{fig:decomp}(a) illustrates the
complete direct mapping flow.  As described above, a LUT function is
represented in terms of a multi-level AIG.  In order to obtain a
2-level ESOP expression for the LUT function, one needs to collapse
the network.  This process is called \emph{cover extraction} and two
techniques called \emph{AIG extract} and \emph{BDD extract} will be
described in this section.  The number of product terms in the
resulting ESOP expression is typically far from optimal and is reduced
using ESOP minimization.  The optimized ESOP expression is first
translated into a reversible network with multiple-controlled Toffoli
gates as described in Section~\ref{sec:mapping} and then each
multiple-controlled Toffoli gate is mapped into a Clifford+$T$ with
the mapping described in~\cite{Maslov16}.

\subsubsection*{ESOP cover extraction} There are several ways to
extract an ESOP expression from an AIG that represents the same
function.  Our implementation uses two different methods.  The choice
of the method has an influence on the initial ESOP expression and
therefore affects both the runtime of the algorithm and the number of
$T$ gates in the final network.

The method \emph{AIG extract} computes an ESOP for each node in the
AIG in topological order.  The final ESOP can then be read from the
output node.  First, all primary inputs $x_i$ are assigned the ESOP
expression $x_i$.  The ESOP expression of an AND gate is computed by
conjoining both ESOP expressions of the children, taking into
consideration possible complementation.  Therefore, the number of
product terms for the AND gate can be as large as the product of the
number of terms of the children.  The final ESOP can be preoptimized
by removing terms that occur twice, e.g.,
$x_1\bar x_2x_3 \oplus x_1\bar x_2x_3 = 0$, or by merging terms that
differ in a single literal, e.g.,
$x_1x_3 \oplus x_1\bar x_2x_3 = x_1x_2x_3$~\cite{MP01}.  \emph{AIG
  extract} is implemented similar to the command `\emph{\&esop}' in
ABC~\cite{BM10}.  We were able to increase the performance of our
implementation by limiting the number of inputs to 32 bits, which is
sufficient in our application, and by using cube
hashing~\cite{SMK+17}.

The method \emph{BDD extract} first expresses the LUT function in
terms of an AIG, again by translating each node into a BDD in
topological order.  From the BDD a Pseudo-Kronecker
expression~\cite{DDT78,Sasao93} is extracted using the algorithm
presented in~\cite{Drechsler99}.  A Pseudo-Kronecker expression is a
special case of an ESOP expression.  For the extracted expression, it
can be shown that it is minimum in the number of product terms with
respect to a chosen variable order.  Therefore, it provides a good
starting point for ESOP minimization.  In our experiments we noticed
that \emph{AIG extract} was always superior and \emph{BDD extract} did
not show any advantage.  \emph{BDD extract} may be advantageous for
larger LUT sizes.

\subsubsection*{ESOP minimization}
In our implementation we use \emph{exorcism}~\cite{MP01} to minimize
the number of terms in the ESOP expression.  Exorcism is a heuristic
minimization algorithm that applies local rewriting of product terms
using the \emph{EXORLINK} operation~\cite{Song92}.  In order to
introduce this operation, we need to define the notation of distance.
For each product term a variable can either appear as positive
literal, as negative literal, or not at all.  The \emph{distance} of
two product terms is the number of variables with different appearance
in each term.  For example, the two product terms $x_1x_3$ and
$x_1\bar x_2x_3$ have distance 1, since $x_2$ does not appear in the
first product term and appears as negative literal in the second one.
It can be shown that two product terms with distance $k$ can be
rewritten as an equivalent ESOP expression with $k$ product terms in
$k!$ different ways.  The EXORLINK-$k$ operation is a procedure to
enumerate all $k!$ replacements for a product term pair with distance
$k$.  Applying the EXORLINK operation to product term pairs with a
distance of less than 2 immediately leads to a reduction of the number
of product terms in an ESOP expression.  In fact, as described above,
such checks are already applied when creating the initial cover.
Applying the EXORLINK-2 operation does not increase the number of
product terms but can decrease it, if product terms in the replacement
can be combined with other terms.  The same applies for distances
larger than 2, but it can also lead to an increase in the number of
product terms.  This can sometimes be helpful to escape local minima.
Exorcism implements a default minimization heuristic, referred to as
\emph{def} in the following, that applies different combinations and
sequences of EXORLINK-$k$ operations for $2 \le k \le 4$.  We have
modified the heuristic by just omitting the EXORLINK-4 operations,
referred to as \emph{def\_wo4} in the following.

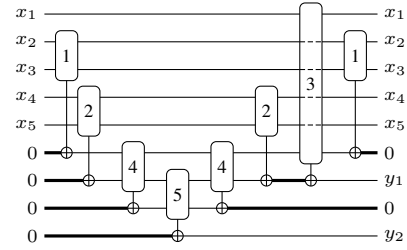
\begin{figure}[t]
  \centering
  \scriptsize
  \begin{tikzpicture}[scale=0.700000,x=1pt,y=1pt]
\filldraw[color=white] (0.000000, -7.500000) rectangle (180.000000, 127.500000);
\draw[color=black] (0.000000,120.000000) -- (180.000000,120.000000);
\draw[color=black] (0.000000,120.000000) node[left] {$x_1$};
\draw[color=black] (0.000000,105.000000) -- (180.000000,105.000000);
\draw[color=black] (0.000000,105.000000) node[left] {$x_2$};
\draw[color=black] (0.000000,90.000000) -- (180.000000,90.000000);
\draw[color=black] (0.000000,90.000000) node[left] {$x_3$};
\draw[color=black] (0.000000,75.000000) -- (180.000000,75.000000);
\draw[color=black] (0.000000,75.000000) node[left] {$x_4$};
\draw[color=black] (0.000000,60.000000) -- (180.000000,60.000000);
\draw[color=black] (0.000000,60.000000) node[left] {$x_5$};
\draw[color=black,very thick] (0.000000,45.000000) -- (12.000000,45.000000);
\draw[color=black,thin] (12.000000,45.000000) -- (168.000000,45.000000);
\draw[color=black,very thick] (168.000000,45.000000) -- (180.000000,45.000000);
\draw[color=black] (0.000000,45.000000) node[left] {$0$};
\draw[color=black,very thick] (0.000000,30.000000) -- (24.000000,30.000000);
\draw[color=black,thin] (24.000000,30.000000) -- (120.000000,30.000000);
\draw[color=black,very thick] (120.000000,30.000000) -- (144.000000,30.000000);
\draw[color=black,thin] (144.000000,30.000000) -- (180.000000,30.000000);
\draw[color=black] (0.000000,30.000000) node[left] {$0$};
\draw[color=black,very thick] (0.000000,15.000000) -- (48.000000,15.000000);
\draw[color=black,thin] (48.000000,15.000000) -- (96.000000,15.000000);
\draw[color=black,very thick] (96.000000,15.000000) -- (180.000000,15.000000);
\draw[color=black] (0.000000,15.000000) node[left] {$0$};
\draw[color=black,very thick] (0.000000,0.000000) -- (72.000000,0.000000);
\draw[color=black,thin] (72.000000,0.000000) -- (180.000000,0.000000);
\draw[color=black] (0.000000,0.000000) node[left] {$0$};
\draw (12.000000,105.000000) -- (12.000000,45.000000);
\begin{scope}[rounded corners=2pt]
\begin{scope}
\draw[fill=white] (12.000000, 97.500000) +(-45.000000:8.485281pt and 19.091883pt) -- +(45.000000:8.485281pt and 19.091883pt) -- +(135.000000:8.485281pt and 19.091883pt) -- +(225.000000:8.485281pt and 19.091883pt) -- cycle;
\clip (12.000000, 97.500000) +(-45.000000:8.485281pt and 19.091883pt) -- +(45.000000:8.485281pt and 19.091883pt) -- +(135.000000:8.485281pt and 19.091883pt) -- +(225.000000:8.485281pt and 19.091883pt) -- cycle;
\draw (12.000000, 97.500000) node {{1}};
\end{scope}
\end{scope}
\begin{scope}
\draw[fill=white] (12.000000, 45.000000) circle(3.000000pt);
\clip (12.000000, 45.000000) circle(3.000000pt);
\draw (9.000000, 45.000000) -- (15.000000, 45.000000);
\draw (12.000000, 42.000000) -- (12.000000, 48.000000);
\end{scope}
\draw (24.000000,75.000000) -- (24.000000,30.000000);
\begin{scope}[rounded corners=2pt]
\begin{scope}
\draw[fill=white] (24.000000, 67.500000) +(-45.000000:8.485281pt and 19.091883pt) -- +(45.000000:8.485281pt and 19.091883pt) -- +(135.000000:8.485281pt and 19.091883pt) -- +(225.000000:8.485281pt and 19.091883pt) -- cycle;
\clip (24.000000, 67.500000) +(-45.000000:8.485281pt and 19.091883pt) -- +(45.000000:8.485281pt and 19.091883pt) -- +(135.000000:8.485281pt and 19.091883pt) -- +(225.000000:8.485281pt and 19.091883pt) -- cycle;
\draw (24.000000, 67.500000) node {{2}};
\end{scope}
\end{scope}
\begin{scope}
\draw[fill=white] (24.000000, 30.000000) circle(3.000000pt);
\clip (24.000000, 30.000000) circle(3.000000pt);
\draw (21.000000, 30.000000) -- (27.000000, 30.000000);
\draw (24.000000, 27.000000) -- (24.000000, 33.000000);
\end{scope}
\draw (48.000000,45.000000) -- (48.000000,15.000000);
\begin{scope}[rounded corners=2pt]
\begin{scope}
\draw[fill=white] (48.000000, 37.500000) +(-45.000000:8.485281pt and 19.091883pt) -- +(45.000000:8.485281pt and 19.091883pt) -- +(135.000000:8.485281pt and 19.091883pt) -- +(225.000000:8.485281pt and 19.091883pt) -- cycle;
\clip (48.000000, 37.500000) +(-45.000000:8.485281pt and 19.091883pt) -- +(45.000000:8.485281pt and 19.091883pt) -- +(135.000000:8.485281pt and 19.091883pt) -- +(225.000000:8.485281pt and 19.091883pt) -- cycle;
\draw (48.000000, 37.500000) node {{4}};
\end{scope}
\end{scope}
\begin{scope}
\draw[fill=white] (48.000000, 15.000000) circle(3.000000pt);
\clip (48.000000, 15.000000) circle(3.000000pt);
\draw (45.000000, 15.000000) -- (51.000000, 15.000000);
\draw (48.000000, 12.000000) -- (48.000000, 18.000000);
\end{scope}
\draw (72.000000,30.000000) -- (72.000000,0.000000);
\begin{scope}[rounded corners=2pt]
\begin{scope}
\draw[fill=white] (72.000000, 22.500000) +(-45.000000:8.485281pt and 19.091883pt) -- +(45.000000:8.485281pt and 19.091883pt) -- +(135.000000:8.485281pt and 19.091883pt) -- +(225.000000:8.485281pt and 19.091883pt) -- cycle;
\clip (72.000000, 22.500000) +(-45.000000:8.485281pt and 19.091883pt) -- +(45.000000:8.485281pt and 19.091883pt) -- +(135.000000:8.485281pt and 19.091883pt) -- +(225.000000:8.485281pt and 19.091883pt) -- cycle;
\draw (72.000000, 22.500000) node {{5}};
\end{scope}
\end{scope}
\begin{scope}
\draw[fill=white] (72.000000, 0.000000) circle(3.000000pt);
\clip (72.000000, 0.000000) circle(3.000000pt);
\draw (69.000000, 0.000000) -- (75.000000, 0.000000);
\draw (72.000000, -3.000000) -- (72.000000, 3.000000);
\end{scope}
\draw (96.000000,45.000000) -- (96.000000,15.000000);
\begin{scope}[rounded corners=2pt]
\begin{scope}
\draw[fill=white] (96.000000, 37.500000) +(-45.000000:8.485281pt and 19.091883pt) -- +(45.000000:8.485281pt and 19.091883pt) -- +(135.000000:8.485281pt and 19.091883pt) -- +(225.000000:8.485281pt and 19.091883pt) -- cycle;
\clip (96.000000, 37.500000) +(-45.000000:8.485281pt and 19.091883pt) -- +(45.000000:8.485281pt and 19.091883pt) -- +(135.000000:8.485281pt and 19.091883pt) -- +(225.000000:8.485281pt and 19.091883pt) -- cycle;
\draw (96.000000, 37.500000) node {{4}};
\end{scope}
\end{scope}
\begin{scope}
\draw[fill=white] (96.000000, 15.000000) circle(3.000000pt);
\clip (96.000000, 15.000000) circle(3.000000pt);
\draw (93.000000, 15.000000) -- (99.000000, 15.000000);
\draw (96.000000, 12.000000) -- (96.000000, 18.000000);
\end{scope}
\draw (120.000000,75.000000) -- (120.000000,30.000000);
\begin{scope}[rounded corners=2pt]
\begin{scope}
\draw[fill=white] (120.000000, 67.500000) +(-45.000000:8.485281pt and 19.091883pt) -- +(45.000000:8.485281pt and 19.091883pt) -- +(135.000000:8.485281pt and 19.091883pt) -- +(225.000000:8.485281pt and 19.091883pt) -- cycle;
\clip (120.000000, 67.500000) +(-45.000000:8.485281pt and 19.091883pt) -- +(45.000000:8.485281pt and 19.091883pt) -- +(135.000000:8.485281pt and 19.091883pt) -- +(225.000000:8.485281pt and 19.091883pt) -- cycle;
\draw (120.000000, 67.500000) node {{2}};
\end{scope}
\end{scope}
\begin{scope}
\draw[fill=white] (120.000000, 30.000000) circle(3.000000pt);
\clip (120.000000, 30.000000) circle(3.000000pt);
\draw (117.000000, 30.000000) -- (123.000000, 30.000000);
\draw (120.000000, 27.000000) -- (120.000000, 33.000000);
\end{scope}
\draw (144.000000,120.000000) -- (144.000000,30.000000);
\begin{scope}[rounded corners=2pt]
\begin{scope}
\draw[fill=white] (144.000000, 82.500000) +(-45.000000:8.485281pt and 61.518290pt) -- +(45.000000:8.485281pt and 61.518290pt) -- +(135.000000:8.485281pt and 61.518290pt) -- +(225.000000:8.485281pt and 61.518290pt) -- cycle;
\clip (144.000000, 82.500000) +(-45.000000:8.485281pt and 61.518290pt) -- +(45.000000:8.485281pt and 61.518290pt) -- +(135.000000:8.485281pt and 61.518290pt) -- +(225.000000:8.485281pt and 61.518290pt) -- cycle;
\draw (144.000000, 82.500000) node {{3}};
\end{scope}
\end{scope}
\draw[color=black,dash pattern=on 2pt off 1pt] (138.000000, 105.000000) -- (150.000000, 105.000000);
\draw[color=black,dash pattern=on 2pt off 1pt] (138.000000, 90.000000) -- (150.000000, 90.000000);
\draw[color=black,dash pattern=on 2pt off 1pt] (138.000000, 75.000000) -- (150.000000, 75.000000);
\draw[color=black,dash pattern=on 2pt off 1pt] (138.000000, 60.000000) -- (150.000000, 60.000000);
\begin{scope}
\draw[fill=white] (144.000000, 30.000000) circle(3.000000pt);
\clip (144.000000, 30.000000) circle(3.000000pt);
\draw (141.000000, 30.000000) -- (147.000000, 30.000000);
\draw (144.000000, 27.000000) -- (144.000000, 33.000000);
\end{scope}
\draw (168.000000,105.000000) -- (168.000000,45.000000);
\begin{scope}[rounded corners=2pt]
\begin{scope}
\draw[fill=white] (168.000000, 97.500000) +(-45.000000:8.485281pt and 19.091883pt) -- +(45.000000:8.485281pt and 19.091883pt) -- +(135.000000:8.485281pt and 19.091883pt) -- +(225.000000:8.485281pt and 19.091883pt) -- cycle;
\clip (168.000000, 97.500000) +(-45.000000:8.485281pt and 19.091883pt) -- +(45.000000:8.485281pt and 19.091883pt) -- +(135.000000:8.485281pt and 19.091883pt) -- +(225.000000:8.485281pt and 19.091883pt) -- cycle;
\draw (168.000000, 97.500000) node {{1}};
\end{scope}
\end{scope}
\begin{scope}
\draw[fill=white] (168.000000, 45.000000) circle(3.000000pt);
\clip (168.000000, 45.000000) circle(3.000000pt);
\draw (165.000000, 45.000000) -- (171.000000, 45.000000);
\draw (168.000000, 42.000000) -- (168.000000, 48.000000);
\end{scope}
\draw[color=black] (180.000000,120.000000) node[right] {$x_1$};
\draw[color=black] (180.000000,105.000000) node[right] {$x_2$};
\draw[color=black] (180.000000,90.000000) node[right] {$x_3$};
\draw[color=black] (180.000000,75.000000) node[right] {$x_4$};
\draw[color=black] (180.000000,60.000000) node[right] {$x_5$};
\draw[color=black] (180.000000,45.000000) node[right] {$0$};
\draw[color=black] (180.000000,30.000000) node[right] {$y_1$};
\draw[color=black] (180.000000,15.000000) node[right] {$0$};
\draw[color=black] (180.000000,0.000000) node[right] {$y_2$};
\end{tikzpicture}
%
%

  \caption{Reversible network from Fig.~\ref{fig:example}(d).
    Additional lines which have a constant 0 value are drawn thicker.
    These lines can be used as additional resources when mapping
    single-target gates.}
  \label{fig:zero-lines}
\end{figure}
\subsection{LUT-based Mapping}
This section describes a mapping technique that exploits two
observations: (i) when mapping a single-target gate there may be
additional lines available with a constant 0 value; and (ii) for
single-target gates with few control lines (e.g., up to 4) one can
precompute near-optimal Clifford+$T$ networks and store them in a
database.  Fig.~\ref{fig:decomp}(b) illustrates the LUT-based mapping
flow.  The idea is to apply 4-LUT mapping on the control function of
the single-target gate and use available 0-valued lines to store
intermediate results from inner LUTs in the mapping.  If enough
0-values lines are available, each of the single-target gates
resulting from this mapping is directly translated into a near-optimum
Clifford+$T$ network that is looked up in a database.  The size of the
database is compressed by making use of Boolean function
classification based on affine input transformations and output
complementation.  This procedure requires no additional lines being
added to the circuit.  If the procedure cannot be applied due to too
few 0-valued lines, direct mapping is used as fallback.

\subsubsection*{Available 0-ancilla resources}
Fig.~\ref{fig:zero-lines} shows the same reversible network as in
Fig.~\ref{fig:example}(d) that is obtained from the initial $k$-LUT
mapping.  However, additional lines that are 0-valued are drawn
thicker.  We can see that for the realization of the first
single-target gate, there are three 0-valued lines, but for the last
single-target gate there is only one 0-valued line. This information
can easily be obtained from the reversible network resulting from
Alg.~\ref{alg:synth-seq}.

The following holds in general.  Let $g = \T_c(X, t)$ be a
single-target gate with $|X| = k$ and let there be $l$ available
0-valued lines, besides a 0-valued line for target line $t$.  If $c$
can be realized as a 4-LUT network with at most $l+1$ LUTs, then we
can realize it as a reversible network with $2l - 1$ single-target
gates that realizes function $c$ on target line $t$ such that all
inner LUTs compute and uncompute their results on the $l$ available
0-valued lines.  This synthesis procedure similar but much simpler
than Alg.~\ref{alg:synth-seq}, since $c$ is a single-output function.
Therefore, the number of additional lines required for the inner LUTs
cannot be improved by a different topological order and is solely
determined by the number of LUTs in the mapping.

\subsubsection*{Near-optimal 4-input single-target gates}
There exists $2^{2^n}$ Boolean functions over $n$ variables, i.e.,
$4, 16, 256$, and $65\,536$ for $n = 1, 2, 3$, and $4$, respectively.
Consequently, there are as many different single-target gates
$\T_c(\{x_1, \dots, x_n\}, x_{n+1})$.  For this number, it is possible
to precompute optimum or near-optimal Clifford+$T$ networks for each
of the single-target gates using exact or heuristic optimization
methods for Clifford+$T$ gates (see,
e.g.,~\cite{AMMR13,AMM14,MM16,MSD14}), and store them in a database.
Mapping single-target gates resulting from the LUT-based mapping
technique described in this sections is therefore very efficient.
However, the number of functions can be reduced significantly when
using \emph{affine function classification}.  Next, we review affine
function classification and show that two optimum Clifford+$T$
networks for two single-target gates with affine equivalent functions
have the same $T$-count.

For a Boolean function $f(x_1, \dots, x_n)$, let us use the notation
$f(\vec x)$, where $\vec x = (x_1, \dots, x_n)^T$.  We say that two
functions $f$ and $g$ are \emph{affine equivalent}~\cite{Harrison64},
if there exists an $n \times n$ invertible matrix
$A \in \mathrm{GL}_n(\B)$ and a vector $\vec b \in \B^n$ such that
\begin{equation}
  \label{eq:affine}
  g(\vec x) = f(A\vec x + \vec b) \qquad \text{for all $\vec x \in \B^n$.}
\end{equation}
We say that $f$ and $g$ are \emph{affine equivalent under
  negation}~\cite{Harrison63}, if either~\eqref{eq:affine} holds or
$g(\vec x) = \bar f(A\vec x + \vec b)$ for all $\vec x$. For the sake
of brevity, we say that $f$ is \emph{AN-equivalent} to $g$ in the
remainder.  AN-equivalence is an equivalence relation and allows to
partition the set of $2^{2^n}$ into much smaller sets of functions.
For $n = 1, 2, 3$, and $4$, there only $2, 3, 6$, and $18$ classes of
AN-equivalent functions (see,
e.g.,~\cite{Harrison63,Harrison64,ZYHZ16}.  And all $4\,294\,967\,296$
5-input Boolean functions fall into only $206$ classes of
AN-equivalent functions!  The database solution proposed in this
mapping can therefore scale for 5 variables given a fast way to
classify functions.  Before we discuss classification algorithms, the
following lemma shows that AN-equivalent functions preserve $T$-cost.
\begin{lemma}
  Let $f$ and $g$ be two $n$-variable AN-equivalent functions.  Then
  the $T$-count in Clifford+$T$ networks realizing $\T_f$ and $\T_g$
  is the same.
\end{lemma}
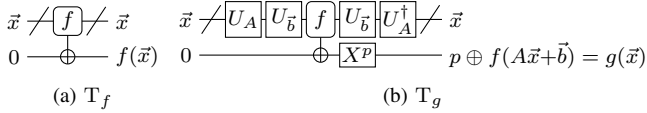
\begin{figure}
  \footnotesize
  \centering
  \subfloat[$\T_f$]{%
\begin{tikzpicture}[scale=0.900000,x=1pt,y=1pt]
\filldraw[color=white] (0.000000, -7.500000) rectangle (34.000000, 22.500000);
\draw[color=black] (0.000000,15.000000) -- (34.000000,15.000000);
\draw[color=black] (0.000000,15.000000) node[left] {${\vec x}$};
\draw[color=black] (0.000000,0.000000) -- (34.000000,0.000000);
\draw[color=black] (0.000000,0.000000) node[left] {$0$};
\draw (1.000000, 9.000000) -- (9.000000, 21.000000);
\draw (17.000000,15.000000) -- (17.000000,0.000000);
\begin{scope}[rounded corners=2pt]
\begin{scope}
\draw[fill=white] (17.000000, 15.000000) +(-45.000000:8.485281pt and 8.485281pt) -- +(45.000000:8.485281pt and 8.485281pt) -- +(135.000000:8.485281pt and 8.485281pt) -- +(225.000000:8.485281pt and 8.485281pt) -- cycle;
\clip (17.000000, 15.000000) +(-45.000000:8.485281pt and 8.485281pt) -- +(45.000000:8.485281pt and 8.485281pt) -- +(135.000000:8.485281pt and 8.485281pt) -- +(225.000000:8.485281pt and 8.485281pt) -- cycle;
\draw (17.000000, 15.000000) node {{$f$}};
\end{scope}
\end{scope}
\begin{scope}
\draw[fill=white] (17.000000, 0.000000) circle(3.000000pt);
\clip (17.000000, 0.000000) circle(3.000000pt);
\draw (14.000000, 0.000000) -- (20.000000, 0.000000);
\draw (17.000000, -3.000000) -- (17.000000, 3.000000);
\end{scope}
\draw (25.000000, 9.000000) -- (33.000000, 21.000000);
\draw[color=black] (34.000000,15.000000) node[right] {${\vec x}$};
\draw[color=black] (34.000000,0.000000) node[right] {${f(\vec x)}$};
\end{tikzpicture}}
  \hfil
  \subfloat[$\T_g$]{%
\begin{tikzpicture}[scale=0.900000,x=1pt,y=1pt]
\filldraw[color=white] (0.000000, -7.500000) rectangle (102.000000, 22.500000);
\draw[color=black] (0.000000,15.000000) -- (102.000000,15.000000);
\draw[color=black] (0.000000,15.000000) node[left] {${\vec x}$};
\draw[color=black] (0.000000,0.000000) -- (102.000000,0.000000);
\draw[color=black] (0.000000,0.000000) node[left] {$0$};
\draw (1.000000, 9.000000) -- (9.000000, 21.000000);
\begin{scope}
\draw[fill=white] (18.500000, 15.000000) +(-45.000000:10.606602pt and 10.606602pt) -- +(45.000000:10.606602pt and 10.606602pt) -- +(135.000000:10.606602pt and 10.606602pt) -- +(225.000000:10.606602pt and 10.606602pt) -- cycle;
\clip (18.500000, 15.000000) +(-45.000000:10.606602pt and 10.606602pt) -- +(45.000000:10.606602pt and 10.606602pt) -- +(135.000000:10.606602pt and 10.606602pt) -- +(225.000000:10.606602pt and 10.606602pt) -- cycle;
\draw (18.500000, 15.000000) node {{$U_A$}};
\end{scope}
\begin{scope}
\draw[fill=white] (35.500000, 15.000000) +(-45.000000:10.606602pt and 10.606602pt) -- +(45.000000:10.606602pt and 10.606602pt) -- +(135.000000:10.606602pt and 10.606602pt) -- +(225.000000:10.606602pt and 10.606602pt) -- cycle;
\clip (35.500000, 15.000000) +(-45.000000:10.606602pt and 10.606602pt) -- +(45.000000:10.606602pt and 10.606602pt) -- +(135.000000:10.606602pt and 10.606602pt) -- +(225.000000:10.606602pt and 10.606602pt) -- cycle;
\draw (35.500000, 15.000000) node {{$U_{\vec b}$}};
\end{scope}
\draw (51.000000,15.000000) -- (51.000000,0.000000);
\begin{scope}[rounded corners=2pt]
\begin{scope}
\draw[fill=white] (51.000000, 15.000000) +(-45.000000:8.485281pt and 10.606602pt) -- +(45.000000:8.485281pt and 10.606602pt) -- +(135.000000:8.485281pt and 10.606602pt) -- +(225.000000:8.485281pt and 10.606602pt) -- cycle;
\clip (51.000000, 15.000000) +(-45.000000:8.485281pt and 10.606602pt) -- +(45.000000:8.485281pt and 10.606602pt) -- +(135.000000:8.485281pt and 10.606602pt) -- +(225.000000:8.485281pt and 10.606602pt) -- cycle;
\draw (51.000000, 15.000000) node {{$f$}};
\end{scope}
\end{scope}
\begin{scope}
\draw[fill=white] (51.000000, 0.000000) circle(3.000000pt);
\clip (51.000000, 0.000000) circle(3.000000pt);
\draw (48.000000, 0.000000) -- (54.000000, 0.000000);
\draw (51.000000, -3.000000) -- (51.000000, 3.000000);
\end{scope}
\begin{scope}
\draw[fill=white] (66.500000, 15.000000) +(-45.000000:10.606602pt and 10.606602pt) -- +(45.000000:10.606602pt and 10.606602pt) -- +(135.000000:10.606602pt and 10.606602pt) -- +(225.000000:10.606602pt and 10.606602pt) -- cycle;
\clip (66.500000, 15.000000) +(-45.000000:10.606602pt and 10.606602pt) -- +(45.000000:10.606602pt and 10.606602pt) -- +(135.000000:10.606602pt and 10.606602pt) -- +(225.000000:10.606602pt and 10.606602pt) -- cycle;
\draw (66.500000, 15.000000) node {{$U_{\vec b}$}};
\end{scope}
\begin{scope}
\draw[fill=white] (66.500000, -0.000000) +(-45.000000:10.606602pt and 7.071068pt) -- +(45.000000:10.606602pt and 7.071068pt) -- +(135.000000:10.606602pt and 7.071068pt) -- +(225.000000:10.606602pt and 7.071068pt) -- cycle;
\clip (66.500000, -0.000000) +(-45.000000:10.606602pt and 7.071068pt) -- +(45.000000:10.606602pt and 7.071068pt) -- +(135.000000:10.606602pt and 7.071068pt) -- +(225.000000:10.606602pt and 7.071068pt) -- cycle;
\draw (66.500000, -0.000000) node {{$X^p$}};
\end{scope}
\begin{scope}
\draw[fill=white] (83.500000, 15.000000) +(-45.000000:10.606602pt and 10.606602pt) -- +(45.000000:10.606602pt and 10.606602pt) -- +(135.000000:10.606602pt and 10.606602pt) -- +(225.000000:10.606602pt and 10.606602pt) -- cycle;
\clip (83.500000, 15.000000) +(-45.000000:10.606602pt and 10.606602pt) -- +(45.000000:10.606602pt and 10.606602pt) -- +(135.000000:10.606602pt and 10.606602pt) -- +(225.000000:10.606602pt and 10.606602pt) -- cycle;
\draw (83.500000, 15.000000) node {{$U^{\dagger}_A$}};
\end{scope}
\draw (93.000000, 9.000000) -- (101.000000, 21.000000);
\draw[color=black] (102.000000,15.000000) node[right] {${\vec x}$};
\draw[color=black] (102.000000,0.000000) node[right] {${p \oplus f(A\vec x {+} \vec b)=g(\vec x)}$};
\end{tikzpicture}}

  \caption{A single-target gate for $\T_f$ and a single target gate
    for an $\T_g$, which can be constructed from $\T_f$, since $f$ and
    $g$ are AN-equivalent.}
  \label{fig:affine}
\end{figure}
\begin{IEEEproof}
  Since $f$ and $g$ are AN-equivalent, there exists
  $A \in \mathrm{GL}_n(\B)$, $\vec b \in \B^n$, and $p \in \B$ such
  that $g(\vec x) = p \oplus f(A \vec x + \vec b)$ for all $\vec x$.
  It is possible to create a reversible circuit that takes
  $\vec x \mapsto A\vec x + \vec b$ using only CNOT gates for $A$ and
  NOT gates for $\vec b$ (see, e.g., \cite{SAM16}).  The output can be
  inverted using a NOT gate.  Fig.~\ref{fig:affine} illustrates the
  proof.  The subcircuit $U_A$ realizes the linear transformation
  using CNOT gates, $U_{\vec b}$ realizes the input inversions using
  NOT gates, and $X^p$ represents a NOT gate, if $p=1$, otherwise the
  identity.
\end{IEEEproof}

In order to make use of the algorithm we need to compute an optimum or
near-optimal circuit for one representative in each equivalence class
for up to 4 variables.  To match an arbitrary single-target gate with
a control function of up to 4 variables in the database, one needs to
derive it's representative.  Algorithms as presented in~\cite{SKM17}
can be used for this purpose.

\begin{figure}[t!]
  \centering
  \includegraphics{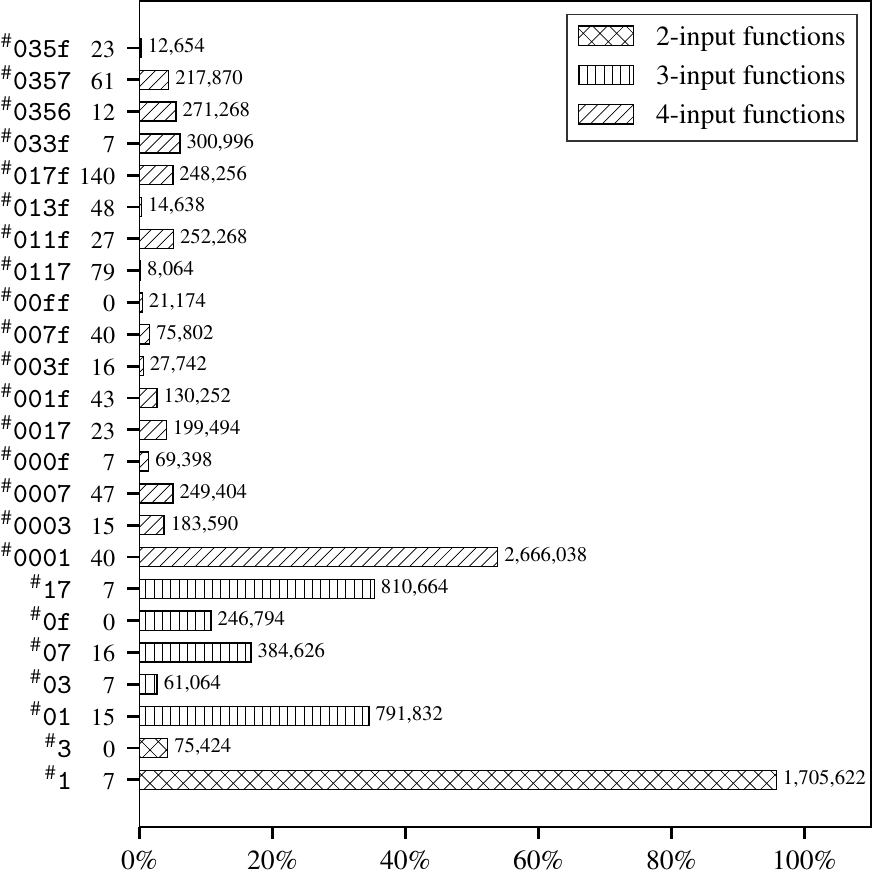}
  \caption{This plot shows the distribution of AN-equivalence classes
    among all 4-LUTs that have been discovered in all our experimental
    results (see Sect.~\ref{sec:results}).  The $y$-axis shows from
    the bottom to the top all 2, 5, and 17 non-constant AN-equivalence
    classes of 2, 3, and 4 variables.  The number right to the truth
    table is the $T$-count in the best known Clifford+$T$ realization
    of the corresponding single-target gate.  The $x$-axis shows the
    frequency percentage with respect to other classes that have the
    same number of variables.  The small number next to the bar shows
    the frequency in absolute values.}
  \label{fig:classes}
\end{figure}

Fig.~\ref{fig:classes} lists all the AN-equivalent classes for 2--4
variables; the class containing the constant functions is omitted.  It
shows how often they are discovered in a 4-LUT in all our experimental
evaluations.  Also, it shows the number of $T$ gates in the best-known
Clifford+$T$ networks of the corresponding single-target gate.  The
classes \hexf{1}, \hexf{01}, and \hexf{0001} occur most frequently.
These classes contain among others $x_1x_2$, $x_1x_2x_3$, and
$x_1x_2x_3x_4$, and are therefore control functions of the
multiple-controlled Toffoli gates.  The table can guide to the classes
which benefit most from optimizing their $T$-count.  Reversible and
Clifford+$T$ networks of the best-known realizations can be found at
\emph{quantumlib.stationq.com}.

\begin{table}[t]
  \caption{Parameters for LHRS}
  \label{tbl:params}
  \begin{tabularx}{\linewidth}{lp{1.5cm}X}
    \toprule
    Parameter & Values & Description \\
    \midrule
    \multicolumn{3}{l}{\textit{Parameters $p_{\mathrm{Q}}$ that affect the number of qubits (and $T$ gates)}} \\[4pt]
    LUT size & $\{3, \dots, 32\}$ & Maximum number of inputs to LUTs in the LUT mapping, default: 16 \\
    Topo.\ order & \emph{topo\_def}, \emph{topo\_tfi\_sort} & Heuristic in which order to traverse LUTs in the LUT mapping (Sect.~\ref{sec:implementation}), default: \emph{topo\_def} \\
    area iters.\ (init) & $\{1, \dots, 10\}$ & Number of iterations for global area recovery heuristic based on exact area~\cite{MCB07,CD94b}, default: 2 \\
    flow iters.\ (init) & $\{1, \dots, 10\}$ & Number of iterations for local area recovery heursitic based on area flow~\cite{CWD99,MDV04}, default: 1 \\
    \midrule
    \multicolumn{3}{l}{\textit{Parameters $p_{\mathrm{T}}$ that only affect the number of $T$ gates}} \\[4pt]
    Mapping & \emph{direct}, \emph{hybrid} & Mapping method (Sect.~\ref{sec:synth-gates}), default: \emph{hybrid} \\
    ESOP extraction & \emph{AIG extract}, \emph{BDD extract} & ESOP extraction method (Sect.~\ref{sec:direct-decomp}), default: \emph{AIG extract} \\
    ESOP heuristic & \emph{none}, \emph{def}, \emph{def\_wo4} & ESOP minimization heuristic (Sect.~\ref{sec:direct-decomp}), \emph{none} means that no optimization is applied, default: \emph{def\_wo4} \\
    \multicolumn{3}{l}{Parameters in \emph{hybrid} mapping method} \\
    \quad area iters. & $\{1,\dots, 10\}$ & \emph{see above} \\
    \quad flow iters. & $\{1,\dots, 10\}$ & \emph{see above} \\
    \quad SAT-based opt. & \emph{true}, \emph{false} & Enables whether 4-LUT networks in the hybrid mapping are post-optimized using the SAT-based technique described in~\cite{SMB17}, default: \emph{false} \\
    \bottomrule
  \end{tabularx}
\end{table}

\subsection{Hybrid Mapping}
LUT-based mapping cannot be applied if the number of available
0-valued lines is insufficient.  As fallback, the 4-LUT network is
omitted and direct mapping is applied on the $k$-LUT network,
therefore not using any of the 0-valued lines at all.  The idea of
hybrid synthesis is to merge 4-LUTs into larger LUTs.  By merging 2
LUTs in the network, the number of LUTs is decreased by 1 and
therefore one fewer 0-valued line is required.  Our algorithm for
hybrid synthesis merges the output LUT with one of its children,
thereby generating a larger output LUT.  This procedure is repeated
until the LUT network is small enough to match the number of 0-valued
lines.  The topmost LUT is then mapped using direct mapping, while the
remaining LUTs are translated using the LUT-mapping technique.

\section{Experimental Evaluation}
\label{sec:results}
We have implemented LHRS as command `\emph{lhrs}' in the open source
reversible logic synthesis framework
RevKit~\cite{SFWD12}.\footnote{The source code can be found at
  \emph{github.com/msoeken/cirkit}} All experiments have been carried
out on an Intel Xeon CPU E5-2680 v3 at 2.50 GHz with 64 GB of main
memory running Linux 4.4 and gcc 5.4.

\begin{table}[t]
  \caption{EPFL arithmetic benchmarks}
  \label{tbl:epfl}
  \def\tabcolsep{4.5pt}
  \begin{tabularx}{\linewidth}{Xrrrrrr}
    \toprule
    Benchmark & Inputs & Outputs & \multicolumn{2}{c}{Original} & \multicolumn{2}{c}{Best-LUT} \\
    \midrule
    & & & \scriptsize AIG nodes & \scriptsize Levels & \scriptsize LUTs & \scriptsize Levels \\[4pt]
    Adder          & 256 & 129 & 1020    & 255    & 192    & 64    \\
    Barrel shifter & 135 & 128 & 3,336   & 12     & 512    & 4     \\
    Divisor        & 128 & 128 & 44,762  & 4,470  & 3268   & 1,208 \\
    Hypotenuse     & 256 & 128 & 214,335 & 24,801 & 40,406 & 4,532 \\
    Log2           & 32  & 32  & 32,060  & 444    & 6,574  & 119   \\
    Max            & 512 & 130 & 2,865   & 287    & 523    & 189   \\
    Multiplier     & 128 & 128 & 27,062  & 274    & 4,923  & 90    \\
    Sine           & 24  & 25  & 5,416   & 225    & 1,229  & 55    \\
    Square-root    & 128 & 64  & 24,618  & 5,058  & 3,077  & 1,106 \\
    Square         & 64  & 128 & 18,484  & 250    & 3,246  & 74    \\
    \bottomrule
  \end{tabularx}
\end{table}

\begin{table*}
  \caption{Experimental evaluation of LHRS on the EPFL arithmetic benchmarks}
  \label{tbl:epfl-experiments}
  \def\tabcolsep{4.5pt}
  \begin{tabularx}{\linewidth}{lrXrrrrrrrrr}
    \toprule
    Bechmark & \multicolumn{2}{l}{LUT size} & & \multicolumn{2}{c}{\emph{def}, \emph{direct}} & \multicolumn{2}{c}{\emph{def\_wo4}, \emph{direct}} & \multicolumn{2}{c}{\emph{def}, \emph{hybrid}} & \multicolumn{2}{c}{\emph{def\_wo4}, \emph{hybrid}} \\
    \midrule
    & & & qubits & $T$-count & runtime & $T$-count & runtime & $T$-count & runtime & $T$-count & runtime \\[4pt]
Adder          & 6  & Best-LUT & 448    & 14,411     & 0.13      & 20,487     & 0.11      & 12,623     & 0.18     & 12,721     & 0.18     \\
               &    & Original & 505    & 6,670      & 0.09      & 6,510      & 0.08      & 2,066      & 0.05     & 2,066      & 0.05     \\
               & 10 & Best-LUT & 445    & 18,675     & 0.14      & 24,320     & 0.14      & 13,576     & 0.23     & 13,674     & 0.21     \\
               &    & Original & 490    & 21,357     & 0.15      & 20,313     & 0.14      & 2,860      & 0.05     & 2,860      & 0.04     \\
               & 16 & Best-LUT & 443    & 45,577     & 0.68      & 50,946     & 0.37      & 14,699     & 0.29     & 14,797     & 0.28     \\
               &    & Original & 463    & 313,142    & 20.19     & 309,018    & 19.73     & 5,122      & 0.07     & 5,122      & 0.05     \\[3pt]
Barrel shifter & 6  & Best-LUT & 840    & 32,512     & 0.22      & 32,512     & 0.22      & 17,024     & 0.05     & 17,024     & 0.04     \\
               &    & Original & 584    & 50,944     & 0.16      & 50,944     & 0.15      & 76,883     & 0.46     & 76,883     & 0.46     \\
               & 10 & Best-LUT & 740    & 52,986     & 0.21      & 52,986     & 0.20      & 42,656     & 0.35     & 42,656     & 0.36     \\
               &    & Original & 584    & 50,944     & 0.15      & 50,944     & 0.16      & 76,883     & 0.44     & 76,883     & 0.45     \\
               & 16 & Best-LUT & 595    & 114,690    & 0.24      & 114,690    & 0.25      & 79,530     & 0.38     & 79,530     & 0.37     \\
               &    & Original & 582    & 52,480     & 0.16      & 52,480     & 0.14      & 78,671     & 0.54     & 78,671     & 0.50     \\[3pt]
Divisor        & 6  & Best-LUT & 3,399  & 346,010    & 2.77      & 341,583    & 2.41      & 435,568    & 4.58     & 435,575    & 4.56     \\
               &    & Original & 12,389 & 754,587    & 10.99     & 754,587    & 11.40     & 819,918    & 4.15     & 819,918    & 4.03     \\
               & 10 & Best-LUT & 3,226  & 450,381    & 3.11      & 445,567    & 2.57      & 510,188    & 5.29     & 510,195    & 5.27     \\
               &    & Original & 12,055 & 875,819    & 11.73     & 875,847    & 11.63     & 1,073,427  & 5.36     & 1,073,427  & 5.27     \\
               & 16 & Best-LUT & 3,017  & 3,815,286  & 298.92    & 3,891,364  & 167.28    & 956,116    & 7.21     & 956,123    & 4.17     \\
               &    & Original & 11,827 & 1,415,585  & 20.79     & 1,417,387  & 18.73     & 1,296,742  & 5.65     & 1,296,742  & 5.48     \\[3pt]
Hypotenuse     & 6  & Best-LUT & 40,611 & 3,807,147  & 153.97    & 3,672,932  & 152.82    & 5,050,507  & 140.24   & 5,050,521  & 140.68   \\
               &    & Original & 47,814 & 2,344,226  & 128.06    & 2,314,392  & 126.26    & 3,571,061  & 53.98    & 3,571,071  & 65.70    \\
               & 10 & Best-LUT & 36,443 & 5,902,745  & 166.84    & 5,834,737  & 157.12    & 7,893,544  & 153.64   & 7,893,427  & 153.90   \\
               &    & Original & 43,871 & 4,345,646  & 123.00    & 4,342,966  & 98.62     & 5,430,785  & 74.84    & 5,430,737  & 69.24    \\
               & 16 & Best-LUT & 32,336 & 20,239,832 & 52,944.40 & 20,919,129 & 12,382.30 & 11,713,691 & 53.25    & 11,716,038 & 48.19    \\
               &    & Original & 39,324 & 22,132,834 & 79245.00  & 23,300,693 & 20,595.10 & 8,148,322  & 19.26    & 8,148,289  & 18.78    \\[3pt]
Log2           & 6  & Best-LUT & 6,625  & 664,450    & 5.85      & 660,068    & 4.28      & 1,363,770  & 19.69    & 1,363,770  & 19.92    \\
               &    & Original & 7,611  & 501,749    & 4.59      & 501,789    & 3.80      & 996,739    & 3.64     & 996,739    & 3.65     \\
               & 10 & Best-LUT & 3,038  & 2,036,814  & 45.87     & 2,050,844  & 35.18     & 20,342,942 & 49.31    & 20,343,371 & 48.84    \\
               &    & Original & 2,875  & 3,284,753  & 70.05     & 3,314,383  & 54.05     & 25,548,190 & 47.13    & 25,549,307 & 44.61    \\
               & 16 & Best-LUT & 2,052  & 56,589,962 & 54,824.70 & 59,587,462 & 16,412.00 & 44,644,140 & 1,661.47 & 44,823,085 & 770.26   \\
               &    & Original & 2,315  & 61,071,767 & 94,192.60 & 64,461,123 & 27,220.50 & 64,368,103 & 1,471.36 & 64,447,809 & 448.72   \\[3pt]
Max            & 6  & Best-LUT & 1,036  & 56,290     & 0.30      & 55,970     & 0.25      & 19,849     & 0.31     & 19,849     & 0.29     \\
               &    & Original & 1,233  & 71,198     & 0.37      & 69,484     & 0.32      & 64,933     & 0.90     & 64,933     & 0.90     \\
               & 10 & Best-LUT & 840    & 195,367    & 1.05      & 193,683    & 0.77      & 25,270     & 0.34     & 25,270     & 0.34     \\
               &    & Original & 901    & 229,669    & 1.05      & 222,761    & 0.70      & 88,221     & 0.88     & 88,221     & 0.84     \\
               & 16 & Best-LUT & 725    & 1,082,574  & 57.52     & 1,077,303  & 49.81     & 34,488     & 0.40     & 34,400     & 0.39     \\
               &    & Original & 796    & 748,389    & 36.09     & 745,389    & 30.24     & 94,758     & 1.21     & 94,702     & 1.19     \\[3pt]
Multiplier     & 6  & Best-LUT & 5,048  & 683,648    & 5.32      & 680,190    & 4.65      & 868,688    & 6.83     & 868,638    & 7.03     \\
               &    & Original & 5,806  & 386,999    & 3.61      & 387,113    & 3.09      & 733,615    & 1.79     & 733,615    & 1.87     \\
               & 10 & Best-LUT & 3,418  & 1,148,166  & 8.98      & 1,144,532  & 6.42      & 1,368,820  & 8.69     & 1,368,802  & 8.27     \\
               &    & Original & 3,105  & 2,377,198  & 18.69     & 2,377,251  & 11.41     & 2,519,515  & 5.22     & 2,519,355  & 5.21     \\
               & 16 & Best-LUT & 2,552  & 9,928,094  & 2,742.34  & 10,422,808 & 1,118.93  & 3,281,867  & 28.53    & 3,294,871  & 12.72    \\
               &    & Original & 2,852  & 8,072,619  & 1,264.52  & 8,283,999  & 489.36    & 3,189,854  & 13.34    & 3,192,826  & 7.65     \\[3pt]
Sine           & 6  & Best-LUT & 1,277  & 141,885    & 0.72      & 141,456    & 0.64      & 211,143    & 10.36    & 211,143    & 10.13    \\
               &    & Original & 1,468  & 87,624     & 0.76      & 87,872     & 0.60      & 134,664    & 2.24     & 134,664    & 2.29     \\
               & 10 & Best-LUT & 557    & 362,108    & 3.55      & 365,175    & 2.39      & 715,314    & 13.20    & 716,079    & 12.67    \\
               &    & Original & 714    & 408,996    & 3.76      & 411,850    & 2.50      & 1,074,871  & 5.45     & 1,074,905  & 5.30     \\
               & 16 & Best-LUT & 418    & 3,730,177  & 6,579.01  & 3,953,575  & 2,210.36  & 1,759,815  & 79.38    & 1,791,503  & 44.60    \\
               &    & Original & 518    & 3,820,348  & 1,104.20  & 3,913,954  & 390.94    & 3,007,519  & 6.86     & 3,008,373  & 6.55     \\[3pt]
Square-root    & 6  & Best-LUT & 3,204  & 368,301    & 2.76      & 357,593    & 2.42      & 447,805    & 5.39     & 447,798    & 5.40     \\
               &    & Original & 8,212  & 279,275    & 2.95      & 279,275    & 2.97      & 749,687    & 1.35     & 749,687    & 1.35     \\
               & 10 & Best-LUT & 2,874  & 549,624    & 4.34      & 542,394    & 3.32      & 566,272    & 7.28     & 566,433    & 7.32     \\
               &    & Original & 7,892  & 323,882    & 3.01      & 323,656    & 3.02      & 833,700    & 1.87     & 833,716    & 1.83     \\
               & 16 & Best-LUT & 2,632  & 6,158,349  & 392.85    & 6,247,981  & 264.42    & 1,404,786  & 16.26    & 1,408,346  & 13.51    \\
               &    & Original & 7,816  & 1,524,733  & 948.33    & 1,729,343  & 550.70    & 1,028,329  & 1.39     & 1,028,401  & 1.34     \\[3pt]
Square         & 6  & Best-LUT & 3,309  & 299,986    & 2.54      & 295,325    & 2.28      & 489,565    & 6.92     & 489,617    & 6.85     \\
               &    & Original & 4,058  & 195,290    & 1.54      & 195,312    & 1.78      & 768,725    & 1.98     & 768,725    & 1.35     \\
               & 10 & Best-LUT & 2,882  & 532,854    & 2.93      & 531,211    & 2.20      & 826,100    & 7.17     & 826,121    & 7.72     \\
               &    & Original & 3,355  & 464,024    & 2.14      & 464,320    & 1.24      & 1,243,876  & 3.38     & 1,243,988  & 2.35     \\
               & 16 & Best-LUT & 2,303  & 3,964,800  & 26,098.40 & 4,142,416  & 7,888.93  & 2,303,327  & 8,299.69 & 2,355,942  & 2,785.77 \\
               &    & Original & 2,664  & 4,249,919  & 29,866.40 & 4,470,766  & 7,080.61  & 3,075,569  & 8,800.60 & 3,156,642  & 1,947.58 \\
    \bottomrule
  \end{tabularx}
\end{table*}

\subsection{LHRS Configuration}
\label{sec:params}
Table~\ref{tbl:params} gives an overview of all parameters that can be
given as input to LHRS.  The parameters are split into two groups.
Parameters in the upper half have mainly an influence on the number of
lines and are used to synthesize the initial reversible network that
is composed of single-target gates (Sect.~\ref{sec:synth-mapping}).
Parameters in the lower half only influence the number of $T$ gates by
changing how single-target gates are mapped into Clifford+$T$
networks.  Each parameter is shown with possible values and some
explanation text.

\subsection{Benchmarks}
We used both academic and industrial benchmarks for our evaluation.
As academic benchmarks we used the 10 arithmetic instances of the EPFL
combinational logic synthesis benchmarks~\cite{AGM15}, which are
commonly used to evaluate logic synthesis algorithms.  In order to
investigate the influence of the initial logic representation, we used
different realizations of the benchmarks: (i)~the original benchmark
description in terms of an AIG, called \emph{Original}, and (ii)~the
best known 6-LUT network wrt.\ the number of lines, called
\emph{Best-LUT}.\footnote{see \emph{lsi.epfl.ch/benchmarks}, version
  2017.1} Further statistics about the benchmarks are given in
Table~\ref{tbl:epfl}.  All experimental results and synthesis outcomes
for the academic benchmarks can be viewed and downloaded
from~\emph{quantumlib.stationq.com}.

As commercial benchmarks we used Verilog netlists of several
arithmetic floating point designs in half (16-bit), single (32-bit),
and double (64-bit) precision.  For synthesis all Verilog files were
translated into AIGs and optimized for size using ABC's
`\emph{resyn2}' script.

\begin{table*}[ht!]
  \centering
  \caption{Experimental evaluation of LHRS on IEEE compliant floating-point designs}
  \label{tab:fpl}
  \begin{tabularx}{\linewidth}{lrrXrrrrrrrrr}
    \toprule
    Benchmark &&&& \multicolumn{3}{c}{$k = 6$} & \multicolumn{3}{c}{$k = 10$} & \multicolumn{3}{c}{$k = 16$} \\
    \midrule
    & \scriptsize size & \scriptsize logic depth && \scriptsize qubits & \scriptsize $T$ gates & \scriptsize runtime & \scriptsize qubits & \scriptsize $T$ gates & \scriptsize runtime & \scriptsize qubits & \scriptsize $T$ gates & \scriptsize runtime \\[4pt]
    add-16 &   788 &   81 & &    230 &     18,351 &     1.03 &    186 &     24,067 &     1.18 &    156 &     33,521 &      1.35 \\
    add-32 &  1763 &  137 & &    526 &     40,853 &     1.43 &    410 &     53,060 &     1.69 &    368 &     66,463 &      1.96 \\
    add-64 &  3934 &  252 & &  1,194 &     77,843 &     1.74 &    960 &    109,164 &     2.30 &    867 &    130,990 &      2.41 \\
    cmp-16 &   110 &   17 & &     65 &      3,720 &     0.16 &     48 &      7,959 &     0.15 &     40 &     30,426 &      0.77 \\
    cmp-32 &   202 &   29 & &    126 &      9,261 &     0.20 &     95 &     16,800 &     0.16 &     81 &     29,335 &      0.21 \\
    cmp-64 &   374 &   59 & &    245 &     16,519 &     0.23 &    181 &     34,372 &     0.19 &    163 &     38,967 &      0.25 \\
    div-16 &  1381 &  310 & &    300 &     12,244 &     0.58 &    223 &     28,639 &     0.82 &    144 &    589,721 &     82.14 \\
    div-32 &  6098 & 1299 & &  1,260 &     32,391 &     0.59 &  1,106 &     58,536 &     0.92 &    935 &    289,978 &     30.04 \\
    div-64 & 28807 & 5938 & &  5,876 &    123,912 &     0.90 &  5,514 &    177,343 &     1.28 &  5,149 &    300,721 &      1.61 \\
    exp-16 &  4240 &  156 & &  1,371 &    141,210 &     1.82 &    978 &    252,299 &     2.46 &     32 &  1,193,083 & 95,269.40 \\
    exp-32 & 16546 &  339 & &  4,636 &    488,579 &     3.11 &  3,489 &    662,350 &     3.85 &  3,019 &    792,008 &      4.49 \\
invsqrt-16 &  4017 &  456 & &    899 &     39,410 &     2.03 &    781 &    119,574 &     3.34 &     32 &    169,282 &    430.99 \\
invsqrt-32 & 19495 & 1915 & &  4,242 &    118,973 &     3.01 &  4,008 &    349,414 &     4.99 &  3,609 &    703,324 &      6.22 \\
invsqrt-64 & 97242 & 8830 & & 20,874 &    408,652 &     6.21 & 20,274 &    886,327 &     7.89 & 19,536 &  2,009,862 &     10.39 \\
     ln-16 &  2601 &   86 & &    867 &    139,456 &     3.43 &    303 &    317,543 &     3.92 &     32 &  1,623,461 &   1115.16 \\
     ln-32 & 11096 &  274 & &  3,275 &    334,303 &     2.63 &  1,317 &  5,672,890 &     8.58 &  1,033 & 15,357,188 &    594.79 \\
     ln-64 & 44929 & 8800 & & 13,150 &    254,749 &     4.69 & 12,551 &    370,890 &     5.07 & 12,031 &  1,305,192 &     16.52 \\
   log2-16 &  2592 &   69 & &    937 &    110,827 &     3.04 &    312 &    317,921 &     3.71 &     32 &    850,331 &    273.23 \\
   log2-32 & 14102 &  261 & &  4,008 &    436,039 &     2.35 &  1,711 &  8,079,605 &    14.81 &  1,244 & 17,600,310 &    496.30 \\
   log2-64 & 23660 & 3266 & &  6,413 &    278,109 &     4.02 &  6,021 &    397,363 &     4.76 &  5,862 &    735,507 &      5.57 \\
   mult-16 &  1923 &   80 & &    499 &     43,447 &     1.63 &    381 &     72,307 &     2.25 &    267 &    141,657 &      3.03 \\
   mult-32 &  5843 &  146 & &  1,536 &    157,040 &     2.41 &  1,020 &    423,648 &     3.32 &    862 &    623,438 &      4.10 \\
   mult-64 & 21457 &  265 & &  5,495 &    598,285 &     2.91 &  3,253 &  1,919,743 &     4.14 &  2,941 &  2,266,026 &      5.09 \\
  recip-16 &  2465 &  113 & &    623 &     63,452 &     2.80 &     43 &     89,497 &    43.40 &     32 &    198,167 &     37.40 \\
  recip-32 &  7605 &  228 & &  1,914 &    201,730 &     3.93 &  1,194 &    519,091 &     5.56 &    916 &  1,963,459 &      7.58 \\
  recip-64 & 42500 &  555 & & 10,277 &  1,101,791 &     5.55 &  7,111 &  1,966,791 &     8.59 &  5,856 &  4,025,162 &     11.16 \\
 sincos-16 &   935 &   76 & &    367 &     25,061 &     0.94 &    278 &     40,579 &     1.20 &     34 &    452,129 &    159.16 \\
 sincos-32 &  5893 &  201 & &  1,740 &    148,104 &     2.25 &  1,438 &    184,706 &     3.14 &  1,284 &    226,328 &      3.46 \\
 square-16 &   479 &   39 & &    113 &     12,223 &     0.49 &     35 &     73,381 &     3.20 &     32 &    133,489 &      6.57 \\
 square-32 &  2260 &   81 & &    564 &     34,821 &     0.86 &    412 &     70,945 &     1.24 &    269 &  1,195,946 &    377.31 \\
 square-64 & 11183 &  166 & &  2,788 &    134,370 &     1.25 &  2,251 &    227,490 &     2.04 &  1,803 &  1,054,537 &    176.85 \\
   sqrt-16 &   563 &  143 & &    131 &     10,676 &     0.63 &     76 &     30,880 &     0.73 &     32 &    165,545 &     59.43 \\
   sqrt-32 &  2759 &  618 & &    597 &     26,342 &     0.72 &    448 &     77,226 &     0.99 &    320 &  6,780,943 & 41,636.45 \\
   sqrt-64 & 13719 & 2895 & &  2,855 &     80,194 &     0.90 &  2,498 &    222,445 &     1.37 &  2,059 &  3,655,540 & 15,194.50 \\
    sub-16 &   789 &   81 & &    231 &     17,972 &     1.01 &    185 &     23,164 &     1.24 &    156 &     33,626 &      1.32 \\
    sub-32 &  1765 &  120 & &    528 &     40,310 &     1.55 &    405 &     54,273 &     1.76 &    363 &     71,079 &      2.03 \\
    sub-64 &  3935 &  216 & &  1,191 &     78,388 &     1.98 &    963 &    109,029 &     2.46 &    874 &    135,901 &      2.49 \\
    \bottomrule
  \end{tabularx}
\end{table*}

\subsection{Experiments for EPFL Benchmarks}
We evaluated LHRS for both realizations (\emph{Original} and
\emph{Best-LUT}) for all 10 arithmetic benchmarks with a LUT size of
6, 10, and 16.  Table~\ref{tbl:epfl-experiments} lists all results.
We chose LUT size 6, because it is a typical choice for FPGA mapping,
and therefore we expect that LUT mapping algorithms perform well for
this size.  We chose 16, since we noticed in our experiments that it
is the largest LUT size for which LHRS performs reasonably fast for
most of the benchmarks.  LUT size 10 has been chosen, since it is
roughly in between the other two.  These configurations allow to
synthesize 6 different initial single-target gate networks for each
benchmark, therefore spanning 6 optimization points in a Pareto set.
For each of these configurations, we chose 4 different configurations
of parameters in $p_{\mathrm{T}}$, based on values to the mapping
method and the ESOP heuristic
($\{\text{\emph{direct}},
\text{\emph{hybrid}}\}\times\{\text{\emph{def}},
\text{\emph{def\_wo4}}\} $).

The experiments confirm the observation of
Sect.~\ref{sec:role-lut-size}: A larger LUT size leads to a smaller
number qubits.  In some cases, e.g.,~\emph{Log2}, this can be quite
significant.  A larger LUT size also leads to a larger $T$-count,
again very considerable, e.g., for~\emph{Log2}.

Slightly changing the ESOP minimization heuristic (note that
\emph{def} to \emph{def\_wo4} are very similar) has no strong
influence on the number of $T$ gates.  There are examples for both the
case in which the $T$-count increases and in which it decreases.
However, the runtime can be significantly reduced.  The choice of the
mapping method has a stronger influence.  For example, in case of the
\emph{Adder}, the \emph{hybrid} mapping method is far superior
compared to the \emph{direct} method.  In many cases, the
\emph{hybrid} method is advantageous only for a LUT size of 16, but
not for the smaller ones.  Also, the initial representation of the
benchmark has a considerable influence.  In many cases, the
\emph{Best-LUT} realizations of the benchmarks require fewer qubits,
which---as expected---results in higher $T$-count.  The effect is
particularly noticeable for the \emph{Divisor} and \emph{Square-root}.
In some cases, better results in both qubits and $T$-count can be
achieved, e.g., for~\emph{Log2} as \emph{Best-LUT} with a LUT size of
16, and for the~\emph{Barrel shifter} as \emph{Original} with a LUT
size of 16.

\subsection{Experiments for Floating point Library}
We reevaluated the LHRS algorithm in its improved implementation and
new parameters on the commercial floating point designs, which were
already used for the evaluation in~\cite{SRWM17b}.  The numbers are
listed in Table~\ref{tab:fpl} for LUT sizes 6, 10, and 16, mapping
method \emph{hybrid}, and ESOP heuristic \emph{def\_wo4}.  Due to the
changes and improvements in the implementation and different
parameters, the numbers vary.  In the vast majority of the cases the
numbers improve for qubits, $T$-count, and runtime.  Consequently, we
did not repeat the comparison to the hierarchical reversible logic
synthesis algorithm presented in~\cite{SC16}, as the previous numbers
already have shown an improvement.

Note that for all benchmarks with 16 inputs (\emph{exp-16},
\emph{invsqrt-16}, \emph{ln-16}, \emph{log-16}, \emph{recip-16},
\emph{sincos-16}, \emph{square-16}, and \emph{sqrt-16}), a LUT size 16
leads to qubit-optimum quantum networks, since every output can be
represented by a single LUT.  Note that LHRS is not aware of this
situation, and will realize every output separately.  A better
runtime, and potentially a better $T$-count, can be achieved by
generating the ESOP cover and optimizing it for all outputs at once.

\subsection{Compositional Functions}
\label{sec:composition}
\begin{figure}[t]
  \centering
  \footnotesize
  \subfloat[Direct translation.]{%
\begin{tikzpicture}[scale=0.800000,x=1pt,y=1pt]
\filldraw[color=white] (0.000000, -6.500000) rectangle (234.000000, 240.500000);
\draw[color=black] (0.000000,234.000000) -- (234.000000,234.000000);
\draw[color=black] (0.000000,234.000000) node[left] {$a$};
\draw[color=black] (0.000000,221.000000) -- (234.000000,221.000000);
\draw[color=black] (0.000000,221.000000) node[left] {$b$};
\draw[color=black] (0.000000,208.000000) -- (234.000000,208.000000);
\draw[color=black] (0.000000,208.000000) node[left] {$c$};
\draw[color=black] (0.000000,195.000000) -- (234.000000,195.000000);
\draw[color=black] (0.000000,195.000000) node[left] {$x$};
\draw[color=black] (0.000000,182.000000) -- (234.000000,182.000000);
\draw[color=black] (0.000000,182.000000) node[left] {$0$};
\draw[color=black] (0.000000,169.000000) -- (234.000000,169.000000);
\draw[color=black] (0.000000,169.000000) node[left] {$0$};
\draw[color=black] (0.000000,156.000000) -- (234.000000,156.000000);
\draw[color=black] (0.000000,156.000000) node[left] {$0$};
\draw[color=black] (0.000000,143.000000) -- (234.000000,143.000000);
\draw[color=black] (0.000000,143.000000) node[left] {$0$};
\draw[color=black] (0.000000,130.000000) -- (234.000000,130.000000);
\draw[color=black] (0.000000,130.000000) node[left] {$0$};
\draw[color=black] (0.000000,117.000000) -- (234.000000,117.000000);
\draw[color=black] (0.000000,117.000000) node[left] {$0$};
\draw[color=black] (0.000000,104.000000) -- (234.000000,104.000000);
\draw[color=black] (0.000000,104.000000) node[left] {$-2$};
\draw[color=black] (0.000000,91.000000) -- (234.000000,91.000000);
\draw[color=black] (0.000000,91.000000) node[left] {$0$};
\draw[color=black] (0.000000,78.000000) -- (234.000000,78.000000);
\draw[color=black] (0.000000,78.000000) node[left] {$0$};
\draw[color=black] (0.000000,65.000000) -- (234.000000,65.000000);
\draw[color=black] (0.000000,65.000000) node[left] {$0$};
\draw[color=black] (0.000000,52.000000) -- (234.000000,52.000000);
\draw[color=black] (0.000000,52.000000) node[left] {$0$};
\draw[color=black] (0.000000,39.000000) -- (234.000000,39.000000);
\draw[color=black] (0.000000,39.000000) node[left] {$0$};
\draw[color=black] (0.000000,26.000000) -- (234.000000,26.000000);
\draw[color=black] (0.000000,26.000000) node[left] {$0$};
\draw[color=black] (0.000000,13.000000) -- (234.000000,13.000000);
\draw[color=black] (0.000000,13.000000) node[left] {$0$};
\draw[color=black] (0.000000,0.000000) -- (234.000000,0.000000);
\draw[color=black] (0.000000,0.000000) node[left] {$0$};
\draw (9.000000,221.000000) -- (9.000000,169.000000);
\begin{scope}[rounded corners=2pt]
\begin{scope}
\draw[fill=white] (9.000000, 195.000000) +(-45.000000:8.485281pt and 45.254834pt) -- +(45.000000:8.485281pt and 45.254834pt) -- +(135.000000:8.485281pt and 45.254834pt) -- +(225.000000:8.485281pt and 45.254834pt) -- cycle;
\clip (9.000000, 195.000000) +(-45.000000:8.485281pt and 45.254834pt) -- +(45.000000:8.485281pt and 45.254834pt) -- +(135.000000:8.485281pt and 45.254834pt) -- +(225.000000:8.485281pt and 45.254834pt) -- cycle;
\draw (9.000000, 195.000000) node {{\rotatebox{-90}{sub}}};
\end{scope}
\end{scope}
\draw[color=black,dash pattern=on 2pt off 1pt] (3.000000, 208.000000) -- (15.000000, 208.000000);
\draw (27.000000,182.000000) -- (27.000000,143.000000);
\begin{scope}[rounded corners=2pt]
\begin{scope}
\draw[fill=white] (27.000000, 162.500000) +(-45.000000:8.485281pt and 36.062446pt) -- +(45.000000:8.485281pt and 36.062446pt) -- +(135.000000:8.485281pt and 36.062446pt) -- +(225.000000:8.485281pt and 36.062446pt) -- cycle;
\clip (27.000000, 162.500000) +(-45.000000:8.485281pt and 36.062446pt) -- +(45.000000:8.485281pt and 36.062446pt) -- +(135.000000:8.485281pt and 36.062446pt) -- +(225.000000:8.485281pt and 36.062446pt) -- cycle;
\draw (27.000000, 162.500000) node {{\rotatebox{-90}{square}}};
\end{scope}
\end{scope}
\draw[color=black,dash pattern=on 2pt off 1pt] (21.000000, 169.000000) -- (33.000000, 169.000000);
\draw (45.000000,208.000000) -- (45.000000,117.000000);
\begin{scope}[rounded corners=2pt]
\begin{scope}
\draw[fill=white] (45.000000, 162.500000) +(-45.000000:8.485281pt and 72.831998pt) -- +(45.000000:8.485281pt and 72.831998pt) -- +(135.000000:8.485281pt and 72.831998pt) -- +(225.000000:8.485281pt and 72.831998pt) -- cycle;
\clip (45.000000, 162.500000) +(-45.000000:8.485281pt and 72.831998pt) -- +(45.000000:8.485281pt and 72.831998pt) -- +(135.000000:8.485281pt and 72.831998pt) -- +(225.000000:8.485281pt and 72.831998pt) -- cycle;
\draw (45.000000, 162.500000) node {{\rotatebox{-90}{square}}};
\end{scope}
\end{scope}
\draw[color=black,dash pattern=on 2pt off 1pt] (39.000000, 195.000000) -- (51.000000, 195.000000);
\draw[color=black,dash pattern=on 2pt off 1pt] (39.000000, 182.000000) -- (51.000000, 182.000000);
\draw[color=black,dash pattern=on 2pt off 1pt] (39.000000, 169.000000) -- (51.000000, 169.000000);
\draw[color=black,dash pattern=on 2pt off 1pt] (39.000000, 156.000000) -- (51.000000, 156.000000);
\draw[color=black,dash pattern=on 2pt off 1pt] (39.000000, 143.000000) -- (51.000000, 143.000000);
\draw (63.000000,130.000000) -- (63.000000,78.000000);
\begin{scope}[rounded corners=2pt]
\begin{scope}
\draw[fill=white] (63.000000, 104.000000) +(-45.000000:8.485281pt and 45.254834pt) -- +(45.000000:8.485281pt and 45.254834pt) -- +(135.000000:8.485281pt and 45.254834pt) -- +(225.000000:8.485281pt and 45.254834pt) -- cycle;
\clip (63.000000, 104.000000) +(-45.000000:8.485281pt and 45.254834pt) -- +(45.000000:8.485281pt and 45.254834pt) -- +(135.000000:8.485281pt and 45.254834pt) -- +(225.000000:8.485281pt and 45.254834pt) -- cycle;
\draw (63.000000, 104.000000) node {{\rotatebox{-90}{mult}}};
\end{scope}
\end{scope}
\draw[color=black,dash pattern=on 2pt off 1pt] (57.000000, 117.000000) -- (69.000000, 117.000000);
\draw (81.000000,156.000000) -- (81.000000,52.000000);
\begin{scope}[rounded corners=2pt]
\begin{scope}
\draw[fill=white] (81.000000, 104.000000) +(-45.000000:8.485281pt and 82.024387pt) -- +(45.000000:8.485281pt and 82.024387pt) -- +(135.000000:8.485281pt and 82.024387pt) -- +(225.000000:8.485281pt and 82.024387pt) -- cycle;
\clip (81.000000, 104.000000) +(-45.000000:8.485281pt and 82.024387pt) -- +(45.000000:8.485281pt and 82.024387pt) -- +(135.000000:8.485281pt and 82.024387pt) -- +(225.000000:8.485281pt and 82.024387pt) -- cycle;
\draw (81.000000, 104.000000) node {{\rotatebox{-90}{div}}};
\end{scope}
\end{scope}
\draw[color=black,dash pattern=on 2pt off 1pt] (75.000000, 143.000000) -- (87.000000, 143.000000);
\draw[color=black,dash pattern=on 2pt off 1pt] (75.000000, 130.000000) -- (87.000000, 130.000000);
\draw[color=black,dash pattern=on 2pt off 1pt] (75.000000, 117.000000) -- (87.000000, 117.000000);
\draw[color=black,dash pattern=on 2pt off 1pt] (75.000000, 104.000000) -- (87.000000, 104.000000);
\draw[color=black,dash pattern=on 2pt off 1pt] (75.000000, 78.000000) -- (87.000000, 78.000000);
\draw (99.000000,65.000000) -- (99.000000,26.000000);
\begin{scope}[rounded corners=2pt]
\begin{scope}
\draw[fill=white] (99.000000, 45.500000) +(-45.000000:8.485281pt and 36.062446pt) -- +(45.000000:8.485281pt and 36.062446pt) -- +(135.000000:8.485281pt and 36.062446pt) -- +(225.000000:8.485281pt and 36.062446pt) -- cycle;
\clip (99.000000, 45.500000) +(-45.000000:8.485281pt and 36.062446pt) -- +(45.000000:8.485281pt and 36.062446pt) -- +(135.000000:8.485281pt and 36.062446pt) -- +(225.000000:8.485281pt and 36.062446pt) -- cycle;
\draw (99.000000, 45.500000) node {{\rotatebox{-90}{exp}}};
\end{scope}
\end{scope}
\draw[color=black,dash pattern=on 2pt off 1pt] (93.000000, 52.000000) -- (105.000000, 52.000000);
\draw (117.000000,234.000000) -- (117.000000,0.000000);
\begin{scope}[rounded corners=2pt]
\begin{scope}
\draw[fill=white] (117.000000, 117.000000) +(-45.000000:8.485281pt and 173.948268pt) -- +(45.000000:8.485281pt and 173.948268pt) -- +(135.000000:8.485281pt and 173.948268pt) -- +(225.000000:8.485281pt and 173.948268pt) -- cycle;
\clip (117.000000, 117.000000) +(-45.000000:8.485281pt and 173.948268pt) -- +(45.000000:8.485281pt and 173.948268pt) -- +(135.000000:8.485281pt and 173.948268pt) -- +(225.000000:8.485281pt and 173.948268pt) -- cycle;
\draw (117.000000, 117.000000) node {{\rotatebox{-90}{mult}}};
\end{scope}
\end{scope}
\draw[color=black,dash pattern=on 2pt off 1pt] (111.000000, 221.000000) -- (123.000000, 221.000000);
\draw[color=black,dash pattern=on 2pt off 1pt] (111.000000, 208.000000) -- (123.000000, 208.000000);
\draw[color=black,dash pattern=on 2pt off 1pt] (111.000000, 195.000000) -- (123.000000, 195.000000);
\draw[color=black,dash pattern=on 2pt off 1pt] (111.000000, 182.000000) -- (123.000000, 182.000000);
\draw[color=black,dash pattern=on 2pt off 1pt] (111.000000, 169.000000) -- (123.000000, 169.000000);
\draw[color=black,dash pattern=on 2pt off 1pt] (111.000000, 156.000000) -- (123.000000, 156.000000);
\draw[color=black,dash pattern=on 2pt off 1pt] (111.000000, 143.000000) -- (123.000000, 143.000000);
\draw[color=black,dash pattern=on 2pt off 1pt] (111.000000, 130.000000) -- (123.000000, 130.000000);
\draw[color=black,dash pattern=on 2pt off 1pt] (111.000000, 117.000000) -- (123.000000, 117.000000);
\draw[color=black,dash pattern=on 2pt off 1pt] (111.000000, 104.000000) -- (123.000000, 104.000000);
\draw[color=black,dash pattern=on 2pt off 1pt] (111.000000, 91.000000) -- (123.000000, 91.000000);
\draw[color=black,dash pattern=on 2pt off 1pt] (111.000000, 78.000000) -- (123.000000, 78.000000);
\draw[color=black,dash pattern=on 2pt off 1pt] (111.000000, 65.000000) -- (123.000000, 65.000000);
\draw[color=black,dash pattern=on 2pt off 1pt] (111.000000, 52.000000) -- (123.000000, 52.000000);
\draw[color=black,dash pattern=on 2pt off 1pt] (111.000000, 26.000000) -- (123.000000, 26.000000);
\draw (135.000000,65.000000) -- (135.000000,26.000000);
\begin{scope}[rounded corners=2pt]
\begin{scope}
\draw[fill=white] (135.000000, 45.500000) +(-45.000000:8.485281pt and 36.062446pt) -- +(45.000000:8.485281pt and 36.062446pt) -- +(135.000000:8.485281pt and 36.062446pt) -- +(225.000000:8.485281pt and 36.062446pt) -- cycle;
\clip (135.000000, 45.500000) +(-45.000000:8.485281pt and 36.062446pt) -- +(45.000000:8.485281pt and 36.062446pt) -- +(135.000000:8.485281pt and 36.062446pt) -- +(225.000000:8.485281pt and 36.062446pt) -- cycle;
\draw (135.000000, 45.500000) node {{\rotatebox{-90}{exp$^\dagger$}}};
\end{scope}
\end{scope}
\draw[color=black,dash pattern=on 2pt off 1pt] (129.000000, 52.000000) -- (141.000000, 52.000000);
\draw (153.000000,156.000000) -- (153.000000,52.000000);
\begin{scope}[rounded corners=2pt]
\begin{scope}
\draw[fill=white] (153.000000, 104.000000) +(-45.000000:8.485281pt and 82.024387pt) -- +(45.000000:8.485281pt and 82.024387pt) -- +(135.000000:8.485281pt and 82.024387pt) -- +(225.000000:8.485281pt and 82.024387pt) -- cycle;
\clip (153.000000, 104.000000) +(-45.000000:8.485281pt and 82.024387pt) -- +(45.000000:8.485281pt and 82.024387pt) -- +(135.000000:8.485281pt and 82.024387pt) -- +(225.000000:8.485281pt and 82.024387pt) -- cycle;
\draw (153.000000, 104.000000) node {{\rotatebox{-90}{div$^\dagger$}}};
\end{scope}
\end{scope}
\draw[color=black,dash pattern=on 2pt off 1pt] (147.000000, 143.000000) -- (159.000000, 143.000000);
\draw[color=black,dash pattern=on 2pt off 1pt] (147.000000, 130.000000) -- (159.000000, 130.000000);
\draw[color=black,dash pattern=on 2pt off 1pt] (147.000000, 117.000000) -- (159.000000, 117.000000);
\draw[color=black,dash pattern=on 2pt off 1pt] (147.000000, 104.000000) -- (159.000000, 104.000000);
\draw[color=black,dash pattern=on 2pt off 1pt] (147.000000, 78.000000) -- (159.000000, 78.000000);
\draw (171.000000,130.000000) -- (171.000000,78.000000);
\begin{scope}[rounded corners=2pt]
\begin{scope}
\draw[fill=white] (171.000000, 104.000000) +(-45.000000:8.485281pt and 45.254834pt) -- +(45.000000:8.485281pt and 45.254834pt) -- +(135.000000:8.485281pt and 45.254834pt) -- +(225.000000:8.485281pt and 45.254834pt) -- cycle;
\clip (171.000000, 104.000000) +(-45.000000:8.485281pt and 45.254834pt) -- +(45.000000:8.485281pt and 45.254834pt) -- +(135.000000:8.485281pt and 45.254834pt) -- +(225.000000:8.485281pt and 45.254834pt) -- cycle;
\draw (171.000000, 104.000000) node {{\rotatebox{-90}{mult$^\dagger$}}};
\end{scope}
\end{scope}
\draw[color=black,dash pattern=on 2pt off 1pt] (165.000000, 117.000000) -- (177.000000, 117.000000);
\draw (189.000000,208.000000) -- (189.000000,117.000000);
\begin{scope}[rounded corners=2pt]
\begin{scope}
\draw[fill=white] (189.000000, 162.500000) +(-45.000000:8.485281pt and 72.831998pt) -- +(45.000000:8.485281pt and 72.831998pt) -- +(135.000000:8.485281pt and 72.831998pt) -- +(225.000000:8.485281pt and 72.831998pt) -- cycle;
\clip (189.000000, 162.500000) +(-45.000000:8.485281pt and 72.831998pt) -- +(45.000000:8.485281pt and 72.831998pt) -- +(135.000000:8.485281pt and 72.831998pt) -- +(225.000000:8.485281pt and 72.831998pt) -- cycle;
\draw (189.000000, 162.500000) node {{\rotatebox{-90}{square$^\dagger$}}};
\end{scope}
\end{scope}
\draw[color=black,dash pattern=on 2pt off 1pt] (183.000000, 195.000000) -- (195.000000, 195.000000);
\draw[color=black,dash pattern=on 2pt off 1pt] (183.000000, 182.000000) -- (195.000000, 182.000000);
\draw[color=black,dash pattern=on 2pt off 1pt] (183.000000, 169.000000) -- (195.000000, 169.000000);
\draw[color=black,dash pattern=on 2pt off 1pt] (183.000000, 156.000000) -- (195.000000, 156.000000);
\draw[color=black,dash pattern=on 2pt off 1pt] (183.000000, 143.000000) -- (195.000000, 143.000000);
\draw (207.000000,182.000000) -- (207.000000,143.000000);
\begin{scope}[rounded corners=2pt]
\begin{scope}
\draw[fill=white] (207.000000, 162.500000) +(-45.000000:8.485281pt and 36.062446pt) -- +(45.000000:8.485281pt and 36.062446pt) -- +(135.000000:8.485281pt and 36.062446pt) -- +(225.000000:8.485281pt and 36.062446pt) -- cycle;
\clip (207.000000, 162.500000) +(-45.000000:8.485281pt and 36.062446pt) -- +(45.000000:8.485281pt and 36.062446pt) -- +(135.000000:8.485281pt and 36.062446pt) -- +(225.000000:8.485281pt and 36.062446pt) -- cycle;
\draw (207.000000, 162.500000) node {{\rotatebox{-90}{square$^\dagger$}}};
\end{scope}
\end{scope}
\draw[color=black,dash pattern=on 2pt off 1pt] (201.000000, 169.000000) -- (213.000000, 169.000000);
\draw (225.000000,221.000000) -- (225.000000,169.000000);
\begin{scope}[rounded corners=2pt]
\begin{scope}
\draw[fill=white] (225.000000, 195.000000) +(-45.000000:8.485281pt and 45.254834pt) -- +(45.000000:8.485281pt and 45.254834pt) -- +(135.000000:8.485281pt and 45.254834pt) -- +(225.000000:8.485281pt and 45.254834pt) -- cycle;
\clip (225.000000, 195.000000) +(-45.000000:8.485281pt and 45.254834pt) -- +(45.000000:8.485281pt and 45.254834pt) -- +(135.000000:8.485281pt and 45.254834pt) -- +(225.000000:8.485281pt and 45.254834pt) -- cycle;
\draw (225.000000, 195.000000) node {{\rotatebox{-90}{sub$^\dagger$}}};
\end{scope}
\end{scope}
\draw[color=black,dash pattern=on 2pt off 1pt] (219.000000, 208.000000) -- (231.000000, 208.000000);
\draw[color=black] (234.000000,234.000000) node[right] {$a$};
\draw[color=black] (234.000000,221.000000) node[right] {$b$};
\draw[color=black] (234.000000,208.000000) node[right] {$c$};
\draw[color=black] (234.000000,195.000000) node[right] {$x$};
\draw[color=black] (234.000000,182.000000) node[right] {$0$};
\draw[color=black] (234.000000,169.000000) node[right] {$0$};
\draw[color=black] (234.000000,156.000000) node[right] {$0$};
\draw[color=black] (234.000000,143.000000) node[right] {$0$};
\draw[color=black] (234.000000,130.000000) node[right] {$0$};
\draw[color=black] (234.000000,117.000000) node[right] {$0$};
\draw[color=black] (234.000000,104.000000) node[right] {$-2$};
\draw[color=black] (234.000000,91.000000) node[right] {$0$};
\draw[color=black] (234.000000,78.000000) node[right] {$0$};
\draw[color=black] (234.000000,65.000000) node[right] {$0$};
\draw[color=black] (234.000000,52.000000) node[right] {$0$};
\draw[color=black] (234.000000,39.000000) node[right] {$0$};
\draw[color=black] (234.000000,26.000000) node[right] {$0$};
\draw[color=black] (234.000000,13.000000) node[right] {$f$};
\draw[color=black] (234.000000,0.000000) node[right] {$0$};
\end{tikzpicture}}

  \subfloat[Overview of results.]{%
    \begin{tabularx}{.8\linewidth}{Xrrr}
      \toprule
      Design & qubits & $T$ gates & runtime \\
      \midrule
      Direct (a) & 6,355 & 8,960,228 & \emph{manual} \\
      Resynthesis ($k = 23$) & 6,283 & 1,850,001 & 11.28 \\
      Resynthesis ($k = 9$) & 8,124 & 982,417 & 9.67 \\
      \bottomrule
    \end{tabularx}
  }
  \caption{Direct implementation of the 32-bit Gaussian compared to resynthesis.}
  \label{fig:composition}
\end{figure}
The synthesis results of the floating point benchmarks can be used to
cost quantum algorithms.  This alone provides a useful tool to quantum
algorithm designers.  However, we show below that using these results
to synthesize a composed function can be sub-optimal.  This is
significant because several quantum algorithms require compositions
of several arithmetic
functions~\cite{HHL09,BBK+16b,Childs10,CNW:2010,BC12,ORR13}.  We show that by
using automatic synthesis with LHRS better quantum networks can be
found relative to na\"ively summing the costs of the constituent
functions in Table~\ref{tab:fpl}.  To demonstrate this effect, we use
a 32-bit implementation of the Gaussian
\begin{equation}
  \label{eq:guassian}
  f(x) = ae^{-\frac{(x-b)^2}{2c^2}},
\end{equation}
where besides $x$ also $a, b$, and $c$ are 32-bit inputs to the
quantum circuit.  We focus on this function because of its importance
to quantum chemistry simulation algorithms, wherein the best known
simulation methods for solving electronic structure problems within a
Gaussian basis need to be able to reversibly compute
\eqref{eq:guassian}~\cite{BBK+16b}.  The overheads from compiling
Gaussians using conventional techniques has, in particular, rendered
these simulation methods impractical.  This makes improved synthesis
methods critical for such applications.

The function has been used in the design of quantum algorithms.  We
can implement the Gaussian by combining the 32-bit versions of the
floating point components (synthesized using $k=16$) in
Table~\ref{tab:fpl}.  This leads to a quantum network as shown in
Fig.~\ref{fig:composition}(a).  Note that we use the multiplication
component with a constant input to realize the denominator in the
argument of the exponent.  Also, we make sure to uncompute all helper
lines.  This leads to a quantum network with 6\,355 qubits and
8\,960\,228 $T$ gates.

We also implemented the Gaussian directly in Verilog, optimized the
resulting design with logic synthesis (as we did for the individual
components in Table~\ref{tab:fpl}), and synthesized it using LHRS.  To
match the quality of the design in Fig.~\ref{fig:composition}(a), we
used `synthesize\_mapping' (see Alg.~\ref{alg:lhrs}) to find the
smallest $k$ that leads to a number of qubits smaller than 6\,355.  In
this case, $k$ is 23.  With this $k$, we synthesize the composed
Guassian design using the same parameters $p_{\mathrm{T}}$ as used for
synthesizing the components.  The result is a quantum network with
6,283 qubits and 1,850,001 $T$ gates, which can be synthesized within
11.28 seconds (see also Fig.~\ref{fig:composition}(b), which
summarizes all results of this experiment).

The experiment demonstrated that by resynthesizing composed functions,
better networks and better cost estimates can be achieved.  The
approach also easily enables design space exploration. For example, if
one is interested in a quantum network with less than 1,000,000 $T$
gates, one can find a realization with 8,124 qubits and 982,417 $T$
gates, after 9.67 seconds by setting $k$ to~$9$.

It can be seen that LHRS finds quantum circuits with much better
qubits/$T$ gates tradeoff.  Further, LHRS allows for a better
selection of results by using the LUT size as a parameter.  One strong
advantage is that in LHRS one can quickly obtain a skeleton for the
final circuit in terms of single-target gates that already has the
final number of qubits.  If this number matches the design
constraints, one can start the computational more challenging task of
finding good quantum circuits for each LUT function.  Here, several
synthesis passes trying different parameter configurations are
possible in order to optimize the result.  Also post-synthesis
optimization techniques likely help to significantly reduce the number
of $T$ gates.

\section{Conclusion}
\label{sec:conclusion}
We presented LHRS, a LUT-based approach to hierarchical reversible
circuit synthesis that outperforms existing state-of-the-art
hierarchical methods.  It allows for much higher flexibility
addressing the needs to trade-off qubits to $T$-count when generating
high quality quantum networks.  The benchmarks that we provide give
what is at present the most complete list of costs for elementary
functions for scientific computing.  Apart from simply showing
improvements, these benchmarks provide cost estimates that allow
quantum algorithm designers to provide the first complete cost
estimates for a host of quantum algorithms.  This is an essential step
towards the goal of understanding which quantum algorithms will be
practical in the first generations of quantum computers.

LHRS can be regarded as a synthesis framework since it consists of
several parts that can be optimized separately.  As one example, we
are currently investigating more advanced mapping strategies that map
single-target gates into Clifford+$T$ networks.  Also, most of the
conventional synthesis approaches that are used in the LHRS flow,
e.g., the mapping algorithms to derive the $k$-LUT network, are not
quantum-aware, i.e., they do not explicitly optimize wrt.\ the quality
of the resulting quantum network.  We expect further improvements,
particularly in the number of $T$ gates, by modifying the synthesis
algorithms in that direction.

\subsubsection*{Acknowledgment}
We wish to thank Matthew Amy, Olivia Di Matteo, Vadym Kliuchnikov,
Giulia Meuli, and Alan Mishchenko for many useful discussions.  All
circuits in this paper were drawn with the \emph{qpic} tool
\cite{qpic}.  This research was supported by H2020-ERC-2014-ADG 669354
CyberCare, the Swiss National Science Foundation (200021-169084
MAJesty), and the ICT COST Action IC1405.



\end{document}